\documentclass[a4paper,fleqn,usenatbib]{mnras}

\usepackage{newtxtext,newtxmath}
\usepackage{mathptmx}
\usepackage{txfonts}

\usepackage[T1]{fontenc}
\usepackage{ae,aecompl}

\usepackage{graphicx}	
\usepackage{amsmath}	
\usepackage{amssymb}	
\usepackage{textcomp}
\usepackage{pdflscape}
\usepackage{color}
\usepackage{ulem}
\usepackage{newtxtext,newtxmath}
\usepackage{mathtools}  
\usepackage{nicefrac}  
\usepackage{xcolor}

\title[Early-type group-dominant galaxies: ionised gas]{Spatially-resolved properties of early-type group-dominant galaxies with MUSE: gas content, ionisation mechanisms and metallicity gradients}

\author[P. Lagos et al.]{P. Lagos,$^{1,2}$\thanks{E-mail: Patricio.Lagos@astro.up.pt} S. I. Loubser,$^{2,3}$ T. C. Scott,$^{1}$ E. O'Sullivan,$^{4}$ K. Kolokythas,$^{2}$ A. Babul,$^{5}$
\newauthor A.  Nigoche-Netro,$^{6}$ V. Olivares, $^{7,8}$ and C. Sengupta$^{9}$
\\
$^{1}$Instituto de Astrof\'isica e Ci\^encias do Espa\c{c}o, Universidade do Porto, CAUP, Rua das Estrelas, 4150-762 Porto, Portugal\\
$^{2}$Centre for Space Research, North-West University, Potchefstroom 2520, South Africa\\
$^{3}$National Institute for Theoretical and Computational Sciences (NITheCS), Potchefstroom 2520, South Africa\\
$^{4}$Center for Astrophysics | Harvard \& Smithsonian, 60 Garden Street, Cambridge, MA 02138, USA\\
$^{5}$Department of Physics and Astronomy, University of Victoria, Victoria, BC, V8W 2Y2, Canada\\
$^{6}$Instituto de Astronom\'ia y Meteorolog\'ia, Av, Vallarta 2602. Col. Arcos Vallarta. Guadalajara, Jalisco. C.P. 44130 M\'exico\\
$^{7}$LERMA, Observatoire de Paris, PSL Research Univ., CNRS, Sorbonne Univ., 75014 Paris, France\\
$^{8}$Department of Physics and Astronomy, University of Kentucky, 505 Rose Street, Lexington, KY 40506, USA\\
$^{9}$Purple Mountain Observatory, No. 8 Yuanhua Road, Qixia District, Nanjing 210034, China
}

\date{Accepted XXX. Received YYY; in original form ZZZ}

\pubyear{2015}

\begin{document}
\label{firstpage}
\pagerange{\pageref{firstpage}--\pageref{lastpage}}
\maketitle

\begin{abstract}
With the goal of a thorough investigation of the ionised gas and its origin  
in early-type group-dominant galaxies, we present archival MUSE data for 18 galaxies 
from the Complete Local-Volume Groups Sample (CLoGS). 
This data allowed us to study the spatially-resolved warm gas properties, including
the morphology of the ionised gas, EW(H$\alpha$) and kinematics as well as the gas-phase
metallicity (12 + log(O/H)) of these systems. In order to distinguish between different ionisation
mechanisms, we used the emission-line ratios [O \,{\sc iii}]/H$\beta$ and [N\,{\sc ii}]/H$\alpha$ 
in the BPT diagrams and EW(H$\alpha$). 
We find that the ionisation sources in our sample have variable impacts at
different radii, central regions are more influenced by low-luminosity AGN,
while extended regions of LINER-like emission are ionised by other mechanisms 
with pAGBs photoionisation likely contributing significantly. We classified our sample into three 
H$\alpha$+[N \,{\sc ii}] emission morphology types.
We calculate the gas-phase metallicity assuming several methods and ionisation sources.
In general, 12 + log(O/H) decreases with radius from the centre for all galaxies, 
independently of nebular morphology type, indicating a metallicity gradient in the abundance profiles.
Interestingly, the more extended filamentary structures and all extranuclear star-forming regions 
present shallow metallicity gradients. Within the uncertainties these extended structures
can be considered chemically homogeneous.
We suggest that group-dominant galaxies in our sample likely acquired their cold gas
in the past as a consequence of one or more mechanisms, e.g. gas-clouds or satellite
mergers/accretion and/or cooling flows that contribute to the growth of the ionised gas structures.
\end{abstract}


\section{Introduction}\label{sec:Intro}

Most galaxies are found in groups and clusters. The distinction between them
is made by the number of their members, clusters have $\gtrsim$50 member galaxies and
are more massive than groups. Bright galaxies at the centres of galaxy groups (hereafter BGGs) 
or group-dominant galaxies typically have stellar masses of M$_{*}$ $\gtrsim$10$^{10-13}$
M$_{\odot}$ \citep[e.g.,][]{Gozaliasl2016,OSullivan2018,Kolokythas2022}, 
with some BGGs also displaying X-ray emission \cite[e.g.,][]{OSullivan2017} 
and containing cold H\,{\sc i} and/or molecular gas in the range $\sim$10$^{8-10}$ M$_{\odot}$
\citep[e.g.,][]{OSullivan2015,OSullivan2018}.
A significant fraction \citep[$\sim$25\%;][]{Gozaliasl2016} of BGGs lie on the main sequence of
the star-forming (late-type) galaxies at z $\lesssim$ 0.4, and this fraction increases with
redshift. However, most BGGs are early-type galaxies (ETGs), i.e., elliptical and/or
lenticular (S0) galaxies. 
Early-type BGGs \citep[e.g.,][]{Kolokythas2022}, and many other ETGs in the literature 
\citep[e.g.,][]{Shapiro2010,Fang2012,Gomes2016}, contain warm ionised gas and ongoing 
low-level star-formation (SF) \citep[SFR$_{FUV} \sim$0.01--0.4 M$_{\odot}$ yr$^{-1}$;][]{Kolokythas2022},
while in late-type BGGs the SF can reach $\sim$10 M$_{\odot}$ yr$^{-1}$ \citep{OlivaAltamirano2014,Gozaliasl2016}.
This same activity is also found in a significant fraction of Brightest Cluster Galaxies 
\citep[BCGs; e.g.,][]{Bildfell2008,Pipino2009,McDonald2010,Donahue2011,McDonald2011,Werner2014,Loubser2016}.

ETGs \citep[e.g.,][]{Phillips1986,TrinchieridiSeregoAlighieri1991,Annibali2010}, including some 
early-type BGGs\footnote{See the list of galaxies in this paper.} also display low-ionisation
nuclear emission-line regions \cite[LINERs,][]{Heckman1980}.
The spectra of LINERs exhibit strong low-ionisation optical emission lines such as 
[O \,{\sc iii}]$\lambda$5007, [O \,{\sc ii}]$\lambda\lambda$3726/29, [N \,{\sc ii}]$\lambda$6584, 
[S \,{\sc ii}]$\lambda\lambda$6717,6731 and hydrogen Balmer lines (H$\alpha$, H$\beta$, etc),
commonly with  [N \,{\sc ii}]$\lambda$6584 > H$\alpha$
emission. Several explanations for LINER emission, in galaxies, have been proposed
including shock-ionisation \citep[e.g.,][]{Heckman1980,DopitaSutherland1995,DopitaSutherland1996,Allen2008}, 
photoionisation by ongoing low-level SF \citep[e.g.,][]{Schawinski2007,Fogarty2015,Shapiro2010},
hot evolved post-asymptotic giant branch (pAGB) stars 
\citep[e.g.,][]{Binette1994,Stasinska2008,Annibali2010,Sarzi2010,CidFernandes2010,CidFernandes2011}, 
photoionisation by the "cooling" X--ray--emitting gas \citep{VoitDonahue1990,DonahueVoit1991, Ferland1994,Voit1994,Polles2021}, 
thermal conduction from the hot gas \citep{Sparks1989} and ionisation by low-luminosity 
active galactic nuclei (AGN) \citep[e.g.,][]{FerlandNetzer1983,Barth1998,Ho1999}. 
Observational evidence suggests that more than one mechanism are likely to be at work in LINERs
\cite[e.g,][in BCGs]{Edwards2007,Fogarty2015}. 
For instance, the contribution of AGN to the photoionisation of LINERs may be restricted 
to the nuclear region \citep[e.g.,][]{Sarzi2010}, while pAGB stars are likely responsible 
for the extended gas emission \citep[e.g.,][]{Binette1994,Sarzi2010,Gomes2016}. 
Extended low-ionisation emission line regions \citep[LIERs;][]{Belfiore2016} or LINER-like 
extranuclear diffuse ionised gas (DIG) emission in ETGs has been detected and studied 
\citep[e.g.,][]{Sarzi2006,Gomes2016} as part of the SAURON \citep{Bacon2001} and Calar Alto 
Legacy Integral Field Area \citep[CALIFA,][]{Sanchez2012} surveys. 
These studies suggests low-luminosity AGN are likely contributing to LINER emission in a less 
than dominant way \citep{YanBlanton2012}.

Extended and DIG filaments are observed in BGGs/BCGs 
\citep[e.g.,][]{Heckman1989,Crawford1999,OSullivan2012,Tremblay2018,Olivares2019}. 
These ionised gas filaments seem to be co-spatial with large quantities
of H\,{\sc i} and molecular gas, traced by CO, and soft X--ray emission features \citep{Werner2014}. 
\cite{McDonald2010} found that the extended warm gas in cluster galaxies, 
in general, is spatially coincident with cooling intracluster medium (ICM) flows.
\cite{McDonald2011,McDonald2012} rule out collisional ionisation by cosmic rays, thermal
conduction, photoionisation by ICM X-rays and AGN as strong contributors to the ionisation 
of the warm gas, in both the nuclei and filaments of cool core clusters. 
They argue that the data are adequately described by a composite model of slow shocks and SF. 
According to \cite{McDonald2011}, the H$\alpha$ filaments in BGGs/BCGs are more strongly 
correlated with the cooling properties of the ICM than with the radio properties of the BCGs. 
The molecular gas in the filaments is located along old radio jets, lobes and/or below
X-ray cavities, suggesting that AGN activity supplies a regular input of energy in those systems 
\citep[][]{Russell2016,OSullivan2021}. Generally, the dominant ionisation mechanism, as well as
their locations (i.e., nuclear and/or extended regions), is not clear and different processes may
be at work during the evolution of each type of system. 

Hot (10$^7$ K) gas-phase metallicity is another key property for studying the effect of galactic 
environments and AGN/SF feedback in groups and clusters \citep[for a review see][]{Gastaldello2021}. 
The X-ray emitting intra-group medium (IGrM) shows several emission lines
typical of elements synthesised by stars/supernovae (SNe) and deposited in the IGrM/ICM via
galactic winds from the group/cluster members \citep[e.g.,][]{Liang2016}, ram-pressure 
stripping/tidal interactions \citep[][]{TaylorBabul2001,McCarthy2008} and others 
\citep[for a review see][]{SchindlerDiaferio2008}.
Several studies \cite[e.g.,][and references therein]{Mernier2017,Gastaldello2021} show that 
the average abundance profiles of Fe and others elements (e.g., O, Si, Ar) in the ICM/IGrM increase 
towards the core of cool-core clusters/groups up to values of $\sim$1.0 Z$_{\odot}$, 
and decrease at large radii. This peaked distribution is associated with the release of metals
from the galaxy members via mechanical feedback from AGN/SF activity, gas turbulence
\citep[][]{Rennehan2019} or infalling galaxies through ram-pressure stripping across the evolution 
of the systems. Whereas some studies \cite[e.g.,][]{Werner2006} report a rather flat O and Mg profiles,
non-cool-core clusters and groups do not exhibit clear Fe abundance gradient in their cores. 
Mechanisms such as merger–induced sloshing motions \cite[][and references therein]{OSullivan2014}
can also transport the released metals from the galaxies into the IGrM/ICM. 
On the other hand, the spatial distribution of abundances of the warm (10$^4$ K) gas-phase
in the interstellar medium (ISM) of the galaxies is also important for understanding 
the formation history of these systems.
Studies of local star-forming galaxies have shown a negative metallicity
gradient (the metallicity decreases radially) with increasing galactocentric 
radius \cite[e.g.,][among many others]{MartinRoy1992,vanZee1998}.
However, this radial distribution can present deviations from this simple negative gradient, 
namely a steep decrease of oxygen abundance (12 + log(O/H)) in the nuclear regions 
and a flattening of the gradient in the outer parts 
\cite[see][and references therein]{SanchezMenguiano2018,Sanchez2020}. 
Several mechanisms have been proposed to explain these features such as radial migration, infall 
of gas or satellite accretion. In early-type group-dominant galaxies, the presence of these 
features and the origin of their gas are poorly understood \cite[e.g.,][]{Annibali2010}. 
The cold gas in group and cluster ETGs potentially originates from two main sources: 
internal production \citep[a large fraction of stellar mass is believed to be released in the 
form of cold gas,][]{Davis2011} and external accretion, i.e., mergers, stripping gas from
satellites, gas from the IGrM/ICM \citep[e.g.,][]{Jung2022}. 
A kinematical decoupling between gas and stars as evidence for external gas origin was found in
many ETGs in the SAURON \citep{Sarzi2006,Sarzi2010} and CALIFA \citep{Gomes2016} surveys.

CLoGS \citep{OSullivan2017} is an optically-selected sample of 53 groups at distances 
of less than 80 Mpc and is the first statistically complete sample of nearby galaxy groups 
studied with an extensive multi-wavelength dataset, including X-ray 
\citep[Chandra and XMM-Newton;][]{OSullivan2017}, radio \citep[GMRT and VLA;][]{Kolokythas2018, Kolokythas2019} 
and sub-mm \citep[IRAM-30m and APEX;][]{OSullivan2018} observations. 
In addition, we have analysed spatially-resolved long-slit spectroscopy from 
the 10m Hobby-Ebberly Telescope \citep[][]{vandenBosch2015} for a sub-sample of 23 of the CLoGS
dominant central galaxies \citep[see][]{Loubser2018}. 
Archival \textit{GALEX} FUV and \textit{WISE} photometry for
the group-dominant galaxies are presented in \cite{Kolokythas2022}. 
In this paper, we use archival optical ESO Very Large Telescope (VLT) Integral Field Unit (IFU)
Multi-Unit Spectroscopic Explorer (MUSE) observations of a sample of 18 
early-type group-dominant CLoGS galaxies. 
We aim to investigate the relation between the properties (i.e., emission line ratios,
chemical abundances, etc) and structure of each galaxy's ISM in order to constrain
the ionisation processes, the origin of their gas and its chemical abundance
distribution. In related papers, using the same MUSE dataset, \cite{Olivares2022} presents a
detailed analysis of the ionised gas kinematics, while in \cite{Loubser2022} we study
the stellar kinematics of the fast and slow rotators in this sample.
The paper is organized as follows: Section \ref{sec:data} contains the technical
details regarding the observations, data reduction, spectral fitting and measurement of line
fluxes; Section \ref{sec:results} describes our results on the structure and
physical properties of the ionised gas; Section \ref{sec:discussion} discusses the results.
Finally, in Section \ref{sec:conclusions} we summarise our conclusions.
In the following analysis we adopt as solar abundances Z$_{\odot}$ = 0.0142, X$_{\odot}$ =
0.7154, and [12 + log(O/H)]$_{\odot}$ = 8.69 from \cite{Asplund2009}.

\section{The data and emission line measurements}\label{sec:data}

\subsection{Observations and data reduction}\label{sec:observations}

Our sample consists of 18 group-dominant galaxies selected from the CLoGS 
nearby groups sample \citep{OSullivan2017}. 
Each galaxy has archival ESO VLT/MUSE \citep{Bacon2010} data. 
Four of our targets were previously observed in the CALIFA IFU survey 
(NGC 677, NGC 924, NGC 1060 and NGC 7619). The observations (program ID 097.A-0366(A)) 
were made with the MUSE spectrograph, which used the Nasmyth B focus on Yepun the 8.2m  VLT UT4. 
The observations were made in the wide field mode over a
1$'\times$1$'$ Field of View (FoV) sampled by a system of 24 spectrographs with an spectral
coverage between $\sim$4800 and 9300 $\rm \AA$, a spectral resolution of $\sim$2.6 $\rm \AA$,
spectral sampling of 1.25 $\rm \AA$ and spatial sampling of 0.2$''$ per spaxel, leading to 
a total of 90,000 spectra per exposure. For each galaxy three exposures were taken.
These observations were dithered with a few arcsecond dither pattern in order to cover 
a large portion of each galaxy's ISM. Offset sky observations were used for sky subtraction.
In Table \ref{table:1} we show the general parameters for the sample members and the observation
log. 

The data were reduced using the MUSE pipeline \citep[v2.6.2,][]{Weilbacher2020}, with steps
including: bias correction, wavelength calibration, construction of datacubes from the individual
spectra from the detectors, correction of the spectra to the heliocentric frame, skysubtraction,
and merging of the individual exposures to form a combined datacube with a total of $\sim$200,000 
spectra per each galaxy. Finally, the data cubes were corrected for redshift and
galactic extinction using the reddening function given by \cite{Cardelli1989}.
In Figure \ref{fig:figures_spectra} we show the nuclear spectra of each member of our sample. 
More details in Section \ref{sec:spec_fitting}.

\begin{table*}
\hspace*{-0.7cm}
\begin{minipage}{190mm}
\caption{General parameters of our sample and observing log. Col. (1) gives the galaxy name,
Col. (2) the morphology, Cols. (3) and (4) give their coordinates, Col. (5) presents the galactic
extinction A$_{V}$ (mag) \citep{SchlaflyFinkbeiner2011} and Cols. (6)-(7) gives the distance and
redshift, respectively. D (Local Group) and z from NED assuming H$_o$ = 73.0 km/sec/Mpc. In Col. (8) 
we give X-ray morphology (extent of the gas halo) and core
type as cool-core/non-cool-core (CC/NCC) based on their temperature profiles \citep{OSullivan2017} 
for systems where thermal emission was detected: group-like (GRP, extent >65 kpc), galaxy-like
(gal, extent $\sim$10-65 kpc) and point-like (pnt, unresolved, extent smaller than the XMM-Newton
PSF). Radio morphology \citep{Kolokythas2018}: point-like (pnt, radio source with sizes 
$\lesssim$11 kpc), diffuse emission (diffuse, with no clear jet or lobe structure), small-scale
jets (jet, <20 kpc jets confined within the stellar body of the host galaxy) and large-scale jets
(JET, >20 kpc jets extending beyond the host galaxy and into the IGrM.
Finally, Col. (9)-(11) shows the observation date, exposure time and mean airmass during the
observations.}      
\label{table:1}      
\centering                         
\begin{tabular}{l c c c c c c c c c c c c} 
\hline
Name     & Morph.  & RA      & DEC     & A$_{V}$ & D & z  & X--ray / Radio & Date      & Exp. time    & Mean  & \\
         &         & (J2000) & (J2000) & (mag)   & (Mpc) & & morphology & of observation & (s) &  airmass  & \\
(1)      &  (2)    &  (3)    & (4)     &  (5)    & (6)   & (7) & (8)            & (9)      & (10) & (11)\\
\hline 

NGC 193      & E    &00h39m18.6s & +03d19m52s & 0.062 & 62.7 & 0.014723 & GRP (CC)  / JET & 2016-06-08 & 3$\times$900 & 1.524\\
NGC 410      & E    &01h10m58.9s & +33d09m07s & 0.161 & 75.8 & 0.017659 & GRP (CC)  / pnt & 2016-08-16 & 3$\times$900 & 1.937\\
NGC 584      & E    &01h31m20.7s & -06d52m05s & 0.116 & 25.9 & 0.006011 & \dots     / pnt & 2016-07-01 & 3$\times$900 & 1.243\\
NGC 677      & E    &01h49m14.0s & +13d03m19s & 0.243 & 71.9 & 0.017012 & GRP (CC)  / diffuse & 2016-07-20 & 3$\times$900   & 1.344\\
NGC 777      & E    &02h00m14.9s & +31d25m46s & 0.128 & 71.4 & 0.016728 & GRP (NCC) / pnt & 2016-08-16 & 3$\times$900 & 1.823\\
NGC 924      & S0   &02h26m46.8s & +20d29m51s & 0.467 & 63.1 & 0.014880 & \dots     / pnt  & 2016-08-14 & 3$\times$900 & 1.651\\
NGC 940      & S0   &02h29m27.5s & +31d38m27s & 0.246 & 72.6 & 0.017075 & pnt       / pnt  & 2016-08-21 & 3$\times$900 & 1.814\\
NGC 978      & E/S0 &02h34m47.6s & +32d50m37s & 0.253 & 67.3 & 0.015794 & gal (NCC) / pnt & 2016-08-16 & 3$\times$900 & 1.866\\
NGC 1060     & E/S0 &02h43m15.0s & +32d25m30s & 0.532 & 73.4 & 0.017312 & GRP (CC)  / jet & 2016-08-22 & 3$\times$900 & 1.845\\
NGC 1453     & E    &03h46m27.2s & -03d58m08s & 0.289 & 52.9 & 0.012962 & GRP (NCC) / pnt & 2016-07-22 & 3$\times$900 & 1.358\\
NGC 1587     & E    &04h30m39.9s & +00d39m42s & 0.197 & 50.0 & 0.012322 & GRP (NCC) / diffuse & 2016-08-18 & 3$\times$900 & 1.360\\
NGC 4008     & E    &11h58m17.0s & +28d11m33s & 0.064 & 49.0 & 0.012075 & gal (NCC) / pnt & 2016-05-13 & 3$\times$900 & 1.668\\
NGC 4169     & S0   &12h12m18.8s & +29d10m46s & 0.058 & 51.4 & 0.012622 & pnt       / pnt & 2016-05-19 & 3$\times$900 & 1.728\\
NGC 4261     & E    &12h19m23.2s & +05d49m31s & 0.049 & 28.4 & 0.007378 & GRP (CC)  / JET & 2016-04-17 & 3$\times$880& 1.311\\
ESO0507-G025 & E/S0 &12h51m31.8s & -26d27m07s & 0.245 & 41.1 & 0.010788 &\dots      / diffuse & 2016-04-12 &  3$\times$900 & 1.107\\
NGC 5846     & E    &15h06m29.3s & +01d36m20s & 0.153 & 23.1 & 0.005711 & GRP (CC)  / jet & 2016-05-16 & 3$\times$880& 1.579\\
NGC 6658     & S0   &18h33m55.6s & +22d53m18s & 0.339 & 61.6 & 0.014243 & pnt       / \dots & 2016-04-23 & 3$\times$870& 1.557\\
NGC 7619     & E    &23h20m14.5s & +08d12m22s & 0.224 & 54.6 & 0.012549 & GRP (CC)  / pnt & 2016-05-23 & 3$\times$870& 1.444\\
\hline                                   
\end{tabular}
\end{minipage}
\end{table*}

\subsection{Spectral fitting and emission-line measurement}\label{sec:spec_fitting}

\begin{figure}
\hspace*{-0.45cm}
\includegraphics[width=0.5\textwidth]{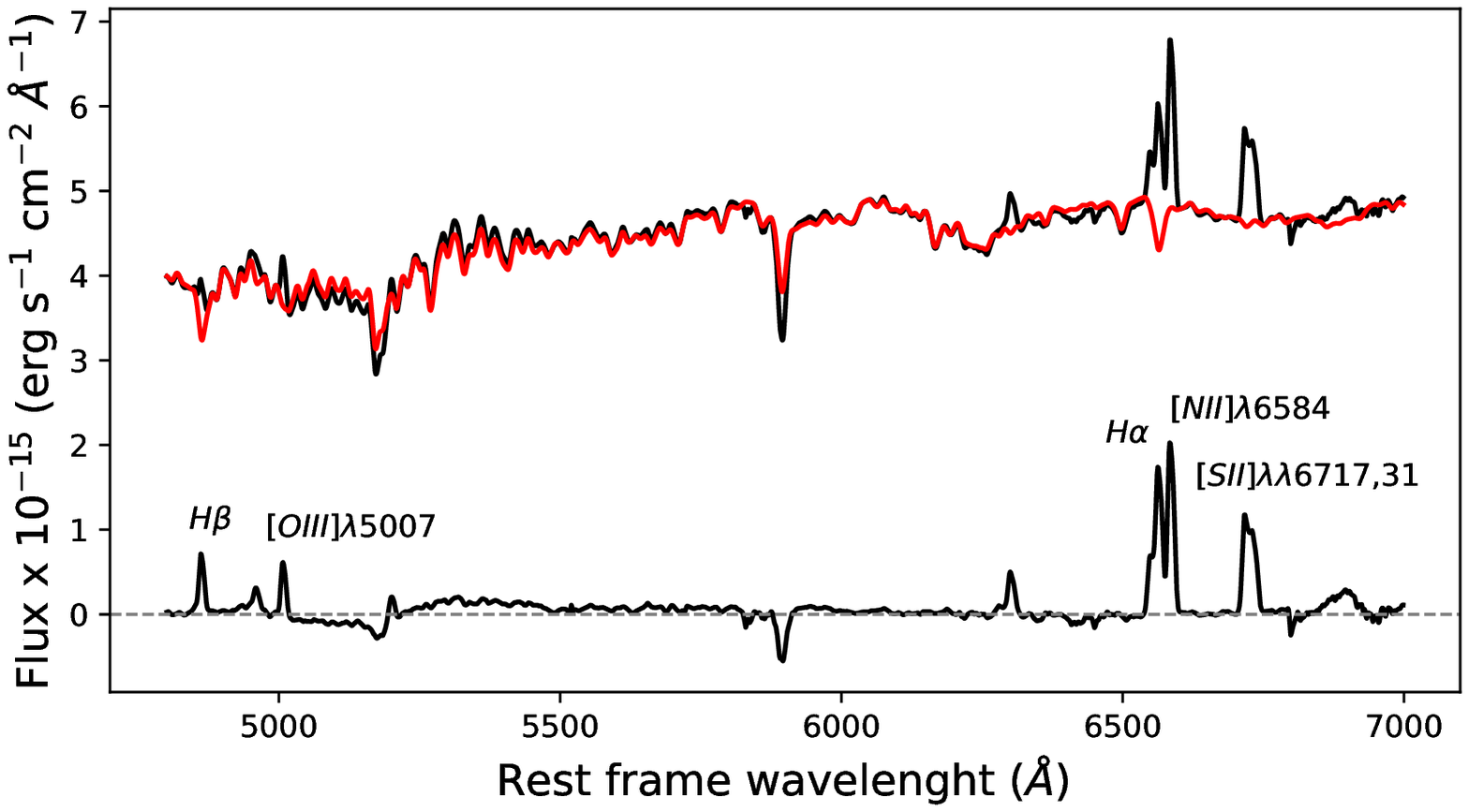}
    \caption{Example of the stellar continuum fit modelling with FADO for the galaxy
    ESO0507-G025 from the nuclear 3" aperture. The black and red upper lines correspond 
    to the observed and modeled stellar spectrum, respectively. 
    By subtracting the latter from the former we obtain the pure nebular emission line spectrum 
    (lower black line). We have labelled the emission lines used in our analysis.}
    \label{fig:Spectrum_example_fit}
\end{figure}

We fitted and removed the stellar component within each spectrum (bin or spaxel), in order to
obtain pure emission line spectra using the spectral synthesis code Fitting Analysis
using Differential evolution Optimization \citep[FADO,][]{GomesPapaderos2017}
version V1.B. Simple stellar population (SSP) templates from the \cite{BruzualCharlot2003} 
libraries were used with metallicities between Z=0.004 and Z=0.05 and stellar ages between 1 Myr
and 15 Gyr, for a Chabrier initial mass function \citep[IMF,][]{Chabrier2003}. 
FADO self-consistently reproduces the nebular characteristics of a galaxy, including 
the hydrogen Balmer-line luminosities, equivalent widths and nebular continuum.
An important advantage of FADO is that its convergence scheme 
employs genetic differential evolution optimization.
This results in improvements with respect to the uniqueness of spectral fits and 
the overall efficiency of the convergence schemes. Artificial intelligence is used to eliminate
redundancies in the SSP base libraries which increases the computational efficiency.
The fit was performed in the 4800-7000 $\AA$ spectral range assuming the \cite{Cardelli1989}
extinction law. The data cubes were tessellated using the Voronoi binning method
\citep{CappellariCopin2003} to achieve an adequate signal to noise 
(S/N; we choose a S/N$\sim$30-50) per bin for the continuum between 6000 to 6200 $\AA$. 
The Voronoi binning method lead to the dilution of the emission lines within the bins, so the shape
of extended and diffuse emission lines was lost 
\citep[see][for a similar approach]{ErrozFerrer2019}. Then, the best fit stellar continuum 
or synthetic Spectral Energy Distribution (SED) was subtracted from the observed spaxels, with S/N $>$ 3 
in the continuum, to obtain pure emission-line data cubes.  
FADO provides as results, among others, mass contributions of individual stellar populations, 
mass- and luminosity-weighted stellar ages and metallicities, emission-line fluxes, equivalent 
widths (EWs), FWHMs and estimates of their uncertainties. 
In Figure \ref{fig:Spectrum_example_fit} we show an example to illustrate the spectral
modelling results with FADO from a nuclear 3" aperture for the galaxy ESO0507-G025.
In order to test the accuracy of these results, we carried out a single-Gaussian fit to each emission line. 
Our fit results were in good agreement with those from FADO, 
consequently for the analysis in this study we create the following emission line maps using 
FADO: H$\alpha$, H$\beta$, [O \,{\sc iii}]$\lambda$5007, [N \,{\sc ii}]$\lambda$6584 and 
[S \,{\sc ii}] $\lambda\lambda$6717, 6731. 
We note that, when deriving the maps we only use spaxels with emission fluxes $>$3$\sigma$ 
above the background. We did not detect significant broad asymmetric emission-line profiles, 
so multiple-component decomposition of the emission lines was not necessary, because each 
emission line is well described by a single Gaussian profile. 
More details about the morphologies of these maps is given in Section \ref{subsec:emission}. 

\begin{figure*}
\vspace*{-0.45cm}
\includegraphics[width=0.9\textwidth]{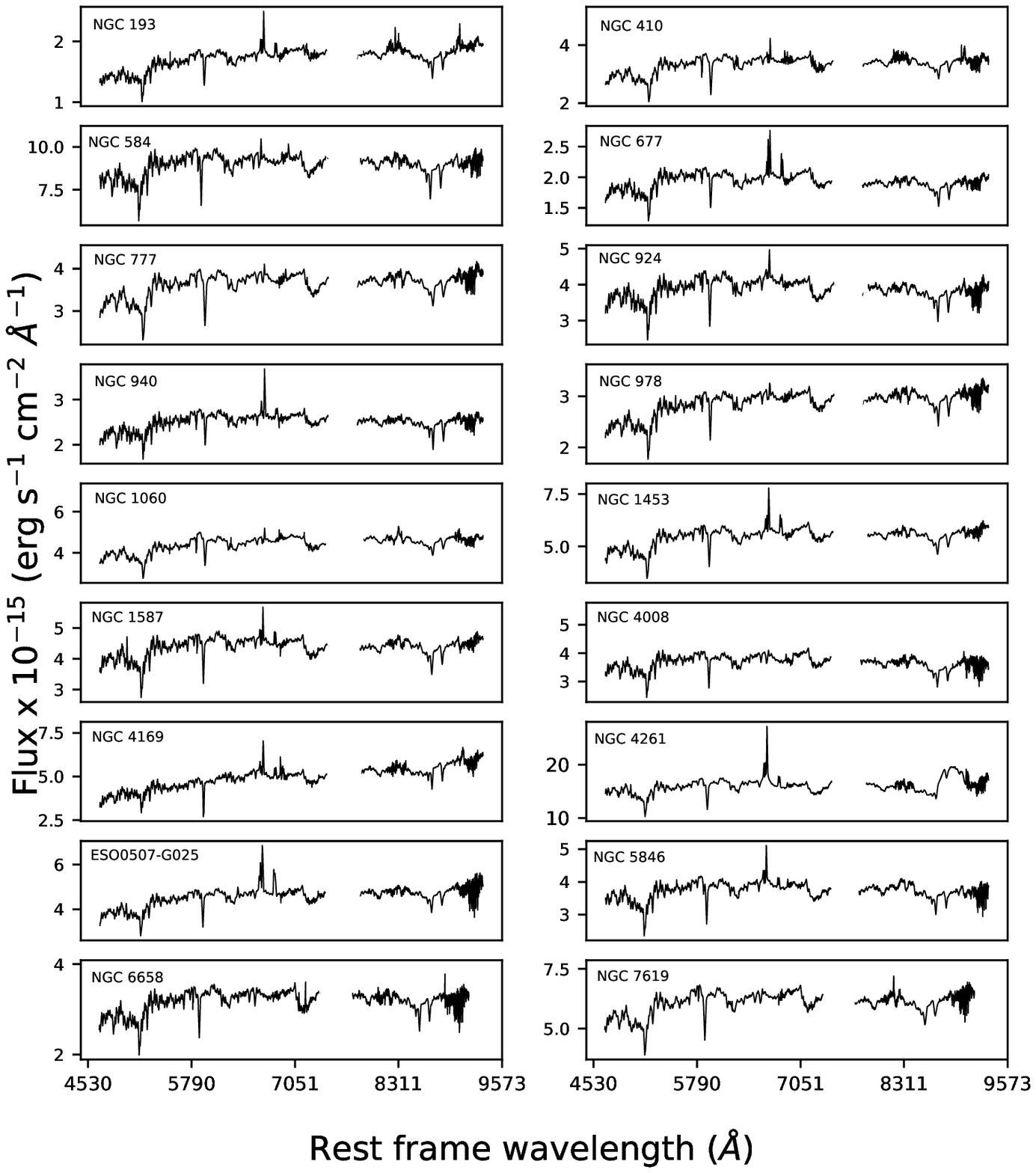}
    \caption{Observed MUSE spectra for our sample of group-dominant galaxies extracted from 
    a nuclear 3" aperture and covering the entire MUSE wavelength range.}
    \label{fig:figures_spectra}
\end{figure*}

Following a similar procedure to \cite{Parkash2019} we extracted for each galaxy a  
spectrum from a circular region at the galaxy continuum centre with a diameter 
of 3" in order to mimic the SDSS spectral fibre aperture. From here on we refer to this circular 
region as the nuclear region and the spectra from this region as the nuclear spectrum 
(see Figure \ref{fig:figures_spectra}). 
The nuclear spectra of the galaxies are characterized by the presence of strong 
[N \,{\sc ii}]$\lambda$6584 emission with [N \,{\sc ii}]$\lambda$6584/H$\alpha>$ 1, which is consistent 
with low-ionisation nuclear emission-line regions. 
Additionally, we extracted an integrated spectrum for each galaxy within an aperture which
encompasses the extended emission in Figure \ref{fig:figures_NB} (see Section  
\ref{subsubsec:Emission_line_morphology}) with [N \,{\sc ii}]$\lambda$6584 emission > 3$\sigma$ per pixel. 
In summary, we find 15/18 objects with intensive nuclear ionised-gas (LINERs) 
with [N \,{\sc ii}]$\lambda$6584 $>$ H$\alpha$. 
\cite{Pagotto2021} found the strongest and most spatially
extended emission, in their sample of massive ETGs in densest clusters of galaxies, 
came from the [N \,{\sc ii}]$\lambda$6584 lines. We find the same result for our sample.
In Table \ref{table:2} we present for each galaxy the 3" nuclear and integrated observed 
F(H$\alpha$) flux, F(H$\alpha$)/F(H$\beta$) multiplied by a factor 
of 100, F([N \,{\sc ii}]]$\lambda$6584) flux and EW(H$\alpha$), respectively.
We note that using the spectral synthesis code FADO \citep{GomesPapaderos2017} 
and the \cite{BruzualCharlot2003} stellar spectral library, we resolved weak
[O \,{\sc iii}]$\lambda$5007, H$\beta$, H$\alpha$ and [N \,{\sc ii}]$\lambda$6584 emission 
in the nuclear region of NGC 6658. However, \cite{Olivares2022} using pPXF \citep{Cappellari2017}
and the Indo-U.S. Coud\'e Feed stellar library \citep{Valdes2004} find no emission lines
in the spectrum of this galaxy using the same MUSE dataset. 
The results for this galaxy are therefore uncertain especially given that no clear emission line
peaks are observed in the integrated spectrum.

\begin{table*}
\centering
\begin{minipage}{170mm}
\caption{H$\alpha$+[N \,{\sc ii}] morphology, observed emission line fluxes and EW(H$\alpha$) for 
our sample galaxies from a simulated 3" diameter and an integrated (int) aperture.
The radial size of the integrated aperture is shown in parentheses to the right of F(H$\alpha$).}      
\label{table:2}                              
\begin{tabular}{l c c c c c c c c c c c c c c c} 
\hline
         & H$\alpha$+[N \,{\sc ii}] & \multicolumn{2}{c}{F(H$\alpha$)} & \multicolumn{2}{c}{(F(H$\alpha$)/F(H$\beta$))$\times$100} & \multicolumn{2}{c}{F([N \,{\sc ii}]$\lambda$6584)} & \multicolumn{2}{c}{EW(H$\alpha$)}\\
         & morphology &\multicolumn{2}{c}{($\times$10$^{-14}$ erg cm$^{-2}$ s$^{-1}$)} & \multicolumn{2}{c}{ } & \multicolumn{2}{c}{($\times$10$^{-14}$ erg cm$^{-2}$ s$^{-1}$)} & \multicolumn{2}{c}{($\AA$)} \\
          &   & \scriptsize{3"} & \scriptsize{int} & \scriptsize{3"} & \scriptsize{int} & \scriptsize{3"} & \scriptsize{int} & \scriptsize{3"} & \scriptsize{int}\\ 
\hline 

NGC 193   & i & 0.43$\pm$0.02 & 1.64$\pm$0.11 (6.0$"$) & 245.16$\pm$0.31 & 134.39$\pm$0.25  & 0.59$\pm$0.06 & 1.70$\pm$0.22  & 2.71  & 1.65\\
NGC 410   & i0 & 0.52$\pm$0.01 & 0.83$\pm$0.02 (4.5$"$) & 146.81$\pm$0.11 & 112.64$\pm$0.14 & 0.52$\pm$0.04 & 0.87$\pm$0.08 & 1.63 & 0.69\\
NGC 584   & i & 0.66$\pm$0.07 & 5.83$\pm$0.67 (21.0$"$)  & 205.26$\pm$0.69 & 131.12$\pm$0.88 & 0.90$\pm$0.08 & 4.1$\pm$0.7  & 0.77 & 0.53\\
NGC 677   & i & 0.63$\pm$0.03 & 4.51$\pm$0.45 (7.5$"$) & 322.66$\pm$0.36 & 313.84$\pm$0.60 & 0.62$\pm$0.07 & 3.87$\pm$0.60  & 3.43  & 3.13 \\
NGC 777   & i0 & 0.10$\pm$0.01 & \dots & 25.55$\pm$0.09 & \dots & 0.19$\pm$0.01 & \dots          & 0.28  & \dots \\
NGC 924   & ii & 0.83$\pm$0.06 & 3.83$\pm$0.30 (24$"$) & 257.83$\pm$0.66 & 246.32$\pm$1.24 & 0.96$\pm$0.06 & 3.48$\pm$0.25 & 2.24 & 1.60 \\
NGC 940   & ii & 0.52$\pm$0.04 & 10.01$\pm$0.24 (13.5$"$) & 436.02$\pm$1.12 & 287.04$\pm$0.31 & 1.14$\pm$0.10 & 6.49$\pm$0.77 & 2.22 & 5.06 \\
NGC 978  & i0 & 0.33$\pm$0.02 & 0.81$\pm$0.08  (4.5$"$) & 148.83$\pm$0.19 & 104.83$\pm$0.21 & 0.28$\pm$0.02 & 0.70$\pm$0.08 & 1.23  & 0.81 \\
NGC 1060  & i & 0.37$\pm$0.05 & 1.37$\pm$0.23 (6.0$"$) & 202.02$\pm$1.96 & 129.84$\pm$0.73  & 0.46$\pm$0.04 & 1.18$\pm$0.19  & 0.86 & 0.63 \\
NGC 1453  & i & 1.35$\pm$0.06 & 8.14$\pm$0.79 (13.5$"$) & 188.92$\pm$0.21 & 91.13$\pm$0.19 & 2.07$\pm$0.08 & 9.43$\pm$0.82 & 2.60 & 2.00 \\
NGC 1587  & i & 0.63$\pm$0.04 & 5.14$\pm$0.18 (15$"$) & 227.91$\pm$0.40 & 161.05$\pm$0.33 & 0.75$\pm$0.04 & 4.13$\pm$0.22 & 1.50 & 1.41 \\
NGC 4008 & i0 & 0.16$\pm$0.02 & \dots & \dots & \dots & 0.14$\pm$0.02 & \dots & 0.46 & \dots \\
NGC 4169  & ii & 0.99$\pm$0.05 & 4.67$\pm$0.28 (10.5$"$) & 325.06$\pm$0.57 & 235.69$\pm$0.76 & 1.71$\pm$0.06 & 6.63$\pm$0.36 & 2.24 & 1.35 \\
NGC 4261  & i0   & 4.31$\pm$0.58 & \dots                   & 450.88$\pm$0.78 & \dots           & 9.44$\pm$0.64 &    \dots           & 2.61  & \dots\\
ESO0507-G025 & ii & 2.59$\pm$0.14 & 20.37$\pm$1.34 (27.0$"$) & 363.09$\pm$0.32 & 291.93$\pm$0.29 & 2.9$\pm$0.18 & 13.68$\pm$1.95 & 5.93  & 3.65\\
NGC 5846   & i  & 0.74$\pm$0.07 & 11.33$\pm$1.27(18$"$)  & 189.27$\pm$0.35 & 139.11$\pm$0.30 & 0.98$\pm$0.09 &11.46$\pm$1.44   &  2.10 & 1.49\\
NGC 6658   & i0  & 0.15$\pm$0.02 & \dots                  & 242.79$\pm$1.89 & \dots           & 0.23$\pm$0.02            &  \dots             & 0.47 & \dots\\
NGC 7619   & i0  & 0.19$\pm$0.05 & \dots & 36.63$\pm$0.13 & \dots & 0.47$\pm$0.04 & \dots & 0.31  & \dots\\
\hline                                   
\end{tabular}
\end{minipage}
\end{table*}

Finally, we explored the data cubes of our target BGGs (see appendix \ref{emission_galaxies_apen})
in order to find additional emission line galaxies in the FoV. 
We found two additional objects in the FoV of NGC 677 and one in the FoVs of NGC 777, NGC 924 and
NGC 1453. These detections are not physically associated (0.1 $\lesssim$ z $\lesssim$ 0.5) with the
main galaxies. In Figure \ref{fig:emission_galaxies} we show the position of these objects in the
FoV of the galaxies, while in Table \ref{table:emission_galaxies} we summarize their main
properties.

\section{Results}\label{sec:results}

\subsection{Ionised gas structure: H$\alpha$ and [N \,{\sc ii}]$\lambda$6584 emission 
line maps and extinction correction}\label{subsec:emission}

\subsubsection{Emission line morphology}\label{subsubsec:Emission_line_morphology}

\begin{figure*}
\includegraphics[width=0.4\textwidth]{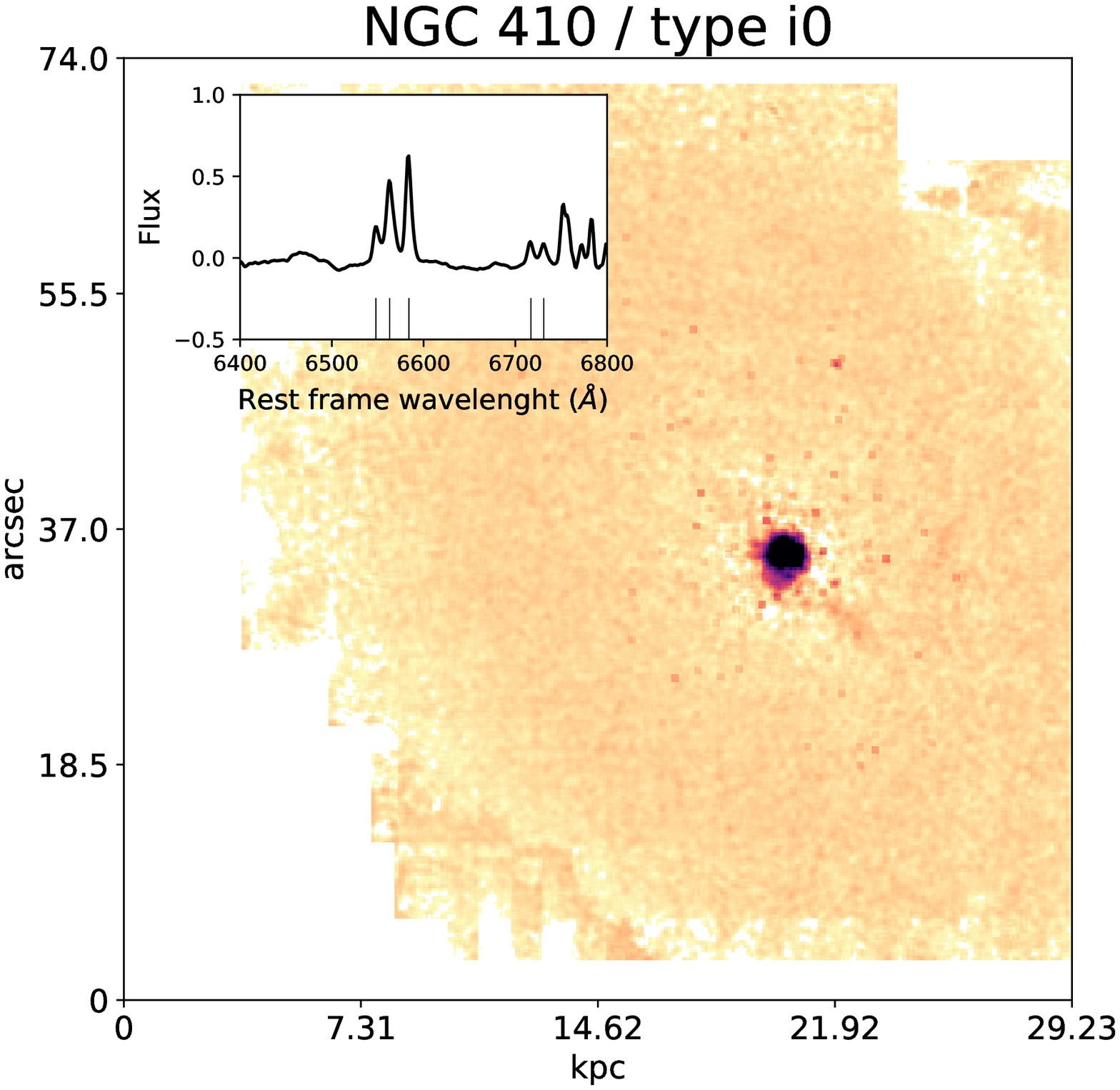}
\includegraphics[width=0.4\textwidth]{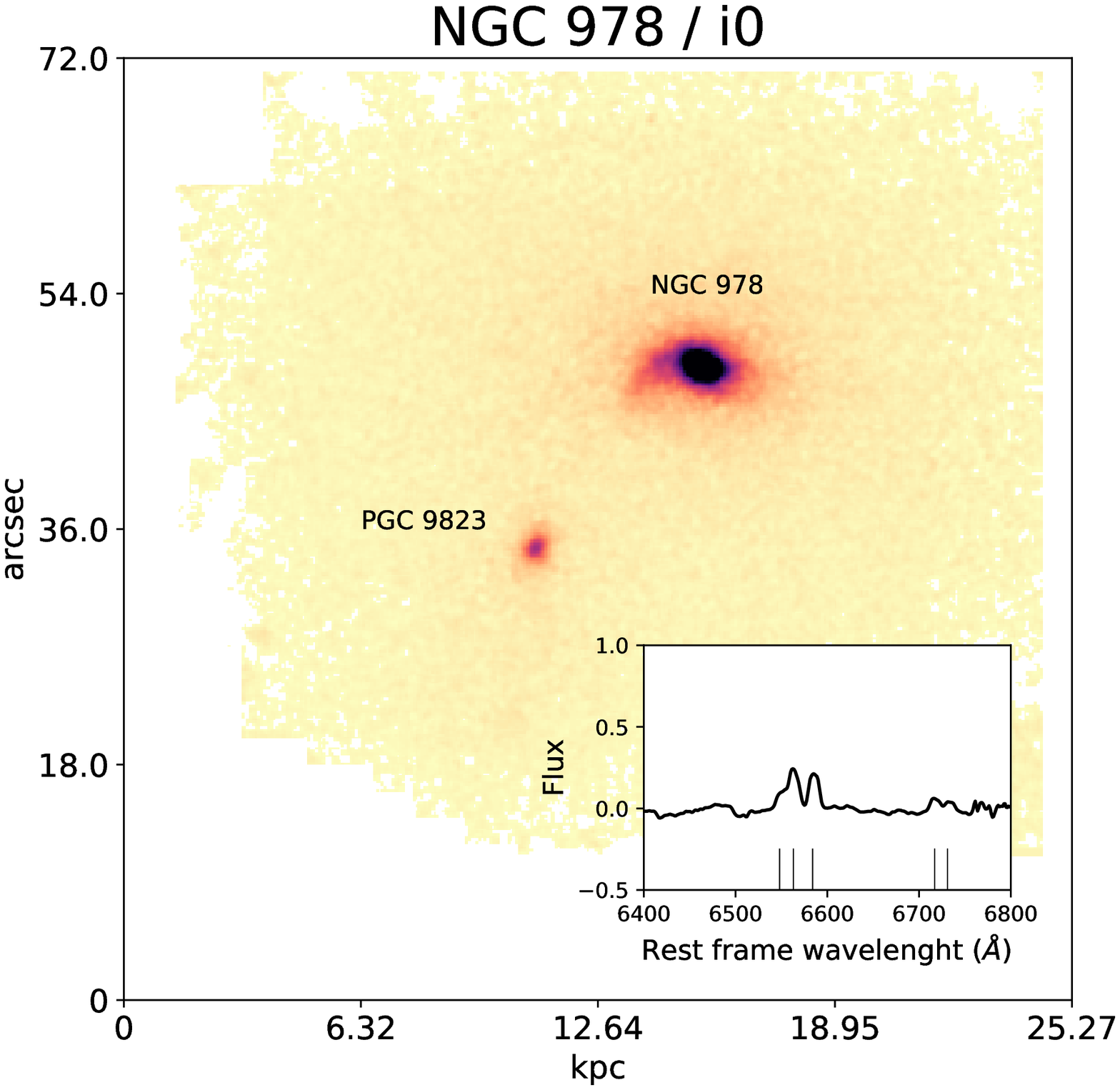}\\
\includegraphics[width=0.4\textwidth]{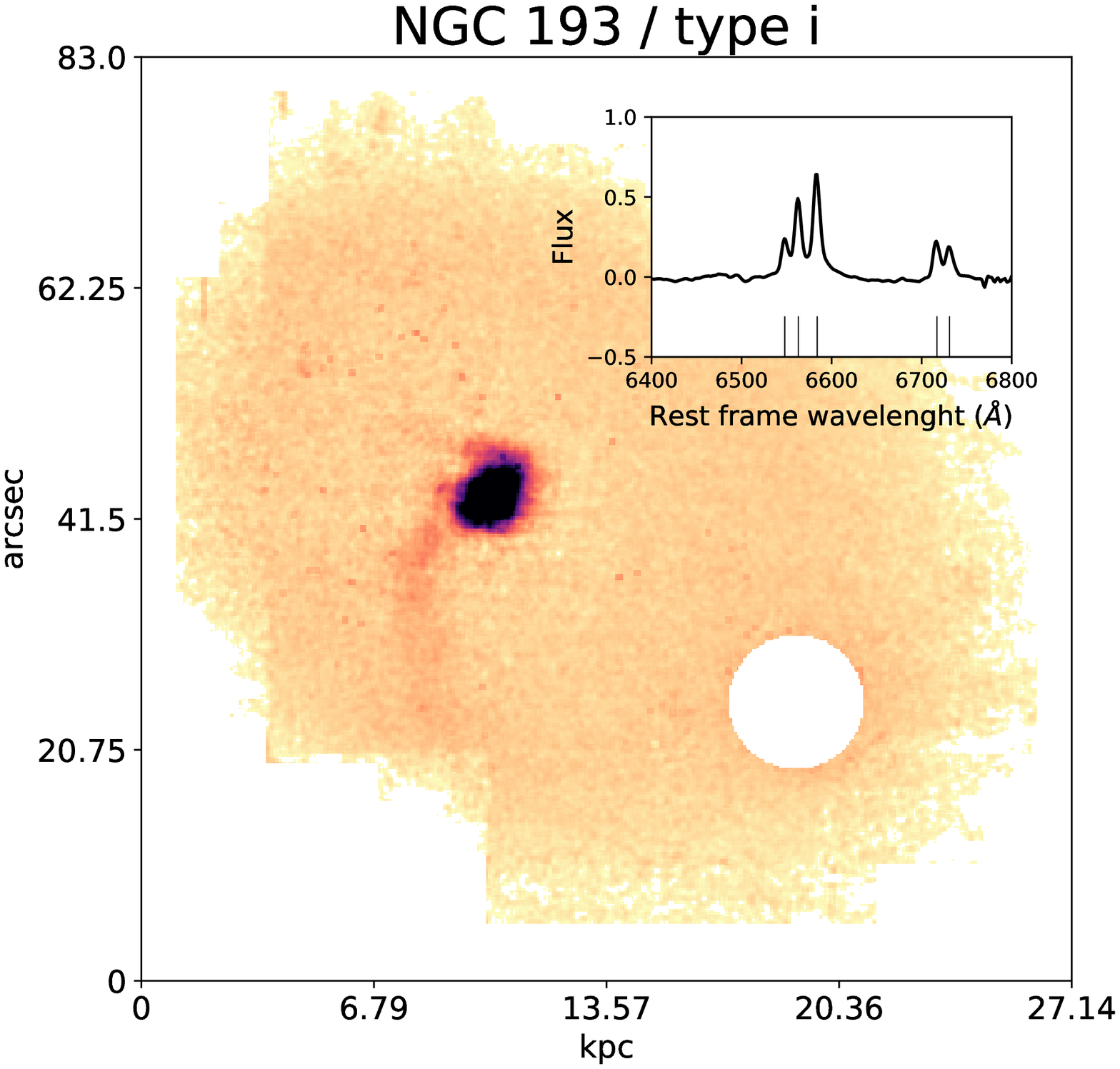}
\includegraphics[width=0.4\textwidth]{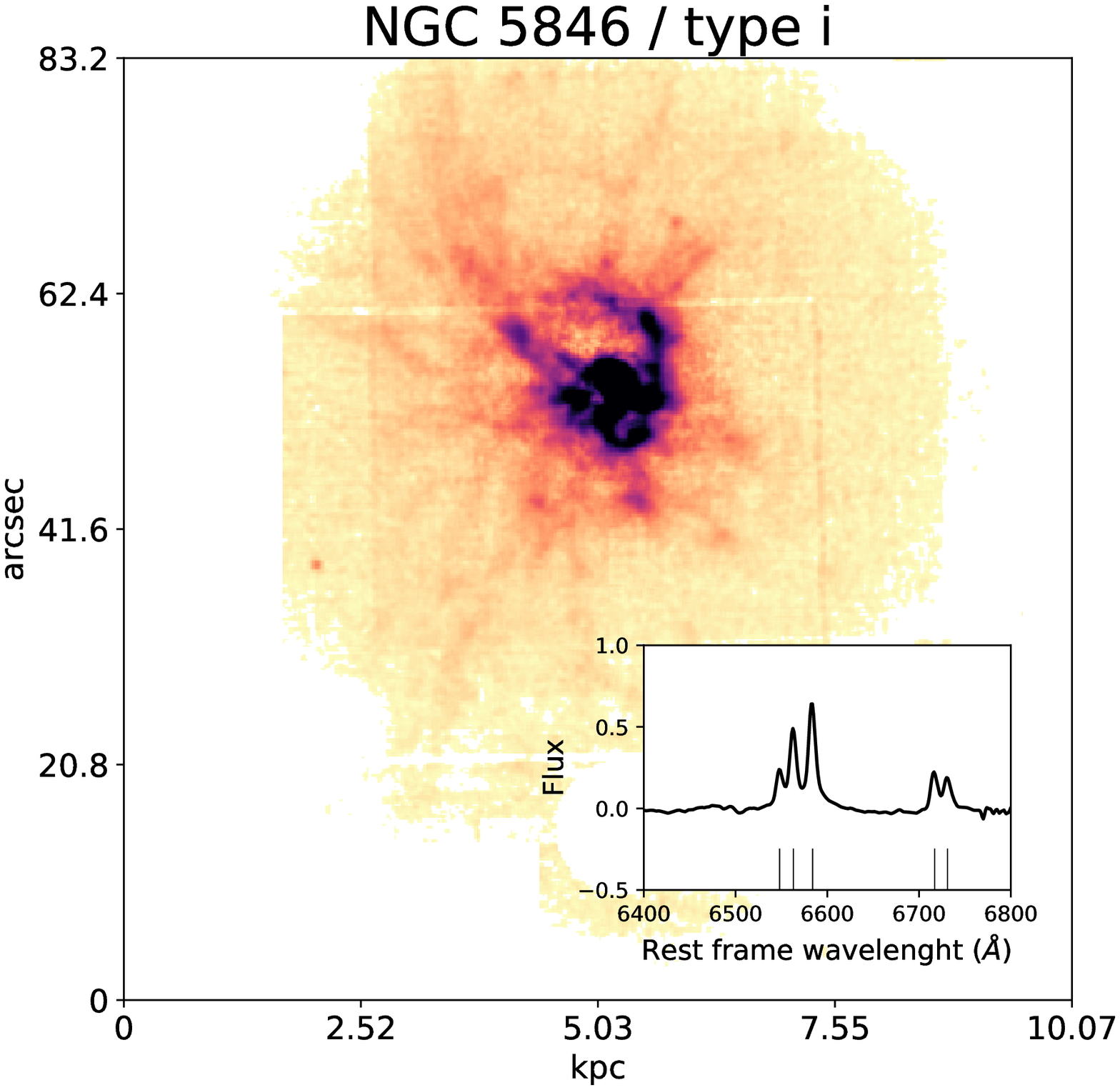}\\
\includegraphics[width=0.4\textwidth]{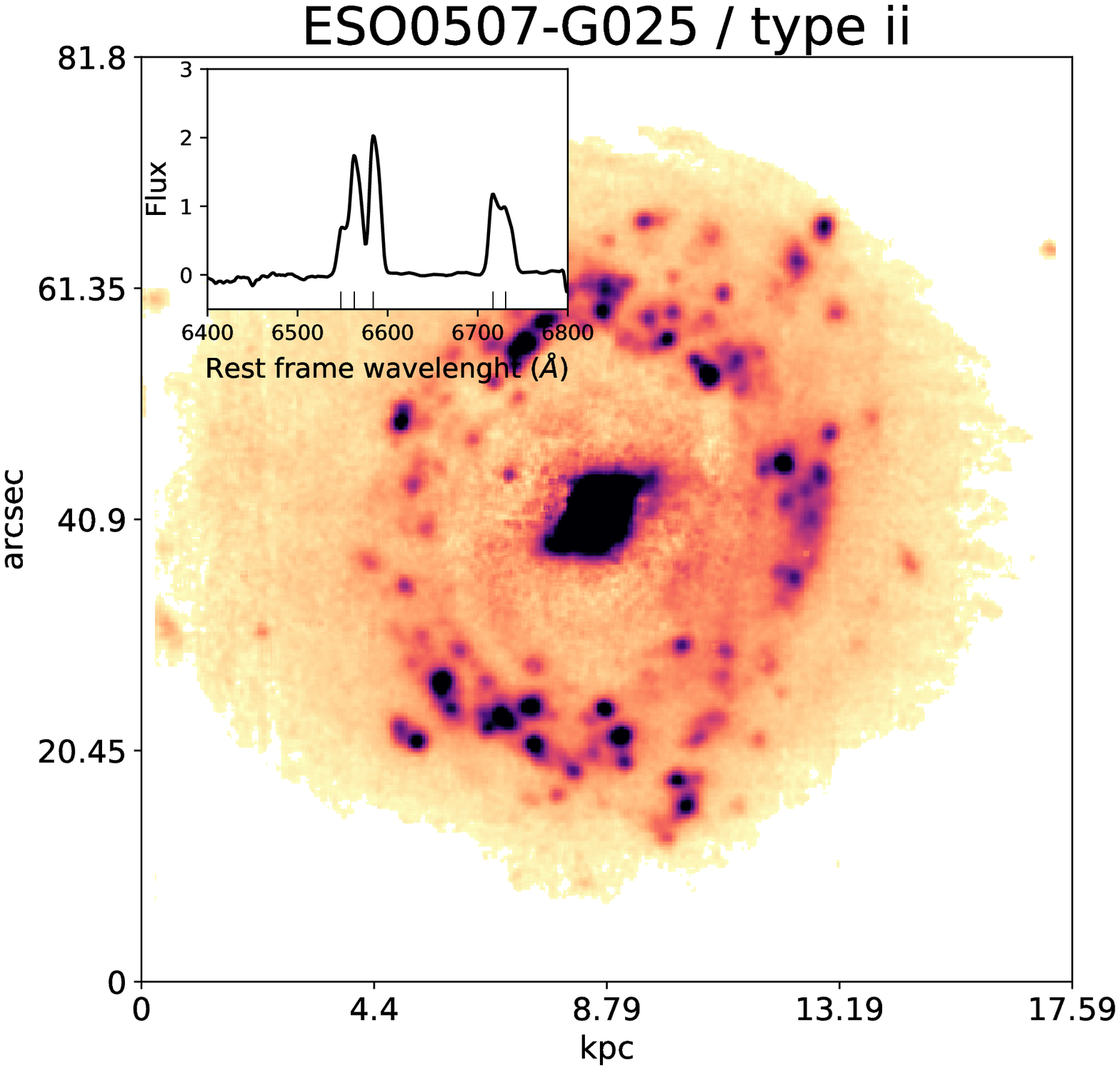}
\includegraphics[width=0.4\textwidth]{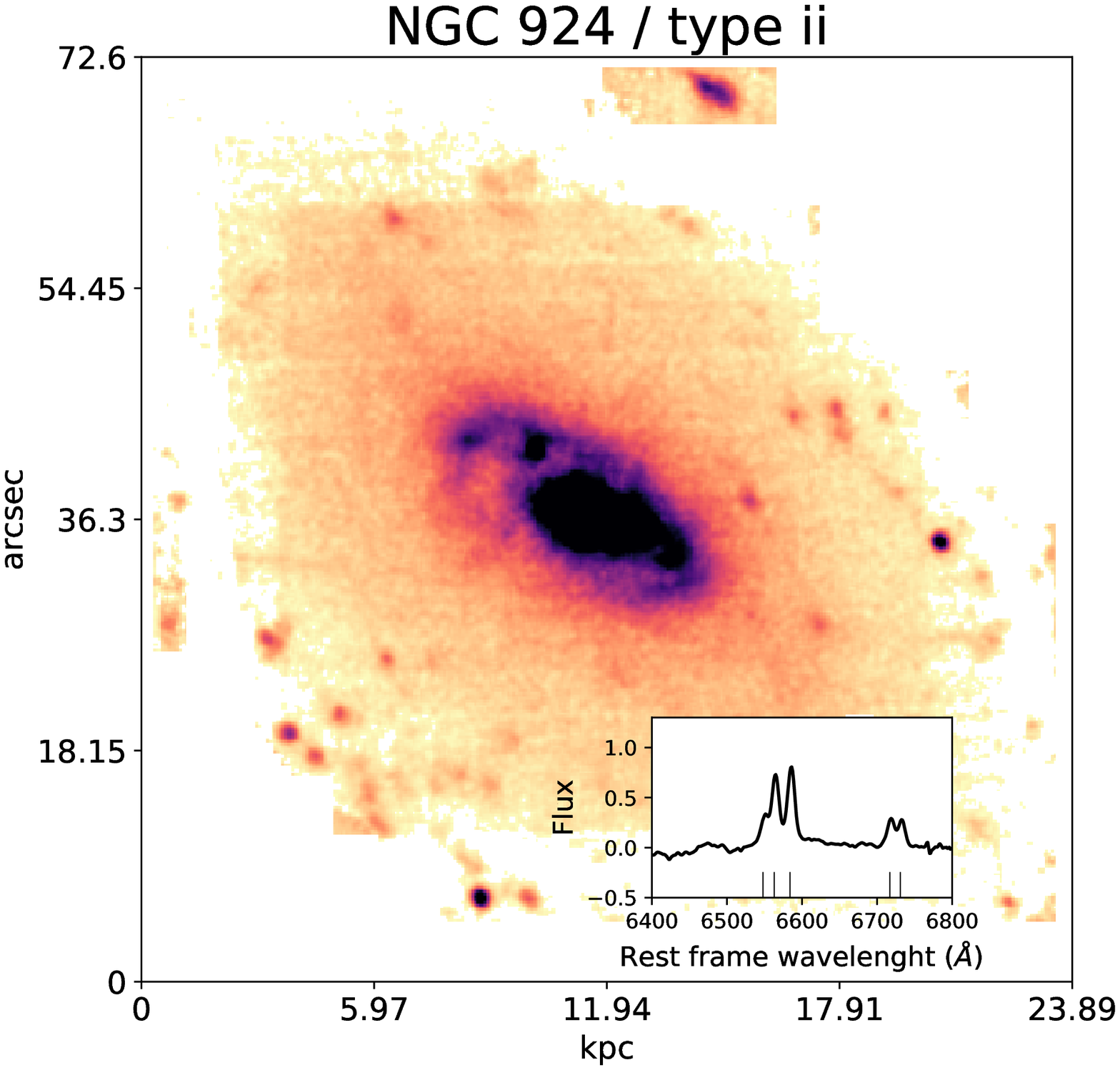}\\
    \caption{Example of H$\alpha$+[N \,{\sc ii}]$\lambda\lambda$6548,6584 emission line maps 
    from our sample. We smoothed the emission line maps using a 3$\times$3 box filter. 
    MUSE continuum-subtracted spectra from a nuclear 3" aperture covering the wavelength range 
    from 6400 $\AA$ to 6800 $\AA$ are shown in the inset panels. The vertical lines in the inset panels
    indicate the wavelengths of the [N \,{\sc ii}]$\lambda$6548, H$\alpha$, 
    [N \,{\sc ii}]$\lambda$6584, [S \,{\sc ii}]$\lambda$6717 and 
    [S \,{\sc ii}]$\lambda$6731 emission lines. Fluxes in units of $\times$10$^{-15}$ 
    (erg s$^{-1}$ cm$^{-2}$ $\AA^{-1}$). In appendix \ref{HaNII_maps_apend} we show
    the emission line maps of all galaxies in our sample. The companion galaxy of NGC 978 is indicated
    in the figure. North is to the top and East to the left.}
    \label{fig:figures_NB}
\end{figure*}

To assist understanding of the physical conditions and detection of the extended and 
diffuse line emitting gas we created continuum-subtracted 
H$\alpha$+[N \,{\sc ii}]$\lambda\lambda$6548,6584 (H$\alpha$+[N \,{\sc ii}]) emission line
images by subtracting the continuum SED models, as explained in Section
\ref{sec:spec_fitting}, and then summing the flux in the data cubes between 6528 
$\rm \AA$ and 6604 $\rm \AA$. In Figures \ref{fig:figures_NB} and \ref{fig:figures_NB_apend} 
(Appendix \ref{HaNII_maps_apend}) we show the maps resulting from this procedure for each galaxy. 
We also show, in these figures, the continuum-subtracted MUSE spectra from the nuclear 3"
aperture covering the wavelength range from 6400 $\AA$ to 6800 $\AA$ as an inset
panel within each image. The DIG and extended filamentary line emitting
H$\alpha$+[N \,{\sc ii}] gas reaches values of $\sim$10$^{-19}$ (e.g., NGC 4008) 
to $\sim$10$^{-17}$ (e.g., NGC 5846) erg cm$^{-2}$ s$^{-1}$ per pixel.

We used the continuum-subtracted H$\alpha$+[N \,{\sc ii}] images to classify the galaxies 
in our sample. To identify the morphology of the line emitting regions and determine 
how much these lines correlate with the overall morphology and properties of our sample galaxies, 
we divided the galaxies into three morphological groups or types: 
\textit{type i0} - strong or diffuse nuclear emission  
(i.e. within a $\sim$1.5" radius from the galaxy centre)
with (or without) unextended ($\lesssim$1 kpc) filamentary structures connected to the nuclear
region, \textit{type i} - strong or diffuse nuclear emission with extended (several kpc)
filamentary structures beyond the nuclear region and \textit{type ii} - i0 or i plus extranuclear
H\,{\sc ii} regions (well-defined or in distorted ring-like structures). 
Using this simple classification scheme we find that 7/18 (NGC 410, NGC 777, NGC 978,
NGC 4008,  NGC 4261, NGC 6658 and NGC 7619) are type i0, another 7/18 are type i
(NGC 193, NGC 584, NGC 677, NGC 1060, NGC 1453, NGC 1587 and NGC 5846)
and 4/18 galaxies are type ii (NGC 924, NGC 940, NGC 4169 and ESO0507-G025).

\subsubsection{Intrinsic reddening}

We computed the extinction corrections with the colour excess E(B-V) from the gas 
using the Balmer decrement H$\alpha$/H$\beta$=2.86 assuming case B recombination 
\citep[][n$_e$=100 cm$^{-3}$ and T$_e$=10$^4$ K]{OsterbrockFerland2006}. 
The E(B-V)$_{gas}$ is calculated as follows

\begin{equation}
E\left(B-V\right)_{gas} $=$ \frac{2.5}{\kappa\left(H\beta\right)-\kappa\left(H\alpha\right)}
\log \left(\frac{\left(H\alpha \diagup H\beta\right)_{o}}{2.86}\right),
\label{eq:extinction}
\end{equation}

\noindent
where $\kappa$(H$\alpha$) and $\kappa$(H$\beta$) are the extinction values from the
\cite{Cardelli1989} curve with R$_V$ =  3.1. (H$\alpha$/H$\beta$)$_{o}$ is
the observed ratio between H$\alpha$ and H$\beta$.
Therefore, we corrected the emission lines for extinction using 
I$_{\lambda}$=F$_{\lambda,o}$ 10$^{0.4 E\left(B-V\right)_{gas} \kappa\left(\lambda\right)}$.
The reddening parameters were set to 0.0 for unrealistic values of 
(H$\alpha$/H$\beta$)$_{o}$ < 2.86 and when H$\beta$ was not detected. 
The assumption of (H$\alpha$/H$\beta$)$_{o}$ = 2.86 instead of 3.1
in spectra with AGN(-like) features does not significantly affect the results.
Galaxies with (H$\alpha$/H$\beta$)$_{o}$ > 2.86 have low-reddening
E(B-V)$_{gas}$ values from 0.13 to 0.46 mag. This result is not surprising as low-balmer decrements 
are also found in other ETGs \citep[e.g.,][]{Annibali2010} and in the nuclei and
nebular filaments of cool-core clusters \citep{McDonald2012}. 
On the other hand, $\sim$72\% (13/18) and $\sim$83\% (15/18) of our galaxies have unrealistic
nuclear and integrated (H$\alpha$/H$\beta$)$_{o}$ values, respectively. 
This indicates that the derived H$\beta$ fluxes are larger for their corresponding H$\alpha$ fluxes. 
For galaxies with (H$\alpha$/H$\beta$)$_{o}$ > 2.86 the E(B-V)$_{nebular}$ are similar
or slightly lower than the E(B-V)$_{stellar}$, while for galaxies with 
(H$\alpha$/H$\beta$)$_{o}$ < 2.86 the E(B-V)$_{stellar}$ is higher.
This might produce an under subtraction of the observed stellar populations.
It is worth noting that unrealistic extinctions are also obtained with other SSP codes
\citep[e.g.,][]{Annibali2010,Herpich2018} in ETGs. 
However, uncertainties in the reddening correction produced by the stellar fitting  
will not significantly affect the emission line ratios nor the properties obtained 
from them in our case. Therefore, large errors in the H$\alpha$/H$\beta$ are due to the uncertainty 
in the H$\beta$ emission.

\subsection{[O \,{\sc iii}]$\lambda$5007/H$\beta$, [S \,{\sc ii}]$\lambda\lambda$6717,6731/H$\alpha$ and [N \,{\sc ii}]$\lambda$6584/H$\alpha$ emission line ratio maps and BPT diagrams}\label{subsec:emission_ratios}

\begin{figure*}
\includegraphics[width=0.95\textwidth]{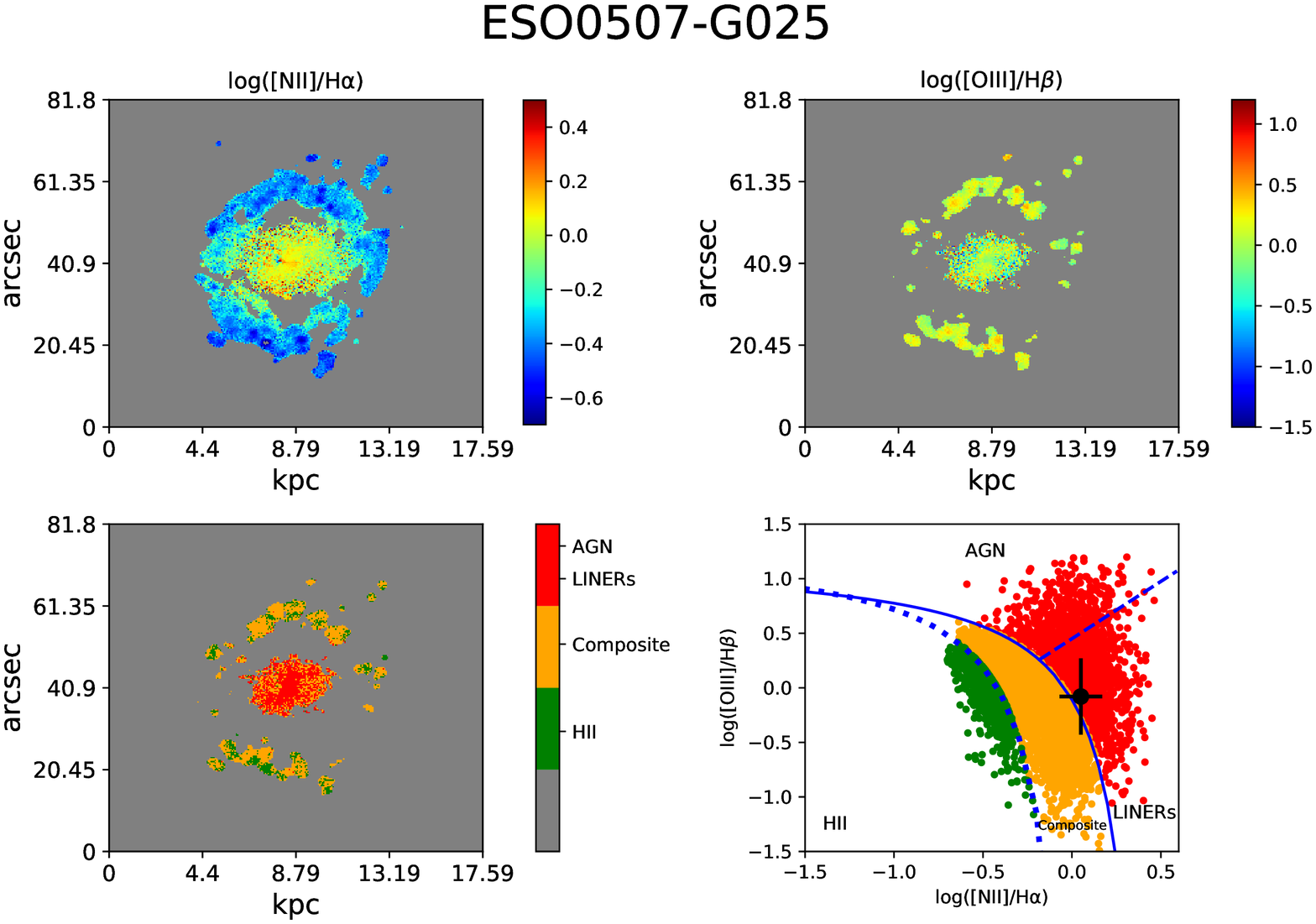}
    \caption{Example of our emission line ratio maps log([N \,{\sc ii}]/H$\alpha$) and 
    log([O \,{\sc iii}]/H$\beta$ (top row), and BPT diagrams (bottom row) for the galaxy 
    ESO0507-G025. In the BPT diagrams we include the Kewley et al. (2001; solid line), 
    Kauffmann et al. (2003; dotted line) and Schawinski et al. (2007; dashed line) model 
    boundary lines which divide regions dominated by star-forming H\,{\sc ii} (green), composite
    (orange data points) and AGN/LINERs (red data points). 
    The filled black data point corresponds to value in the 3" aperture.
    In Appendix \ref{Maps_sample} we show the emission line ratio maps and the BPT diagrams/maps 
    for our entire sample. North is to the top and East to the left.}
    \label{fig:example_figure}
\end{figure*}

Using the information derived in the previous sections it is possible to distinguish 
between different ionisation mechanisms using BPT \citep[][]{Baldwin1981} diagrams. 
To do this, we used the following emission line ratios:
[O\,{\sc iii}] $\lambda$5007/H$\beta$ ([O \,{\sc iii}]/H$\beta$) and 
[N\,{\sc ii}] $\lambda$6584/H$\alpha$ ([N\,{\sc ii}]/H$\alpha$). 
The [O\,{\sc iii}] $\lambda$5007/H$\beta$ emission line ratio is
an excitation indicator and provides information about the available fraction of hard ionising
photons. On the other hand, [N\,{\sc ii}] $\lambda$6584 is a low-ionisation emission line tracer,
which is usually weak in pure H\,{\sc ii} regions. 
Therefore, the emission line ratios [N\,{\sc ii}]/H$\alpha$ and [O \,{\sc iii}]/H$\beta$ 
can effectively discriminate between photoionisation under physical conditions that are typical 
for H\,{\sc ii} regions and other excitation mechanisms (e.g., AGN or shocks) which are likely 
present in our sample. In Figure \ref{fig:example_figure} we show an example of the emission line
ratio maps log([N \,{\sc ii}]/H$\alpha$) and log([O \,{\sc iii}]/H$\beta$,
the log([N \,{\sc ii}]/H$\alpha$) vs log([O \,{\sc iii}]/H$\beta$ BPT diagram and the 2D BPT
map for the galaxy ESO0507-G025.
In this figure, we separate star-forming and AGN-ionised regions (i.e., Seyferts and LINERs) 
with blue demarcation lines from models by \cite{Kauffmann2003} (dotted line), \cite{Kewley2001} 
(solid line) and \cite{Schawinski2007} (dashed line).
We colour coded the data points located in the different areas of the BPT diagram, i.e, 
AGN/LINERs in red, composite in orange and the star-forming H\,{\sc ii} dominated area in green.
In Appendix \ref{Maps_sample} we show the emission 
line ratio maps and the BPT diagrams/maps for our entire sample. 
The BPT maps show that most of our data points/spaxels in the nuclear regions 
(see Figure \ref{fig:example_figure}) are dominated by AGN/LINERs while the SF 
becomes more important in the extended regions. 
We note that, in most cases, we obtain a very small or nonexistent number of data points
($\sim$17\% of the spaxels in ESO0507-G025) in the H\,{\sc ii} area of the BPT diagrams 
(see figures in Appendix \ref{Maps_sample}). Most spaxels lie in the AGN/LINERs region of the
diagrams ($\sim$71\% of the spaxels on average), while only in two cases 
(NGC 978 and ESO0507-G025) are most spaxels in the composite region with 62\% and 55\% of the
spaxels, respectively.

\begin{figure}
\hspace*{-0.25cm}
\includegraphics[width=0.5\textwidth]{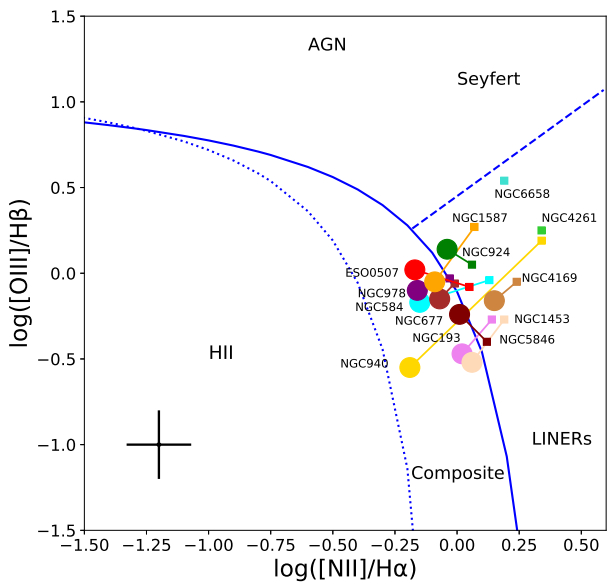}
    \caption{[O \,{\sc iii}]/H$\beta$ vs [N \,{\sc ii}]/H$\alpha$ diagnostic diagram for 
    the nuclear spectra using equivalent 3" apertures to SDSS spectra (squares) and
    integrated emission (circles). The Kewley et al. (2001; solid line), Kauffmann et al.
    (2003; dotted line) and Schawinski et al. (2007; dashed line) model boundary lines in blue
    discriminate between areas of the diagram dominated by SF (H \,{\sc ii}) and AGN/LINERs 
    emission. The mean errors are shown in the lower of the figure.}
    \label{fig:figures_BPT_int}
\end{figure}

In Table \ref{table:2b} we present for each galaxy the emission line ratios 
log([O \,{\sc iii}]/H$\beta$), log([N \,{\sc ii}]/H$\alpha$) and 
log([S \,{\sc ii}]/H$\alpha$) from the 3" and integrated apertures. 
Figure \ref{fig:figures_BPT_int} shows the BPT diagram [O \,{\sc iii}]/H$\beta$ versus 
[N\,{\sc ii}]/H$\alpha$ of the nuclear regions (small squares) 
for galaxies where [O\,{\sc iii}] $\lambda$5007 and H$\beta$ are detected. In the figure,
we compare these with values obtained from the integrated apertures (circles). 
We use the same colour code to identify the galaxies in this diagram throughout the paper.
In most cases, the nuclear values lie in the LINER region of the BPT diagram, while the 
integrated ones are in the composite region. The log([O \,{\sc iii}]/H$\beta$) for most galaxies
decreases significantly when we consider the larger integrated aperture, which changes their
positions in the BPT diagram. In the case of ESO0507-G025, NGC 924 and NGC 5846 
the log([O \,{\sc iii}]/H$\beta$) increases and the log([N \,{\sc ii}]/H$\alpha$) ratio decreases.
Figure \ref{fig:figures_BPT_int} clearly shows that the ionisation structure of our sample
galaxies can not be correctly assessed from small fixed apertures (e.g., SDSS aperture) and the
ionisation sources and structure of the warm gas may be different in different parts of the
galaxies (see Appendix \ref{Maps_sample}). 
In Section \ref{subsec:morphology_aper} and \ref{subsec:morphology_2DBPT} we explore the effect of
using different apertures on the observed properties of the ISM and the ionisation
mechanisms likely present in our sample galaxies.

\begin{table*}
\centering
\hspace*{-1.4cm}
\begin{minipage}{165mm}
\caption{Main properties of our sample galaxies from the simulated SDSS spectral fibre diameter
(3") and integrated (int) apertures. SFR(H$\alpha$) calculated from the L(H$\alpha$) in the
largest aperture int/3".}      
\label{table:2b}                              
\begin{tabular}{l c c c c c c c c c c c c c c c} 
\hline
            & \multicolumn{2}{c}{log ([O \,{\sc iii}]/H$\beta$)}  & \multicolumn{2}{c}{log ([N \,{\sc ii}]/H$\alpha$)} & \multicolumn{2}{c}{log ([S \,{\sc ii}]/H$\alpha$)} & \multicolumn{2}{c}{L(H$\alpha$)} & SFR(H$\alpha$)\\
            &        &    &  & & & &\multicolumn{2}{c}{($\times$10$^{39}$ erg cm$^{-1}$)} &  (M$_{\odot}$ yr$^{-1}$) \\

             & 3" & int & 3" & int & 3" & int & 3" & int  &  \\ 
\hline 

NGC 193    &   -0.27$\pm$0.22 & -0.49$\pm$0.40 & 0.14$\pm$0.16 & 0.02$\pm$0.20 & -0.02$\pm$0.15 & -0.14$\pm$0.22 & 2.01$\pm$0.12 & 7.70$\pm$0.52 & 0.0354$\pm$0.0024\\

NGC 410    & \dots  & \dots & 0.01$\pm$0.11 & 0.02$\pm$0.12 & -0.33$\pm$0.15 & -0.13$\pm$0.17 & 3.55$\pm$0.08 & 5.72$\pm$0.13 & 0.0263$\pm$0.0006\\

NGC 584    &  -0.04$\pm$0.40 & -0.17$\pm$1.03 & 0.13$\pm$0.19 & -0.15$\pm$0.29 & -0.16$\pm$0.34 & -0.44$\pm$0.75 & 0.53$\pm$0.06 & 4.68$\pm$0.54 & 0.0215$\pm$0.0025\\

NGC 677    & -0.06$\pm$0.22 & -0.15$\pm$0.30 & -0.01$\pm$0.16 & -0.07$\pm$0.23 & -0.03$\pm$0.08 & -0.09$\pm$0.15 & 5.22$\pm$0.24 & 34.71$\pm$3.50 & 0.1597$\pm$0.0161\\

NGC 777    & \dots  & \dots & 0.28$\pm$0.22 & \dots & 0.30$\pm$0.28 & \dots & 0.61$\pm$0.09 & \dots & 0.0028$\pm$0.0004\\

NGC 924    & 0.05$\pm$0.37 & 0.14$\pm$0.60 & 0.06$\pm$0.14 & -0.04$\pm$0.15 & -0.10$\pm$0.20 & -0.12$\pm$0.27 & 3.97$\pm$0.28 & 18.24$\pm$1.41 & 0.0839$\pm$0.0065\\

NGC 940    & 0.19$\pm$0.43  & -0.55$\pm$0.36 & 0.34$\pm$0.17 & -0.19$\pm$0.14  & 0.02$\pm$0.25 & -0.45$\pm$0.20 & 8.84$\pm$0.75 & 63.67$\pm$1.50 & 0.2929$\pm$0.0069\\

NGC 978    & -0.09$\pm$0.17 & -0.15$\pm$0.22  & -0.08$\pm$0.14 & -0.07$\pm$0.22 & -0.18$\pm$0.21 & -0.11$\pm$0.42 & 1.81$\pm$0.21 & 4.40$\pm$0.49 & 0.0203$\pm$0.0023\\

NGC 1060    & \dots  & \dots & 0.10$\pm$0.23 & -0.06$\pm$0.33 & 0.09$\pm$0.44 & 0.18$\pm$0.57 & 2.38$\pm$0.33 & 8.80$\pm$1.51 & 0.0405$\pm$0.0069\\

NGC 1453    & -0.27$\pm$0.26 & -0.52$\pm$0.37 & 0.19$\pm$0.08 & 0.06$\pm$0.18 & 0.09$\pm$0.08 & -0.12$\pm$0.17 & 4.52$\pm$0.19 & 27.23$\pm$2.64 & 0.1253$\pm$0.0121\\

NGC 1587    & 0.27$\pm$0.16 & -0.05$\pm$0.34 & 0.07$\pm$0.11 & -0.09$\pm$0.09 & 0.03$\pm$0.16 & -0.11$\pm$0.23 & 1.88$\pm$0.11 & 15.37$\pm$0.52 & 0.0707$\pm$0.0024\\

NGC 4008   & \dots   & \dots & -0.07$\pm$0.25 & \dots &  0.09$\pm$0.41 & \dots & 0.47$\pm$0.06 & \dots & 0.0022$\pm$0.0003\\

NGC 4169   & -0.05$\pm$0.32 & -0.16$\pm$0.51 & 0.24$\pm$0.08 & 0.15$\pm$0.12 & 0.10$\pm$0.16 & 0.10$\pm$0.15 & 4.24$\pm$0.19 & 14.75$\pm$0.90 & 0.0679$\pm$0.0041\\

NGC 4261   & 0.25$\pm$0.48  & \dots & 0.34$\pm$0.20 & \dots & -0.07$\pm$0.29 & \dots & 12.08$\pm$1.62 & \dots & 0.0556$\pm$0.0074\\

ESO0507-G025 & -0.08$\pm$0.35 & 0.02$\pm$0.34 & 0.05$\pm$0.12 & -0.17$\pm$0.21 & 0.07$\pm$0.15 & -0.16$\pm$0.26 & 9.17$\pm$0.52 & 42.56$\pm$2.96 & 0.1958$\pm$0.0136\\

NGC 5846   & -0.40$\pm$0.19 & -0.24$\pm$0.41 & 0.12$\pm$0.18 & 0.01$\pm$0.24 & -0.12$\pm$0.21 & -0.26$\pm$0.43 & 0.47$\pm$0.04 & 7.23$\pm$0.81 & 0.0333$\pm$0.0037\\

NGC 6658   & 0.54$\pm$0.79   & \dots &  0.19$\pm$0.19 & \dots &  \dots & \dots &  0.65$\pm$0.08 & \dots &  0.0030$\pm$0.0004\\

NGC 7619 &  \dots    & \dots &  0.40$\pm$0.36 & \dots &  \dots & \dots &  0.66$\pm$0.17 & \dots &  0.0030$\pm$0.0008 \\
\hline                                   
\end{tabular}
\end{minipage}
\end{table*}

In summary, we find that all galaxies in our sample, with [O\,{\sc iii}] $\lambda$5007 
and H$\beta$ detections, have a dominant AGN/LINER nuclear region.
Extended LINER-like regions are observed in most galaxies with filamentary structures 
(type i galaxies) and ring-like structures and/or extranuclear H\,{\sc ii} regions (type ii galaxies).
Most spaxels in these regions fall in the BPT area of mixed contribution (composite) 
from SF and/or AGN.

\subsection{Velocity field maps}\label{sub:velocity}

\begin{figure*}
\includegraphics[width=1.\textwidth]{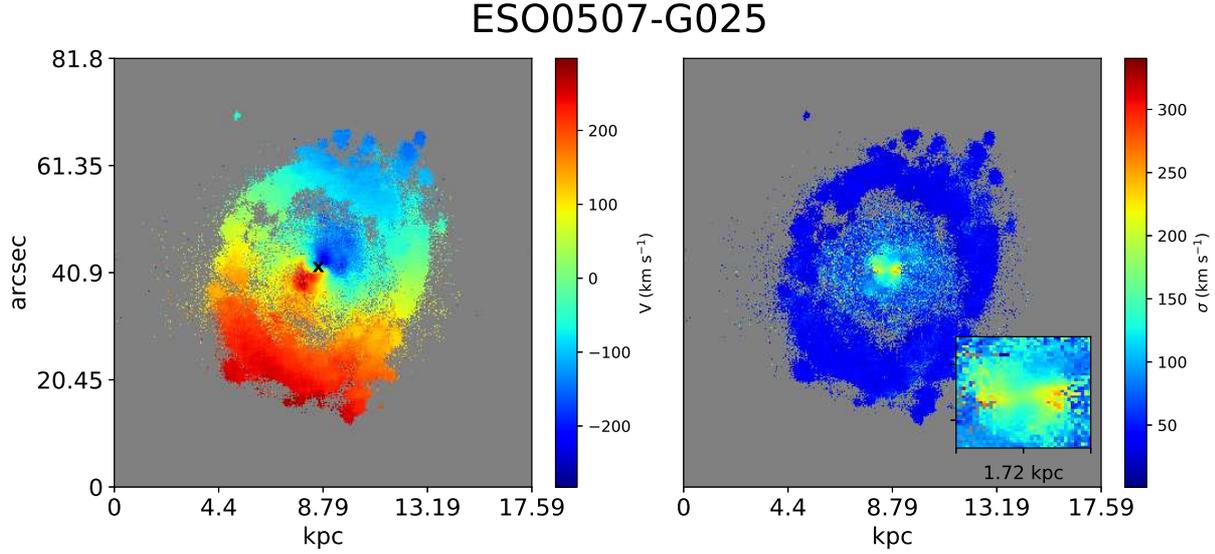}
\caption{Radial velocity V([N\,{\sc ii}]) (left panel) and velocity dispersion 
$\sigma$([N\,{\sc ii}]) (right panel) maps for the galaxy ESO0507-G025.
The position of the continuum maximum is indicated by a X symbol. 
A zoom-in of the central region of the velocity dispersion map is shown in the inset panel.
North is up and east is to the left.}
\label{fig:ESO0507_velocity}
\end{figure*}

We obtained the radial velocity V([N\,{\sc ii}]) and velocity dispersion $\sigma$([N\,{\sc ii}]) 
maps by fitting a single Gaussian to the [N\,{\sc ii}]$\lambda$6584 emission line profiles.
The velocity dispersion $\sigma$ (or FWHM=2.35$\sigma$) was obtained 
as $\sigma^2$ = $\sigma^2_{obs}$ - $\sigma^2_{inst}$ where $\sigma_{inst}$
is the instrumental dispersion, see Section \ref{sec:observations}.
The detailed kinematical analysis of the emitting gas is beyond the scope of this paper, however 
in Figure \ref{fig:ESO0507_velocity} and Appendix \ref{appendix_NII_velocity_maps}
we show the velocity field maps for the type ii galaxies ESO0507-G025 and NGC 924, NGC 940 and 
NGC 4169, respectively. Radial velocity and velocity dispersion maps for our full sample are shown 
in \cite{Olivares2022}. In order to include the kinematical information from the extended and
filamentary structures we considered spaxels with S/N $\gtrsim$ 2.
In Figure \ref{fig:ESO0507_velocity} we see an example of these velocity fields
for the galaxy ESO0507-G025. Clearly, this galaxy has two independent
velocity structures or rotating discs, while the velocity dispersion in the nuclear region
shows a bicone-shaped structure which increases in velocity dispersion with distance from the
centre. The gas in these bicone structures is located in the AGN/LINERs region of the BPT diagram 
(see Figures \ref{fig:example_figure} and \ref{fig:figures_BPT_int}) suggesting an outflow 
perpendicular to the observer.

Our velocity fields (V and $\sigma$) are in agreement with those obtained independently by
\cite{Olivares2022}. In general, the radial velocity fields in our sample galaxies 
(see Appendix \ref{appendix_NII_velocity_maps} and \citealt{Olivares2022}) show rotation or
gradients but with relatively small velocity ranges of around $\pm$200-300 km s$^{-1}$.  
These variations are similar in both, the compact and extended gas emission regions.  
We see the highest $\sigma$([N \,{\sc ii}]) values, in most cases, at positions close to the
continuum flux maximum with values of $\sim$150 to $\sim$300 km s$^{-1}$, 
while extended regions or filaments show lower values. 
In general $\sigma$([N\,{\sc ii}]) values in the filaments vary little across the galaxies.
This trend was also reported by \cite{McDonald2012} for a sample of galaxies in galaxy
clusters. They found that the most extended optical filaments in their sample, likely originated 
from ICM cooling and were experiencing only minor turbulence. 

\subsection{[S \,{\sc ii}]$\lambda$6716 / [S \,{\sc ii}]$\lambda$6731 ratio}\label{subsec:density}

The [S \,{\sc ii}]$\lambda$6716 / [S \,{\sc ii}]$\lambda$6731 intensity ratio was used to
determine the electron density n$_e$([S \,{\sc ii}]). 
We computed the values of n$_e$(S \,{\sc ii}) using the IRAF STS temden package  
assuming t$_e$([O \,{\sc iii}])=10000 K. We set unrealistic (saturated) values of 
[S \,{\sc ii}]$\lambda$6716 / [S \,{\sc ii}]$\lambda$6731 > 1.43 to 
n$_e$([S \,{\sc ii}]) $\sim$1 cm$^{-3}$ and for
[S \,{\sc ii}]$\lambda$6716 / [S \,{\sc ii}]$\lambda$6731 < 0.46 to $\sim$52452 cm$^{-3}$. 
In Table \ref{table:properties_density} we show the 
electron density for the nuclear (3" aperture) regions for our sample.
We note that the [S \,{\sc ii}]$\lambda$6716 / [S \,{\sc ii}]$\lambda$6731 ratio
is greater than 1 for 12/18 galaxies, which indicates a predominantly low density
regime ($\sim$100 cm$^{-3}$). Only 6/18 galaxies show values $\gtrsim$1000 cm$^{-3}$
in the nuclear regions. In Figure \ref{fig:density} (appendix \ref{appendix_SII_maps}) 
we show the [S \,{\sc ii}]$\lambda$6716 / [S \,{\sc ii}]$\lambda$6731 ratio maps for 
our sample galaxies.

\begin{table}
\hspace*{-0.7cm}
\begin{minipage}{95mm}
\centering
\caption{Electron density for the nuclear (3" aperture) regions of our sample of group-dominant galaxies.
We show the unrealistic (saturated) values of [S \,{\sc ii}]$\lambda$6716 / [S \,{\sc ii}]$\lambda$6731 by
$\checkmark$ symbols.}      
\label{table:properties_density}                     
\begin{tabular}{l c c c}
\hline
Name        &  [S \,{\sc ii}]$\lambda$6716 / [S \,{\sc ii}]$\lambda$6731 & Saturated & n$_e$([S \,{\sc ii}])\\
            &                                                            & values &  (cm$^{-3}$)  \\
\hline
NGC 193      & 0.92$\pm$0.16 & & 839.3$\pm$397.4\\
NGC 410      & 1.85$\pm$0.23 & $\checkmark$ & $\lesssim$1.0\\
NGC 584      & 1.46$\pm$0.70 & $\checkmark$ & $\lesssim$1.0\\
NGC 677      & 1.29$\pm$0.09 & & 131.3$\pm$98.0\\
NGC 777      & 1.11$\pm$0.30 & & 377.5$\pm$653.8\\
NGC 924      & 1.38$\pm$0.36 & &  43.7$\pm$257.3\\
NGC 940      & 1.06$\pm$0.34 & & 471.4$\pm$992.9\\
NGC 978      & 1.23$\pm$0.37 & & 200.8$\pm$436.0\\
NGC 1060     & 0.66$\pm$0.37 & & 2788.0$\pm$1126.3\\
NGC 1453     & 1.29$\pm$0.10 & & 131.3$\pm$109.2\\
NGC 1587     & 1.21$\pm$0.24 & & 226.4$\pm$229.5\\
NGC 4008     & 0.69$\pm$0.39 & & 2352.9$\pm$960.4\\
NGC 4169     & 1.07$\pm$0.25 & & 451.4$\pm$584.7\\
NGC 4261     & 1.12$\pm$0.36 & & 360.4$\pm$649.3\\
ESO0507-G025 & 0.46$\pm$0.09 & & 48893.1$\pm$21200.4\\
NGC 5846     & 1.06$\pm$0.24 & & 471.4$\pm$574.6\\
NGC 6658     & 0.29$\pm$0.12 & $\checkmark$ & $\gtrsim$52452.0\\
NGC 7619     & 0.74$\pm$0.11 & & 1823.7$\pm$1122.5\\
\hline               
\end{tabular}
\end{minipage}
\end{table}

Interestingly, the ionisation cones observed in ESO0507-G025 (see the emission line ratio maps 
[O \,{\sc iii}]/H$\beta$ and [N \,{\sc ii}]/H$\alpha$ in Figure \ref{fig:example_figure}
and velocity fields in Figure \ref{fig:ESO0507_velocity}) together
with the high electron density and the shape of the spatially resolved 
[S \,{\sc ii}]$\lambda$6716 / [S \,{\sc ii}]$\lambda$6731 ratio map indicate 
that the central region of this galaxy is almost certainly ionised by an AGN \citep[e.g.,][]{Kakkad2018}.
This clear pattern is not observed in our other galaxies. 
In general, the presence of high electron density in the nuclear regions and extended enhanced
regions with densities $\gtrsim$1000 cm$^{-3}$ could indicate outflowing ISM likely
driven by expanding hot gas heated by SNe, SF stellar winds, AGN activity
and/or the collision/interaction of galaxies \citep{Westmoquette2012}.
We will discuss in detail in Section \ref{sec:discussion} the results obtained 
in this section and their implications for the evolutionary stages of the gas in our sample
galaxies.

\subsection{Star-formation rate}\label{sub:sfr}

Following a common practice, the current star-formation rates (SFRs) were estimated from 
the integrated H$\alpha$ luminosity L(H$\alpha$), assuming the \cite{Kennicutt1998} 
formula, for a solar metallicity, after correction for a \cite{Chabrier2003} initial mass 
function, i.e., SFR(H$\alpha$) (M$_{\odot}$ yr$^{-1}$) = 4.6 $\times$ 10$^{-42}$ $\times$ L(H$\alpha$) 
(erg s$^{-1}$) \citep{Parkash2019}. 
However, it is unlikely in our case that all H$\alpha$ emission results from SF, since 
the BPT diagnostics indicate a LINER or composite nature for most spaxels in our emission line maps. 
Therefore, the derived SFRs are likely to be an overestimates and are included here for comparison
with other studies. Table \ref{table:2b} shows the resultant SFRs, obtained from the integrated apertures 
of the galaxies.

We note that, our average SFRs is $\sim$16\% lower than that found by \cite{Kolokythas2022} 
using GALEX FUV fluxes as SF indicator, while FIR SFRs \citep{OSullivan2015} for some of our
galaxies (NGC 777, NGC 940, NGC 1060 and NGC 5846) are one or two orders of magnitude higher than
our SFR(H$\alpha$). Contamination of the FIR luminosity as a SF tracer can not be discarded.  
Most galaxies in both samples are AGN-dominated at FIR wavelengths and their derived FIR SFRs 
are greater than that expected from SF. However, in the case of NGC 940, they do not confirm 
the presence of a radio AGN  \citep{OSullivan2015}. 
In summary, we find that in this and the aforementioned studies, the SFR 
in type i and type ii group-dominant galaxies, on average, are  higher than in type i0 galaxies.

\section{Discussion}\label{sec:discussion}

\subsection{Aperture effects}\label{subsec:morphology_aper}

As shown in Section \ref{subsec:emission_ratios}, aperture selection and 2D mapping
can impact conclusions about the dominant mechanism ionising the
gas in our sample. This was demostrated in Figure \ref{fig:figures_BPT_int} where we show 
the [O \,{\sc iii}]/H$\beta$ vs. [N \,{\sc ii}]/H$\alpha$ diagnostic diagram for the nuclear and
integrated apertures. The displacement of the data points in this diagram when the aperture is
increased, indicates how our interpretation about the ionising sources can change depending on
aperture size. The limitations of using SDSS spectra for nearby galaxies has also been 
highlighted recently using IFU spectroscopy \citep[see][and references therein]{Gomes2016}.
In our case, we can see that in the inner most part of the galaxies, the gas is likely
dominated by shocks and/or AGN activity given that the nuclear
gas emission line ratios lie in the LINER region of the BPT diagram, however, at large apertures
photoionisation by OB stars likely becomes increasingly more important. 

\begin{figure*}
\hspace*{-1.2cm}
\includegraphics[width=0.75\textwidth]{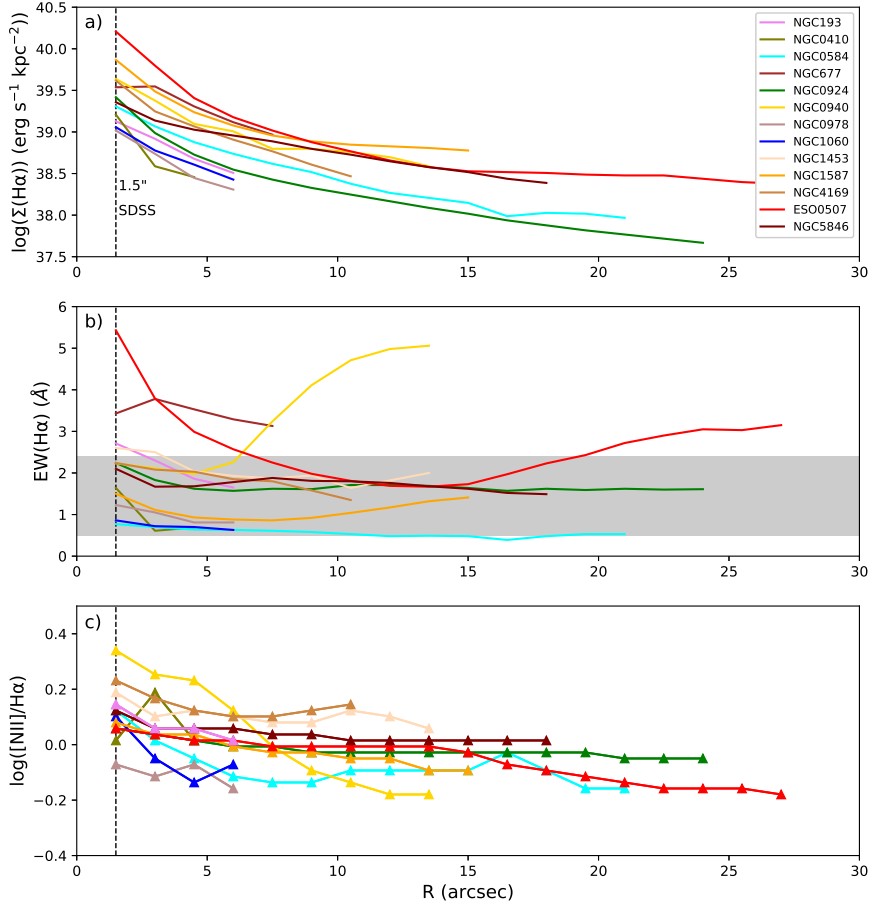}
    \caption{Observed gas properties as a function of 1.5" aperture bins. a) Luminosity surface
    density $\Sigma$(H$\alpha$). b) EW(H$\alpha$), the grey region between 0.5--2.4 $\AA$ 
    is consistent with pure pAGB photoionisation. For points with EW(H$\alpha$) $>$2.4 $\AA$ an
    additional gas excitation source is needed to account for the observed EW(H$\alpha$) 
    (Gomes et al., 2016). c) log([N \,{\sc ii}]/H$\alpha$) emission line ratio. 
    The vertical dotted line indicates the SDSS radial spectra fibre aperture.}
    \label{fig:figures_radial}
\end{figure*}

In Figure \ref{fig:figures_radial} we show the effects of using different apertures 
on the observed properties of our sample. 
In panel a) we show the L(H$\alpha$) surface density
$\Sigma$(H$\alpha$) calculated as L(H$\alpha$)/area, in b) the EW(H$\alpha$) 
and panel c) shows the log([N \,{\sc ii}]/H$\alpha$) emission line ratio. 
In this figure, the apertures were selected to increase in steps of 1.5" of radius.
The $\Sigma$, in panel a) of Figure \ref{fig:figures_radial}, in all cases decreases with 
radius and shows the presence of extended H$\alpha$ emission beyond the SDSS aperture. 
\cite{CidFernandes2011} used the EW(H$\alpha$) as an alternative method (explained below in
Section \ref{subsec:morphology_2DBPT}) to the BPT diagrams (emission-line classification). 
Using this method, \cite{Gomes2016} argue that evolved pAGB stellar background is
sufficient to photoionise the diffuse gas in ETGs and explain the observed EW(H$\alpha$) 
in the range 0.5 - 2.4 $\AA \rm$. Most EW(H$\alpha$) values in panel b) irrespective of aperture
lie within the area (grey region in the panel) which is consistent with pure pAGB
photoionisation \citep{CidFernandes2011,Gomes2016}. 
We find that one type ii galaxy (ESO0507-G025) and three type i galaxies
(NGC 193, NGC 677 and NGC 1453) show EW(H$\alpha$) $>$2.4 $\AA$ at their centres. 
These results indicate that pAGBs alone do not explain the observed values in these regions.
The EW(H$\alpha$) values of remaining galaxies are consistent with pure pAGB emission.
The [N \,{\sc ii}]/H$\alpha$ found in the nuclear regions (SDSS aperture; see panel c) 
and at large radii, in most cases, is consistent with LINER emission since
log([N \,{\sc ii}]/H$\alpha$) $\gtrsim$ 0.0. 
We observe that the log([N \,{\sc ii}]/H$\alpha$) declines with
increasing apertures in almost all cases, although the radial profiles differ significantly.

\subsection{Ionisation mechanisms}\label{subsec:morphology_2DBPT}

In this section we examine the mechanisms responsible for the ionisation which produces 
the optical emission lines in our sample.
For this, we use photoionisation models from the literature, which incorporate the
expected physical conditions in our sample galaxies.

\begin{figure}

\hspace*{-0.5cm}
\includegraphics[width=0.53\textwidth]{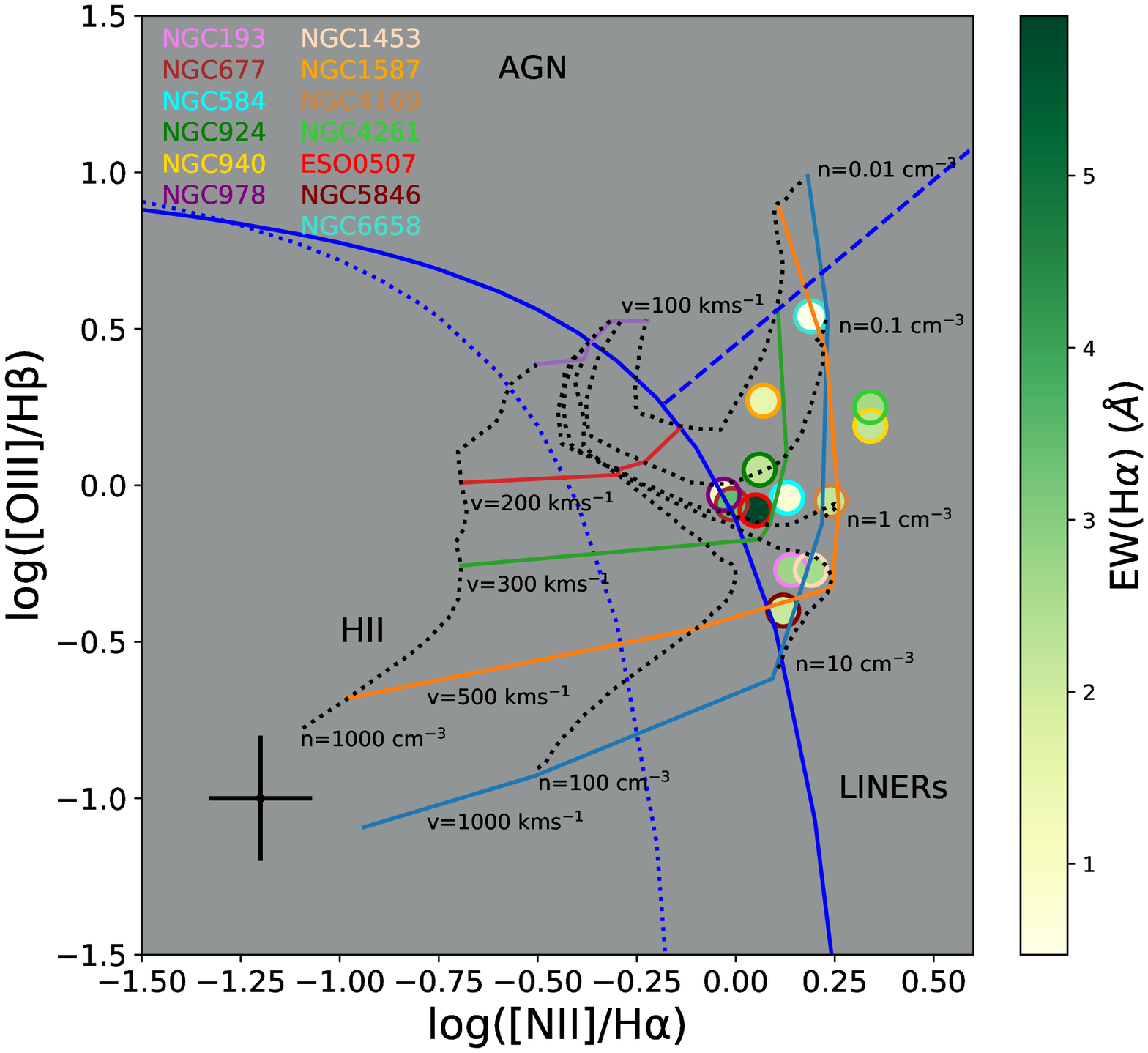}
    \caption{BPT [O \,{\sc iii}]/H$\beta$ vs [N \,{\sc ii}]/H$\alpha$ diagnostic diagram,
    similar to Figure 4, for the nuclear spectra (circles). 
    Superimposed the shock models of Allen et al. (2008) at solar metallicity (more details 
    in the text) and the Kewley et al. (2001; solid line), Kauffmann et al. (2003; dotted line) 
    and Schawinski et al. (2007; dashed line) boundaries as in Figure \ref{fig:figures_BPT_int}. 
    The fill colour of the circles indicates the EW(H$\alpha$) and the edge colours identifies 
    the galaxies from the legend.} 
    \label{fig:figures_BPT_EWHa_Shocks}
\end{figure}

In Figure \ref{fig:figures_BPT_EWHa_Shocks} we show the BPT diagram for the nuclear 
emission for our sample galaxies. The fill colour of the data points indicates the
EW(H$\alpha$) from the colour scale and the edge colour of the circles identifies the galaxies 
from the legend. Superimposed on the plot are shock models from \cite{Allen2008} calculated with 
the MAPPINGS III shock and photoionisation code. 
These shock models have solar metallicity, densities from 
n = 0.01 cm$^{-3}$ to 1000 cm$^{-3}$, velocities from v = 100 kms$^{-1}$ to 1000 kms$^{-1}$,
and magnetic field B = 1 $\mu $G. The models for v = 100 kms$^{-1}$ (purple), 200  kms$^{-1}$ 
(red), 300 kms$^{-1}$ (green), 500 kms$^{-1}$ (orange) and 1000 kms$^{-1}$ (blue) are shown as
coloured lines. The nuclear regions in our sample lie in the velocity range of > 200 kms$^{-1}$
to $\gtrsim$1000 kms$^{-1}$ and density between $\sim$0.1 cm$^{-1}$ to < 100 cm$^{-1}$. 
\cite{Annibali2010} compared the [S \,{\sc ii}] densities obtained for a sample of
ETGs/LINERs with ionised gas with the \cite{Allen2008} models. 
The majority of their galaxies had n$_e \gtrsim$ 100 cm$^{-3}$ for an assumed T = 10,000 K.
But as in our case (see Table \ref{table:properties_density}), a significant fraction of galaxies
have saturated [S \,{\sc ii}] ratios around 1.45 which are consistent with pre-shock densities
(n$_e\approx$ 0.01 to $\lesssim$100 cm$^{-3}$). 
These pre-shock regions are observed in some regions of Figure \ref{fig:density}. 
This could indicate together with the increasing 
[S \,{\sc ii}]$\lambda$6716 / [S \,{\sc ii}]$\lambda$6731 ratios that 
AGN/SF-driven shocks are ionising the gas mainly in the nuclear regions.
This would be consistent with the \cite{Molina2018} conclusion that the line emission 
of LINERs with low-luminosity AGN are predominantly powered by shocked gas due to jets or outflows
\cite[see also][]{Edwards2007}.

\begin{figure*}
\includegraphics[width=0.49\textwidth]{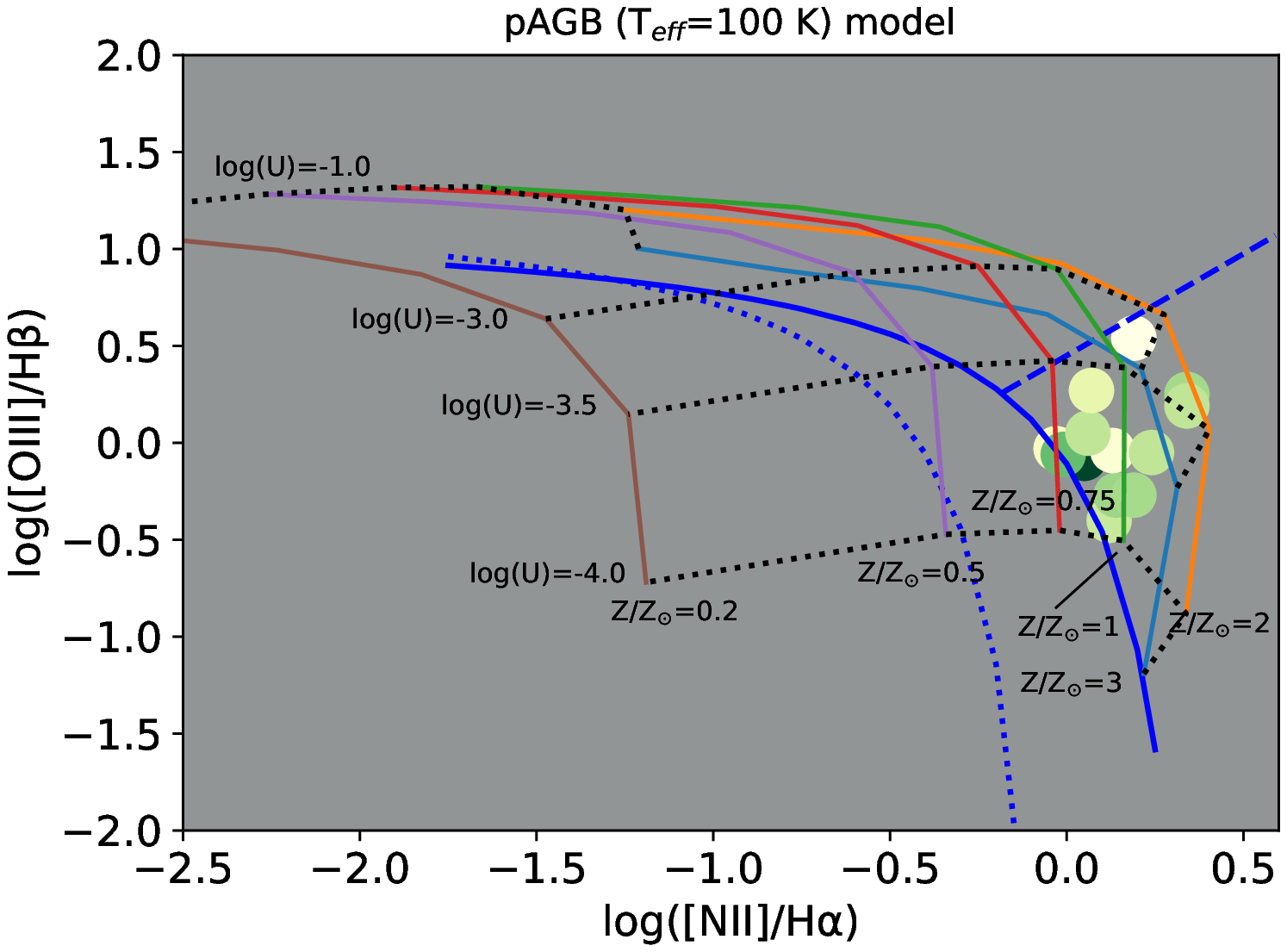}
\includegraphics[width=0.49\textwidth]{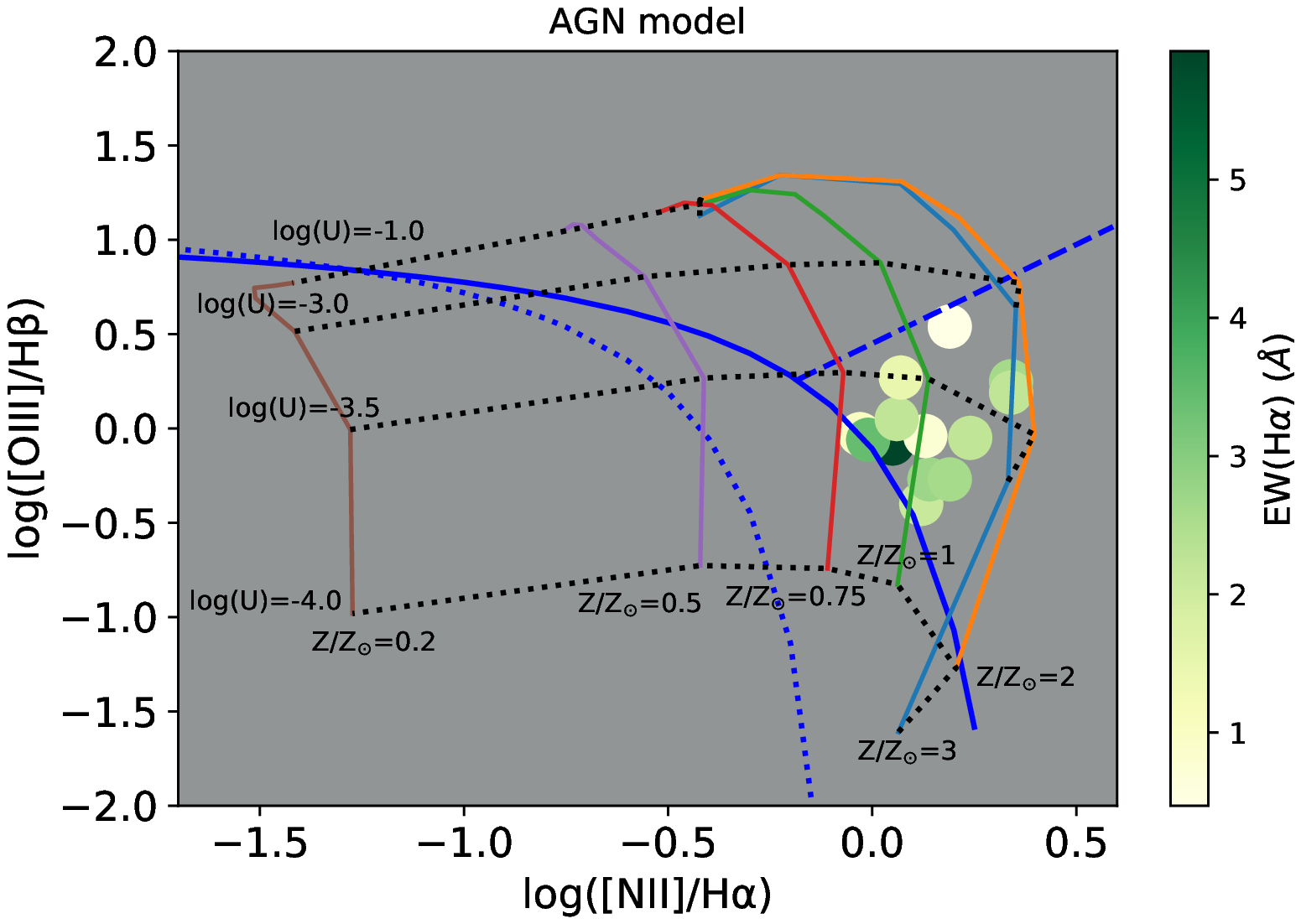}
\includegraphics[width=0.49\textwidth]{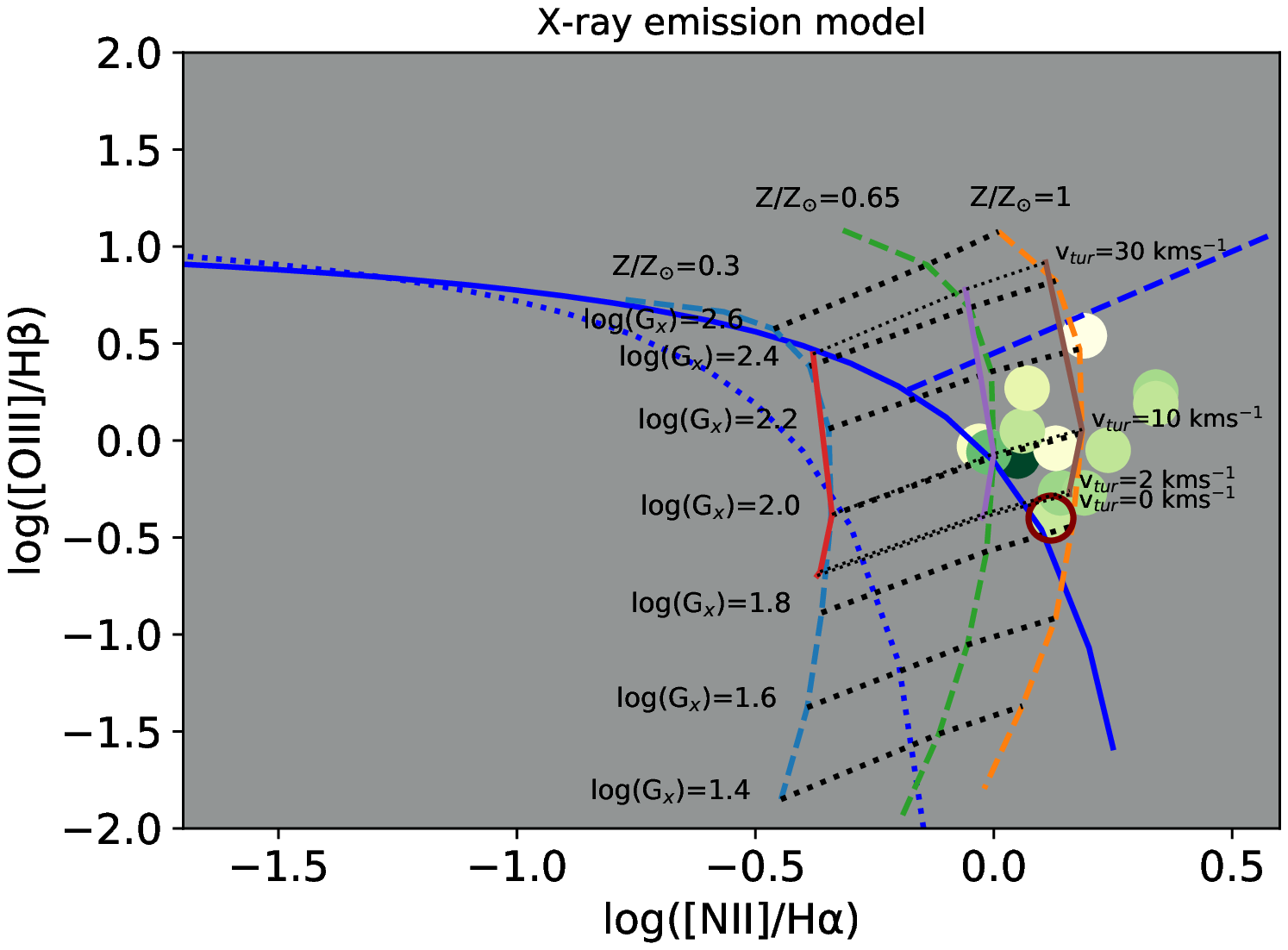}
\caption{BPT [O \,{\sc iii}]/H$\beta$ vs [N \,{\sc ii}]/H$\alpha$ diagnostic diagrams
for several photoionisation models. 
Upper panels: Colored lines represent the photoionisation models from Krabbe et al. (2021)
considering pAGB stars (left panel) and AGN activity (right panel) for metallicities 
from Z/Z$_{\odot}$ = 0.2 to 3. Black dotted lines correspond to different ionisation parameters 
from log(U) = -1 to 4.0. 
Bottom panel: Colored lines represent photoionisation models from Polles et al. (2021)
for three metallicities Z/Z$_{\odot}$ = 0.3, 0.65 and 1 considering the variation of X-ray
emission log(G$_x$) from 2.8 to 1.4 in steps of 0.2 dex (dotted coloured lines and dotted black
lines) and including turbulence velocity v$_{tue}$ = 30, 10, 2 and 0 km s$^{-1}$  
(solid colored lines and small dotted black lines). In these models the optical depth 
A$_V$ = 0.1 mag. The fill colour of the circles indicates the EW(H$\alpha$). 
The position of NGC 5846 is indicated by the open maroon circle.
Superimposed in blue the Kewley et al. (2001; solid line), Kauffmann et al. (2003; dotted line) 
and Schawinski et al. (2007; dashed line) boundaries as in Figure \ref{fig:figures_BPT_int}.}
\label{fig:figures_BPT_int_krabbe}
\end{figure*}

In Figure \ref{fig:figures_BPT_int_krabbe} (upper panels) we show the BPT diagrams, as in Figure
\ref{fig:figures_BPT_EWHa_Shocks}, but including two different photoionisation model grids
from \cite{Krabbe2021}. Their models consider the main (LINER) properties observed in our
sample of group-dominant galaxies for two distinct SEDs. 
One model assumes pAGB stars (upper left panel) with different T$_{eff}$ as the ionising source
and the other considers AGN SED with a multicomponent continuum (upper right panel).
The models were created using the CLOUDY code version 17.0 \citep{Ferland2017} for
metallicities Z/Z$_{\odot}$ = 0.2, 0.5, 0.75, 1.0, 2.0 and 3.0 and ionisation parameters 
U\footnote{The ratio of the ionizing photon density to the particle density.}
in the range of -4.0 $\le$ log(U) $\le$ -0.5 in steps of 0.5 dex. Their assumed density of n$_e$
500 cm$^{-3}$ is typical for BCGs \citep[e.g.,][]{Ciocan2021} and it is compatible with our values 
from Section \ref{subsec:density}. In both cases, most galaxies in our
sample lie in the region between Z/Z$_{\odot}$ = 0.75 (12 + log(O/H)= 8.56) and 
Z/Z$_{\odot}$= 1 (12 + log(O/H) = 8.69). 
However, NGG 4261 and NGC 940 have values compatible with models at 2-3 Z$_{\odot}$
(12 + log(O/H) = 8.99 - 9.17), assuming AGN activity. While for pAGB models, the values for 
NGG 4261, NGC 940, NGC 4169 and NGC 6658 are compatible with models at > 1.0 Z$_{\odot}$
metallicities. These results indicate that the nuclear regions in our sample are consistent,
within the uncertainties, with metallicities slightly (above) solar, independent 
of the ionising sources. In Section \ref{subsec:abundances} we estimate the gas phase
metallicity for each nuclear region in our sample.

In Figure \ref{fig:figures_BPT_int_krabbe} (Bottom pannel) we show the BPT diagram
overlaid with the predicted emission line ratios from CLOUDY models obtained by \cite{Polles2021}, 
which include photoionisation by X-ray emission. We show, in this figure, models for
three values of metallicity Z/Z$_{\odot}$ = 0.3, 0.65 and 1 and several values of X-ray emission
log(G$_x$) from 2.8 to 1.4 in steps of 0.2 dex with the turbulent velocity (produced by e.g., AGN
jets, turbulent mixing between the hot and cold phases and the collisions between filaments) 
fixed to 10 km s$^{-1}$ (dotted colored lines and dotted black lines). 
In the same panel, we added three models for the aforementioned metallicities and turbulence 
velocities v$_{tue}$ = 30, 10, 2 and 0 km s$^{-1}$ (solid colored lines and small dotted black
lines). The X-ray radiation field G$_{x}$ is fixed to 100. In general, these grid of models 
can reproduce the observed values without an excess in X-ray luminosity, even if the optical depth 
A$_V$ increases \citep[see figures E1 and E2 in][]{Polles2021}. 
In the case of NGC 5846, the observed emission line ratios (maroon open circle) are reproduced 
by models at very low (or no) turbulence velocities. 
Also, we find no sign of a clear rotational pattern (disturbed velocity field) in the FoV of the galaxy 
\cite[see the v$_r$ map in][]{Olivares2022} together with extended filaments of low velocity dispersion. 
This is in agreement with a cooling flows scenario, where the cool gas may have cooled
in (or close to) the centre of the group \citep{Temi2018,Jung2022}. In addition, in this galaxy
there is a good correlation between the CO cloud positions, detected by \cite{Temi2018}, and the
warm gas emission.

\begin{figure}
\centering
\hspace*{-0.3cm}
\includegraphics[width=0.5\textwidth]{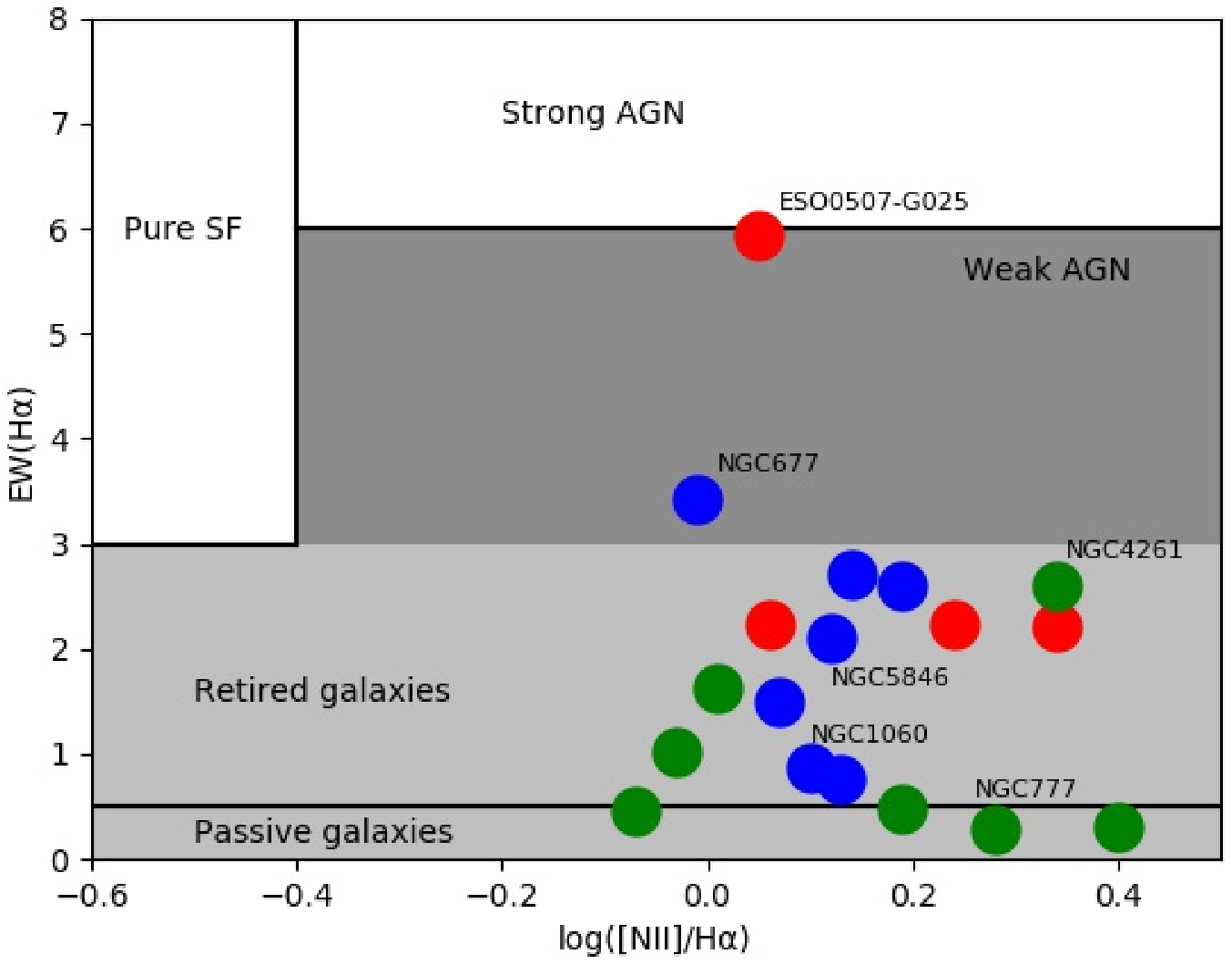}
    \caption{EW(H$\alpha$) vs. log [N \,{\sc ii}]/H$\alpha$ (WHAN) diagram. The light and dark grey
    regions represents the division between gas ionised by weak AGN and retired galaxies 
    (ionisation by pAGBs). The colour of the data points correspond to our H$\alpha$+[N \,{\sc ii}]
    morphological groups: green are type i0, blue are type i
    and red corresponds to type ii galaxies.}
    \label{fig:figures_BPT_WHAM}
\end{figure}

\cite{CidFernandes2011} introduced the WHAN diagram which uses the EW(H$\alpha$) 
(or W$_{H\alpha}$) in order to discriminate low ionisation AGN from galaxies
that are ionised by evolved pAGB stars. This diagram identifies 5 classes of galaxies:
1) pure star-forming galaxies: log([N \,{\sc ii}]/H$\alpha$) $<$-0.4 and EW(H$\alpha$)
$>$3 $\AA$, 2) strong AGN emission: log([N \,{\sc ii}]/H$\alpha$) $>$-0.4 and EW(H$\alpha$)
$>$6 $\AA$, 3) weak AGN: log([N \,{\sc ii}]/H$\alpha$) $>$-0.4 and EW(H$\alpha$) between 3 and
6 $\AA$, 4) retired galaxies: EW(H$\alpha$) $<$3 $\AA$ and 5) passive galaxies: EW(H$\alpha$)
and EW([N \,{\sc ii}]) $<$0.5 $\AA$. According to this classification scheme 
most of our sample (see Figure \ref{fig:figures_BPT_WHAM}) can be classified as retired
galaxies likely ionised by pAGB stars, independent of their position on the BPT diagram 
(see the EW(H$\alpha$) colour scale values in Figures \ref{fig:figures_BPT_EWHa_Shocks}
and \ref{fig:figures_BPT_int_krabbe}) and H$\alpha$+[N \,{\sc ii}] emission morphology.
Only in ESO0507-G025 and NGC 677 do we see clear indications of AGN activity in the
nuclear region from the WHAN diagram, with the EW(H$\alpha$) $\sim$3--6 $\AA$ indicating 
a weak AGN). However, some galaxies in our sample have clear signatures of AGN
activity as reported in studies at other wavelengths. For example, NGC 4261 is the brightest
galaxy in the NGC 4261 group. Hubble Space Telescope (HST) WFPC2 observations \citep{Jaffe1993}
and our MUSE data reveals a bright nuclear optical source surrounded by a disc of gas and dust, 
while radio observations identified two jets perpendicular to the disc
\citep{BirkinshawDavies1985,Scheneider2006,Kolokythas2015}. 
Recent kinematical studies using ALMA CO \citep{Boizelle2021} find a dynamical mass for the central 
supermassive black hole (BH) in NGC 4261 of 1.67$\times$10$^9$ M$_{\odot}$. 
So, the observed LINER properties in this galaxy may be explained 
by an obscured low-luminosity AGN \citep{Zezas2005}.
In addition, 17/18 galaxies in our sample show detected radio continuum emission
\citep{Kolokythas2018}; 4/18 show small ($\lesssim$20 kpc; NGC 1060 and NGC 5846) and large-scale 
($>$20 kpc; NGC 193 and NGC 4261) jets, 3/10 diffuse (NGC 677, NGC 1587 and ESO0507-G025) and
10/18 a point-like or unresolved point source radio continuum morphology
($\lesssim$11 kpc; NGC 410, NGC 584, NGC 777, NGC 924, NGC 940, NGC 978, NGC 1453, NGC 4008, 
NGC 4169 and NGC 7619). 

The results in this section indicate that EW(H$\alpha$), log [N \,{\sc ii}]/H$\alpha$ 
and the BPT diagrams, alone, cannot distinguish between dominant ionising
sources producing the central and extended LINER-like emission in our sample. 
However, the overall log [N \,{\sc ii}]/H$\alpha$  (Figure  \ref{fig:figures_BPT_int}) 
shows that this emission line ratio decreases as the extent of the emission line region increases,
indicating that \textit{the ionisation sources are having different impacts at different radii}.
The same decreasing value with the distance pattern is observed for $\sigma$([N \,{\sc ii}]), the
12 + log(O/H) abundances and in most cases the EW(H$\alpha$). 
Central regions are almost certainly are influenced by low-luminosity AGN, while the extended regions
are ionised by other mechanisms (SF and/or cooling flows shocks and pAGBs).
The EW(H$\alpha$) in the outer parts is in the range of 0.5-2.4 $\AA$ 
(see Figure \ref{fig:figures_radial}), thus indicating that pAGBs are likely contributing significantly 
to the ionization of these regions. Finally, our interpretations are based on the comparison with models. 
Factors like $Ly_{c}$ photon escape and dilution of nuclear EWs \citep{Papaderos2013} have not been
considered. Consequently, the addition of these processes may constitute an important
element in understanding of ETGs with extended optical LINER-like emission.

\subsection{Gas-phase abundances 12 + log(O/H) and metallicity gradients}\label{subsec:abundances}

Warm gas-phase abundances in our sample are difficult to obtain at optical wavelengths 
due to the uncertainty about the ionisation mechanism (see Section \ref{sec:Intro} 
and previous results in Section \ref{subsec:morphology_2DBPT}).
The wavelength range covered by MUSE does not allow the detection of 
[OII]$\lambda$3727 and [OIII]$\lambda$4363. Therefore, the 12+log(O/H) abundance in each nuclear
region was derived using linear interpolation between the models (AGN, pAGBs and X-ray emission)
and their measured emission line ratios (log([O\,{\sc iii}]/H$\beta$ 
and log([N \,{\sc ii}]/H$\alpha$)). Additionally, we used the following calibrators: 
i) AGN calibration (AGN N2 calibrator) proposed by \cite{Carvalho2020}: 

\begin{equation}
12 $ + $ log\left(\frac{O}{H}\right) $ = $ 12 $ + $ log\left((\frac{Z}{Z_{\odot}} \times 10^{log\left(\frac{O}{H}\right)_{\odot}}\right),
\end{equation}

\noindent
where Z/Z$_{\odot}$ = 4.01$^{N2}$ - 0.07 for 0.3 < (Z/Z$_{\odot}$) < 2.0,
and N2=log([N \,{\sc ii}]$\lambda$6584/H$\alpha$). 
Although this calibration was developed for seyfert 2 nuclei, it has been used
to derive the O/H abundance in narrow line AGN regions and LINERs 
\cite[e.g.,][]{Krabbe2021,doNascimento2022}.
ii) most of our extended regions lie to the right of the empirical \cite{Kauffmann2003} boundary
line on the BPT diagrams in Figures \ref{fig:example_figure} and \ref{fig:BPT_maps_1} 
(appendix \ref{Maps_sample}). 
The O3N2 ratio was suggested by \cite{Kumari2019} as a metallicity tracer of DIG and
LI(N)ERs (EW(H$\alpha$)< 3 $\rm \AA$) since these regions are likely ionised by pAGB stars, 
star-forming clusters and weak AGN. Therefore, the abundances in these regions can be estimated
by:

\begin{equation}
12 $ + $ log\left((\frac{O}{H}\right) $ = $ 7.673 $ + $ 0.22 \times \sqrt[]{25.25 - 9.072 \times O3N2} $ + $ 0.127 \times O3,
\label{eq:oxygen_O_Kumari}
\end{equation}
with O3N2 = log([O \,{\sc iii}]$\lambda$5007/H$\beta \times$H$\alpha$/[N \,{\sc ii}]$\lambda$6584) and O3 = log([O \,{\sc iii}]$\lambda$5007/H$\beta$.

\noindent
Finally, iii) we compare those values with the ones inferred from the N2 diagnostic, calibrated
by \cite{Marino2013} (star-forming H\,{\sc ii} calibrator) obtained using empirically 
calibrated direct abundance data (T$_e$-based measurements) from H\,{\sc ii} regions in 
the CALIFA survey. The \cite{Marino2013} calibration is defined as:

\begin{equation}
12 $ + $ log\left(\frac{O}{H}\right) $ = $ 8.743 $ + $ 0.462 \times N2,
\label{eq:oxygen_N2_M13}
\end{equation}

\noindent
We note that the emission line ratios used in any of these relations are not highly affected by reddening.
\noindent

\begin{figure}
\hspace*{-0.4cm} 
\includegraphics[width=0.5\textwidth]{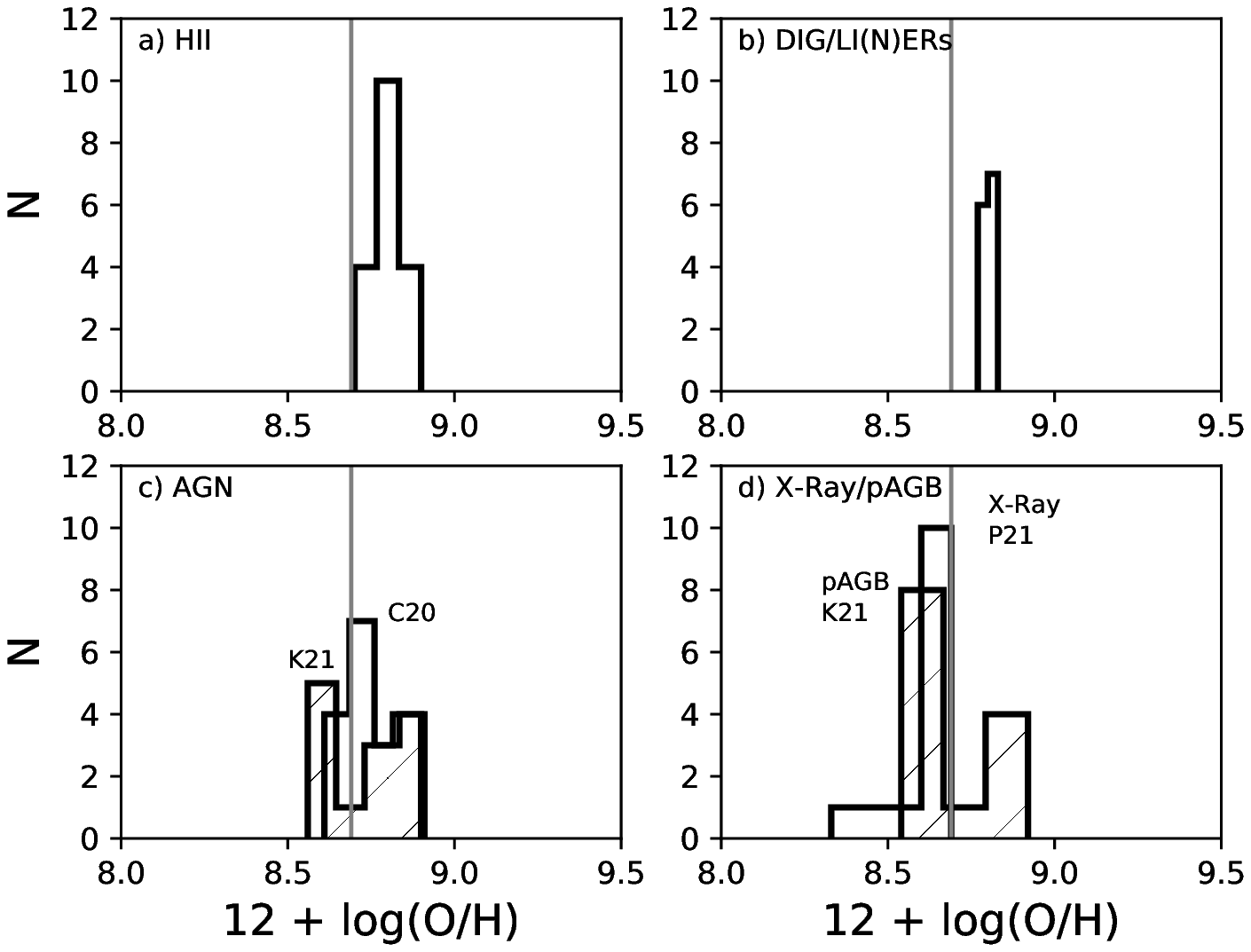}
    \caption{Oxygen abundance distributions for the nuclear SDSS 3" apertures
    in panels a) ionization by stars (H \,{\sc ii}) using the N2  
    calibration by Marino et al. (2013). Panel b) shows the abundances from DIG/LI(N)ERs O3N2 
    calibration by Kumari et al (2019), respectively. 
    Panel c) shows the distribution of abundances from AGN photoionization model 
    (Krabbe et al. 2021; K21) and Carvalho et al. (2020; C20) calibration. 
    Panel d) shows the same but considering pAGB photoioniazation (K21) and X-ray
    (Polles et al. 2021; P21) models. 
    The vertical grey lines indicate the solar abundance 12 + log(O/H)$_{\odot}$ = 8.69.}
    \label{fig:figures_Histogram_OH}
\end{figure}

In Table \ref{table:abundances} we show the oxygen abundance for the nuclear 3" region of
each galaxy using the aforementioned methods and their distributions are shown in Figure \ref{fig:figures_Histogram_OH}. 
In the figure and table we see that the 12 + log(O/H) derived
from the AGN model and the AGN N2 calibrator are in agreement within the uncertainties 
(of $\pm$0.1 dex) with those inferred from the H\,{\sc ii} calibration in $\sim$62\% (8/13) and 
$\sim$89\% (16/18) of the galaxies, respectively. These values drop to $\sim$38\% (5/13) and 
$\sim$0\% (0/13) when compared to pAGB and X--ray emission models, respectively. However, 100\% (13/13)
of the nuclear DIG/LI(N)ERs abundances are in agreement, at 1$\sigma$ level, with the  H\,{\sc ii} values.
We note that nuclear metallicities obtained from pAGB and X-ray emission models are, in most
cases, lower compared to those obtained from the H\,{\sc ii}-based method.
Two of our galaxies are included within the \cite{Annibali2010} sample of ETGs
with ionised gas: NGC 1453 and NGC 5846. 
Those authors found 12 + log(O/H) = 8.55$\pm$0.19 and 8.84$\pm$0.17 using the calibration 
in \cite{Kobulnicky1999} for NGC 1453 and NGC 5846, respectively. 
Our measurements are in reasonable agreement within the errors with those derived 
by \cite{Annibali2010} considering apertures and the intrinsic differences between the calibrations. 

\begin{table*}
\centering
\begin{minipage}{152mm}
\caption{Oxygen abundance determinations for the nuclear region (3" aperture)
assuming pure H\,{\sc ii} regions (Marino et al. (2013; H\,{\sc ii}),
AGN models (Krabbe et al. 2021) and the AGN N2-based calibrator (Carvalho et al. 2021), the DIG/LI(N)ERs O3N2  calibrator (Kumari et al. 2019; DL), and pAGB (Krabbe et al. 2021) and X-ray (Polles et al. 2021) emission models, respectively.}
\label{table:abundances}                              
\begin{tabular}{l c c c c c c c c} 
\hline
             & 12 + log(O/H)$_{HII}$ &  \multicolumn{2}{c}{12 + log(O/H)$_{AGN}$} & 12 +log(O/H)$_{DL}$ & 12 + log(O/H)$_{pAGB}$ & 12 + log(O/H)$_{X-ray}$ \\
             & N2 & model & N2 & O3N2 & model & model\\
\hline 
NGC 193      & 8.81$\pm$0.09 &  8.76$\pm$0.16 & 8.75$\pm$0.10 & 8.82$\pm$0.10 & 8.63$\pm$0.16 & 8.68$\pm$0.10\\
NGC 410      & 8.75$\pm$0.05 &  \dots         & 8.66$\pm$0.07 & \dots         & \dots& \dots\\
NGC 584      & 8.80$\pm$0.10 &  8.71$\pm$0.16 & 8.74$\pm$0.11 & 8.81$\pm$0.16 & 8.60$\pm$0.17 & 8.67$\pm$0.15\\
NGC 677      & 8.74$\pm$0.08 &  8.56$\pm$0.08 & 8.65$\pm$0.10 & 8.78$\pm$0.10 & 8.55$\pm$0.05 & 8.48$\pm$0.27\\
NGC 777      & 8.87$\pm$0.13 &  \dots         & 8.84$\pm$0.13 & \dots         & \dots& \dots\\
NGC 924      & 8.77$\pm$0.07 &  8.57$\pm$0.11 & 8.70$\pm$0.08 & 8.79$\pm$0.15 & 8.55$\pm$0.12 & 8.60$\pm$0.18\\
NGC 940      & 8.90$\pm$0.11 &  8.90$\pm$0.10 & 8.88$\pm$0.10 & 8.83$\pm$0.17 & 8.92$\pm$0.17 & >8.69\\
NGC 978      & 8.71$\pm$0.07 &  8.56$\pm$0.01 & 8.64$\pm$0.08 & 8.77$\pm$0.08 & 8.55$\pm$0.03 & 8.33$\pm$0.25\\
NGC 1060     & 8.79$\pm$0.12 &  \dots         & 8.72$\pm$0.14 & \dots         & \dots& \dots\\
NGC 1453     & 8.83$\pm$0.06 &  8.83$\pm$0.06 & 8.78$\pm$0.05 & 8.83$\pm$0.10 & 8.74$\pm$0.14 & >8.69\\
NGC 1587     & 8.78$\pm$0.06 &  8.59$\pm$0.11 & 8.70$\pm$0.07 & 8.77$\pm$0.08 & 8.54$\pm$0.10 & 8.60$\pm$0.13\\
NGC 4008     & 8.71$\pm$0.12 &  \dots         & 8.61$\pm$0.15 & \dots         & \dots& \dots\\
NGC 4169     & 8.85$\pm$0.06 &  8.85$\pm$0.04 & 8.81$\pm$0.05 & 8.83$\pm$0.12 & 8.81$\pm$0.10 & >8.69\\
NGC 4261     & 8.90$\pm$0.13 &  8.90$\pm$0.13 & 8.88$\pm$0.12 & 8.83$\pm$0.19 & 8.91$\pm$0.26 & >8.69\\
ESO0507-G025 & 8.76$\pm$0.06 &  8.57$\pm$0.08 & 8.69$\pm$0.07 & 8.79$\pm$0.14 & 8.55$\pm$0.07 & 8.59$\pm$0.16\\
NGC 5846     & 8.80$\pm$0.10 &  8.75$\pm$0.18 & 8.74$\pm$0.11 & 8.83$\pm$0.09 & 8.59$\pm$0.16 & 8.67$\pm$0.14\\
NGC 6658     & 8.83$\pm$0.11 &  8.81$\pm$0.28 & 8.78$\pm$0.11 & 8.78$\pm$0.31 & 8.81$\pm$0.32 & >8.69\\
NGC 7619     & 8.93$\pm$0.20 &  \dots         & 8.91$\pm$0.22 & \dots         & \dots& \dots\\
\hline                                   
\end{tabular}
\end{minipage}
\end{table*}

\begin{figure*}
\includegraphics[width=0.8\textwidth]{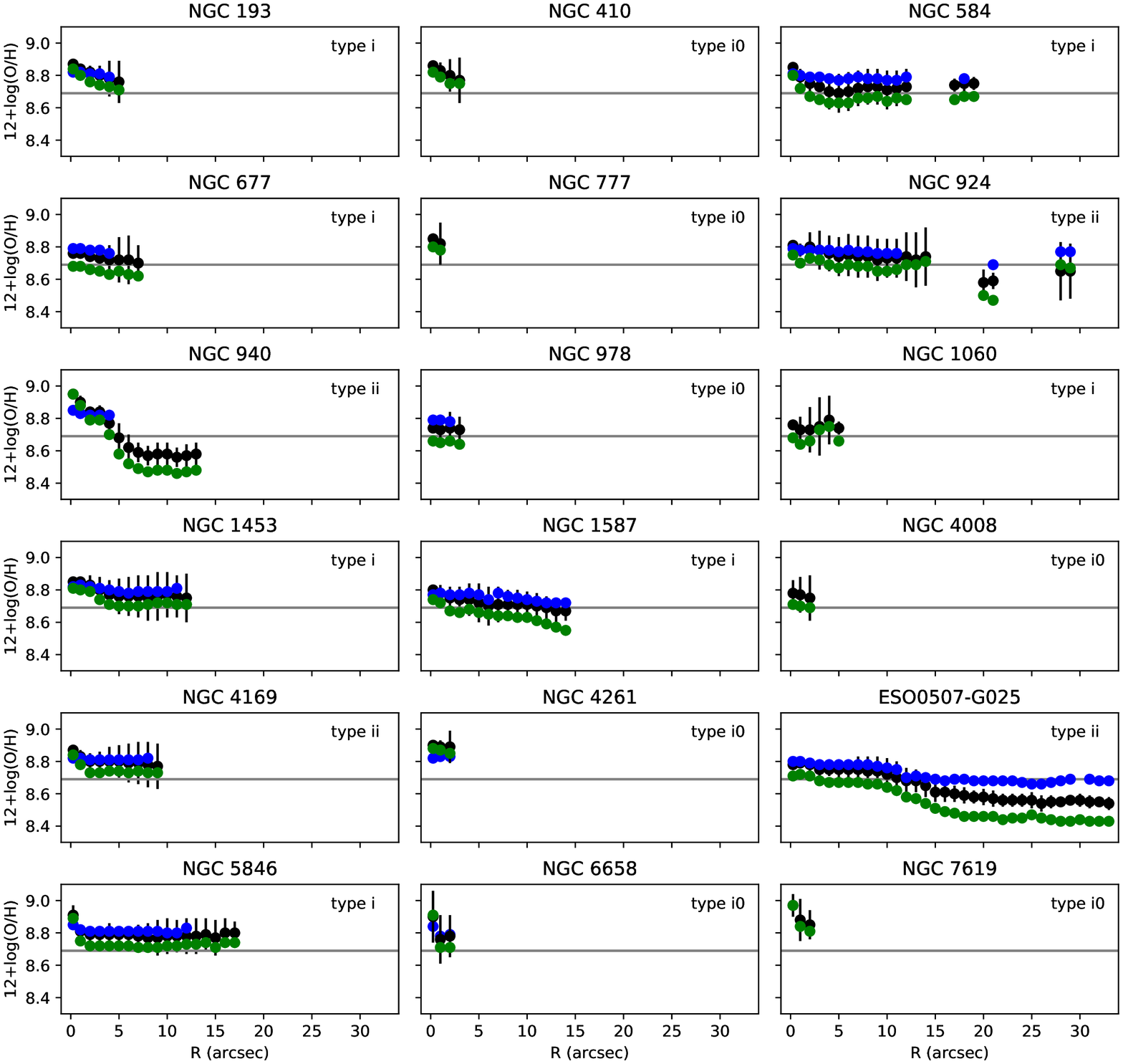}
    \caption{12 + log(O/H) profiles for the sample galaxies by adopting the N2
    calibrations by Marino et al. (2013) (HII; black data points), Carvalho et al. (2020) 
    (AGN; green data points) and Kumari et al. (2019) (DIG/LI(N)ERs; blue data points).
    The data points corresponds to the median 12 + log(O/H) from bins in steps of 1.0" radii,
    except the central bin which is 0.5". The error bars denote the 1$\sigma$ distribution 
    of the y-values per each bin. The solar abundance 12 + log(O/H)$_{\odot}$ = 8.69 is indicated
    by the grey horizontal lines.}
    \label{fig:figures_OH_profiles_pp}
\end{figure*}

It is important to bear in mind that the N2 metallicity diagnostics are known to have a dependency 
on the ionisation parameter and the nitrogen-to-oxygen ratio of the gas,
given that metallicities increase with N-enrichment.
Unfortunately, we are unable to explore the extended metallicity distribution using  
methods that do not have a dependence on the aforementioned parameters, given that only
the [N \,{\sc ii}]$\lambda$6584 and H$\alpha$ emission lines were detected 
in most of our sample galaxies.
On the other hand, the O3N2 method gives nuclear abundances that are in agreement,
within the uncertainties, with the ones found using the N2-based abundances. 
Therefore, we used the N2 and O3N2 indicators, in galaxies with extended 
[N \,{\sc ii}]$\lambda$6584 emission, as a way of obtaining the spatially resolved morphologies
of the ionised gas which can be translated into metallicities in a relative rather than 
absolute way. In Section \ref{subsec:emission_ratios} we showed that most spaxels in our
spatially resolved BPT maps lie in the AGN and composite areas of the diagrams. 
Therefore, we calculated the pixel-by-pixel 12 + log(O/H) abundance
in those regions by adopting the \cite{Carvalho2020} (AGN N2; green dots) and \cite{Kumari2019} 
(DIG/LI(N)ERs O3N2; blue dots) calibrators. While values from the N2 calibration by \cite{Marino2013} 
(H\,{\sc ii}; black dots) are included for comparison. 
In Figure \ref{fig:figures_OH_profiles_pp} we show for each galaxy in our
sample the 12 + log(O/H) abundances as function of the radius from the galaxy's centre. 
The dots are the median values within circular bins of 1.5" radii, except for 
the first bin which has a radius of 0.5". The error bars, in this figure, denote the 1$\sigma$
distribution of the H\,{\sc ii} 12 + log(O/H) per bin.  

In Figure \ref{fig:figures_OH_profiles_pp} we see that the calibrators predict comparable
metallicities but with an small offset ($\lesssim$0.1 dex in most cases) 
in the innermost regions, while for the extended regions this difference can reach values of 
$\sim$0.3 dex in the case of ESO0507-G025.
From this figure it is clear that there is a break in the metallicity gradient slope
with a very steep gradient in the central region which, as indicated 
in Section \ref{subsec:morphology_2DBPT}, is more influenced by low-luminosity AGN.
We calculated the metallicity gradient ($\nabla_{\text{O/H}}$) as the slope of the linear fit to the
median 12 + log(O/H) values separately for the innermost and extended regions for all
the calibrations considered. In Table \ref{table:4} we present for each galaxy the results of our
linear fitting and statistics of all pixels/spaxels used to create the 12 + log(O/H) profiles.
From this table we see that the central metallicity gradients are in all cases negative, while
the extended regions show a flat gradient.

\subsection{Properties of the gas and its origin}\label{subsec:On_the_properties}

\subsubsection{Properties of the gas and metallicity gradients}\label{subsec:properties_gas}

\begin{figure}
\centering
\hspace*{-0.2cm}
\includegraphics[scale=0.5]{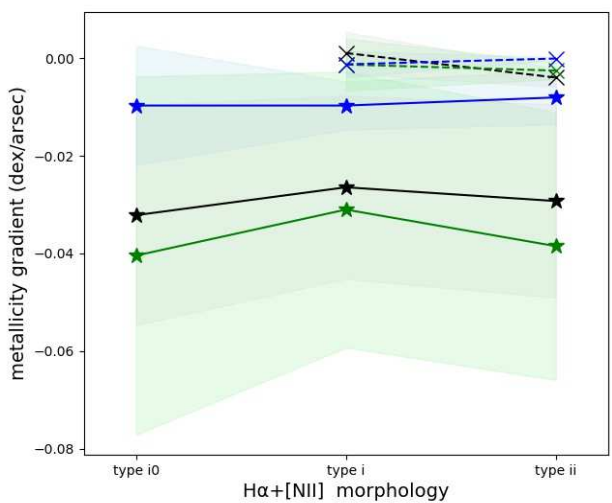}
    \caption{Relationship between metallicity gradients $\nabla_{\text{O/H}}$ and 
    H$\alpha$+[N\,{\sc ii}] morphology of the galaxies. 
    More details about the classes of emission line morphology are given in Section
    \ref{subsec:emission}. Horizontal lines indicate the average of the metallicity
    gradients $\nabla_{\text{O/H}}$ for central (stars) and extended (crossed) fit components
    considering three different calibrators for H\,{\sc ii} regions (black; Marino et al. 2013), 
    AGN (green; Carvalho et al. 2020) and DIG/LIERs (blue; Kumari et al. 2019).
    The standard deviation for each quantity is indicated by the filled regions.
    }
    \label{fig:type_gradient}
\end{figure}

From the previous section we see that the mean gas-phase metallicity in the nuclear and extended 
regions are $\langle$(O/H)$_{nuclear}\rangle$ > $\langle$(O/H)$_{extended}\rangle$
with the 12 + log(O/H) abundance generally decreasing with radius, independently of the ionisation
source or method considered, with a flattening of metallicity gradients for the outermost regions. 
In Figure \ref{fig:type_gradient} we show the relationship between metallicity gradients 
$\nabla_{\text{O/H}}$ for the central and extended regions and H$\alpha$+[N\,{\sc ii}] morphology 
of our sample. While in Table \ref{table:metallicity_gradients} we summarise the average 
metallicity gradients for each region and calibrator considered. 
A weak positive Spearman's correlation $\sim$0.6 between metallicity
gradients and morphological H$\alpha$+[N\,{\sc ii}] types of the galaxies is found
in Figure \ref{fig:type_gradient}. Despite the low number statistics in our study, these results
suggest that the nuclear $\nabla_{\text{O/H}}$ of type i group-dominant galaxies, on average, is 
lightly higher than type i0 and type ii galaxies. However, the intrinsic uncertainties associated
with these values lead us to consider the metallicity gradients may not be statistically
significant. Therefore, we argue that the similarity in the shape of the nuclear metallicity
gradients in all galaxy types and the flattening of the outer regions in our sample of 
group-dominant galaxies with type i (strong nuclear emission plus extend filamentary structures)
and type ii (strong or diffuse nuclear emission plus extranuclear H\,{\sc ii} regions) morphologies 
are a common property in this group of galaxies.
Several groups \cite[e.g.,][and references therein]{Sanchez2014,Belfiore2017} have studied the 
metallicity gradients in local galaxy samples. In \cite{Sanchez2014} and \cite{Sanchez2020} 
they show that a significant fraction of galaxies in the CALIFA sample exhibit shallow metallicity 
slopes in their innermost and/or outermost regions.
In particular, they argue that the flattening in the outer regions is a universal
property of disc galaxies, which is independent of the inclination, mass, luminosity,
morphology and the presence of bars. In this and other similar works, several mechanisms like
radial motion, inside-out growth, metal-poor/rich gas accretion, turbulent transport and
outflow of gas \citep[e.g.,][see references therein]{Kewley2010,SanchezMenguiano2018,Sanchez2020} 
are invoked as the sources of producing the gas metallicity profiles. 

\begin{table*}
\centering
\begin{minipage}{130mm}
\caption{Average and standard deviation (sd) metallicity gradients ($\nabla_{\text{O/H}}$), in units of 
dex/arcsec, for the central and extended regions (left and right values in each column) using the
calibrators H\,{\sc ii} N2 regions (Marino et al. 2013), AGN N2 (Carvalho et al. 2020) and O3N2 DIG/LI(N)ERs 
(Kumari et al. 2019).}      
\label{table:metallicity_gradients}                              
\begin{tabular}{l c c c c c c} 
\hline
        & \multicolumn{2}{c}{H\,{\sc ii} N2} & \multicolumn{2}{c}{AGN N2} & \multicolumn{2}{c}{DIG/LIERs O3N3 }\\
        & central & extended & central & extended & central & extended \\   
        & $\langle\nabla_{\text{(O/H)}}\rangle$/sd& $\langle\nabla_{\text{(O/H)}}\rangle$/sd & $\langle\nabla_{\text{(O/H)}}\rangle$/sd & $\langle\nabla_{\text{(O/H)}}\rangle$/sd& $\langle\nabla_{\text{(O/H)}}\rangle$/sd& $\langle\nabla_{\text{(O/H)}}\rangle$/sd\\
\hline 
type i0 &  -0.032/0.022  & \dots          & -0.040/0.037 & \dots          & -0.010/0.012 & \dots        \\
type i  &  -0.026/0.019  &  0.0011/0.0043 & -0.031/0.028 & -0.0013/0.0053 & -0.010/0.005 & -0.0012/0.0027\\
type ii &  -0.029/0.020  & -0.0039/0.0019 & -0.038/0.027 & -0.0025/0.0017 & -0.008/0.005 & -0.0000/0.0003 \\
\hline                                   
\end{tabular}
\end{minipage}
\end{table*}

The properties of the extended warm ISM in our sample of type i and type ii galaxies
suggest, within the uncertainties, nearly solar ($\pm$0.2 dex) homogeneous 
chemical abundances (see Figure \ref{fig:figures_OH_profiles_pp} and Table \ref{table:4}). 
This likely requires mechanisms (e.g., radial motions, gas-clouds or satellite accretion/interactions, 
AGN/SF-driven outflows) for the efficient gas transport, mixing and radial 
flattening of metallicity into the outer regions of the galaxies in a relative short time scale 
\citep[e.g.,][]{Werk2011,Bresolin2012,Rennehan2019,Rennehan2021}, likely similar to other low
redshift galaxy classes. 

\subsubsection{Cold gas content}\label{subsec:HI_gas}

H\,{\sc i} is a sensitive tracer of external environmental mechanisms in galaxies.
A study of H\,{\sc i} in ETGs by \cite{Serra2012} found that H\,{\sc i} detections were
relatively uncommon near Virgo cluster centre (10\%) but were common (40\%) in field
with the detected H\,{\sc i} mass inversely related to environment density. 
H\,{\sc i} morphology was also found to vary in a continuous way from regular, settled H\,{\sc i}
discs and rings in the field to unsettled gas distributions (including tidal or accretion tails)
at the Virgo cluster centre, with the H\,{\sc i} and CO-richest galaxies found in the poorest
environments where the SF detection rate was also higher. This implies that galaxy group 
processing is involved in evolving pre-existing ETG gas properties.

In our sample, 8/18 galaxies have the H\,{\sc i} properties available 
from the literature and 18/18 have been observed in CO \citep[see our Table \ref{table:properties_HI} 
for a summary;][]{OSullivan2015,OSullivan2018}. 
In Figure \ref{fig:figures_HI} we show single dish H\,{\sc i} spectra from the literature 
for 7 of these galaxies. Excluding the two galaxies in Table \ref{table:properties_HI} which
have a caveat about their H\,{\sc i} properties (see Table \ref{table:properties_HI} note)
the mean H\,{\sc i} in type ii galaxies is 17.2$\times$10$^{9}$ M$_{\odot}$ compared to the
mean H\,{\sc i} in type i galaxies of 1.4$\times$10$^{9}$ M$_{\odot}$, i.e., the type ii have an 
order of magnitude more H\,{\sc i}.
NGC 940 was excluded from the above calculation because of the high uncertainty about 
it's H\,{\sc i} detection, however it has a large M(H$_2$) mass (6.1$\times$10$^{9}$ M$_{\odot}$).
The H$_2$ mass of NGC 940 together with the H\,{\sc i} masses of the other type ii galaxies confirms 
that the type ii galaxies are cold gas rich.
To varying degrees, all the H\,{\sc i} in our galaxies display double horned H\,{\sc i} profiles
which are indicative of rotating discs, with the clearest examples being NGC 924 and NGC 940 
which also have the lowest H\,{\sc i} profile asymmetries as measured by the A$_{flux}$ parameter 
\citep{Espada2011}. 

Its seems likely the cold and warm discs in the type ii galaxies are part of the same kinematical
gas structures, with support for this coming from the H\,{\sc i} profiles and the [N\,{\sc ii}]
velocity fields in Figure \ref{fig:ESO0507_velocity} and Appendix \ref{appendix_NII_velocity_maps}.
In the  [N\,{\sc ii}] velocity fields we see that all of the type ii galaxies show clear
rotating disc patterns, with NGC 940 presenting a highly symmetric case. The high levels
of symmetry in the velocity fields, especially in NGC 940, together with the flattened metallicity
gradients argue against the gas having been recently acquired. 
In particular, using Romulus hydrodynamic cosmological simulations \cite{Jung2022}
examined the re-emergence of gaseous and stellar discs in BGGs following their destruction by mergers.
They find that ordered discs take $\sim$1 Gyrs to be established.
We suggest that our gas-rich systems obtained their cold gas at least $\sim$1 Gyr ago
\citep{Holwerda2011,Jung2022}, i.e., the H\,{\sc i} virialization time scale after gas-clouds
or satellite mergers/accretion.
However, we observe that $\sigma_{gas,central} \gtrsim \sigma_{gas,extended}$,
so the warm gas in the central regions is unlikely to be dynamically relaxed \citep{Olivares2022} 
in the gravitational potential of the galaxy, indicating an AGN outburst contribution.
Those results are in agreement with the morphology of the warm gas distribution and velocity 
fields observed in our sample (Section \ref{sub:velocity}).

\subsubsection{Summary of properties, chemical abundances and possible scenarios for the origin of the gas in group-dominant galaxies}\label{subsec:origin_gas}

The interpretation of internal (stellar mass loss) and external mechanisms 
(cooling from the IGrM or mergers/interactions) for the origin of the gas in 
our sample is difficult, since different mechanisms could be acting at different
evolutionary stages. \cite{Kolokythas2022} examined the relationships between radio power,
SFR(FUV) and stellar mass for CLoGS, which includes our analysed galaxies, and find no correlations 
between these quantities. This suggests a mix of origins for the cool gas in these systems, including 
stellar mass loss, cooling from the IGrM/galaxy halo, settling gas due to mergers and tidal 
or ram pressure stripping \citep[see also][]{Loubser2022}. 
In systems like NGC 978 and NGC 1587 the observed ionised gas could be associated
with an interaction with the companion galaxy. 
Although, no cold gas has been reported in NGC 978 (see Figure \ref{fig:figures_NB_apend}), 
but the interaction debris gas may eventually enter the galaxy's (hot) halo to trigger the SF 
and feed the AGN activity. 
On the other hand, the SF in extranuclear regions (type ii galaxies) is likely a later stage 
after streams gas settling from orbiting satellites \citep{Jung2022}. 
\cite{Kolokythas2022} argued that S0 group-dominant galaxies (type ii galaxies) occupy X-ray 
faint systems and have point-like radio sources \citep{Kolokythas2018}, which indicates that 
the cold gas is more likely to be the result of gas-rich mergers or tidal interactions instead 
of cooling from a hot IGrM. 
While in some type i galaxies (as in the case of NGC 5846 where the ionised gas 
morphology supports a cooling flow) the cooled gas from the IGrM may be an effective mechanism 
forming filaments and rotating discs in the galaxy nuclei 
\citep[][and references therein]{OSullivan2015,Olivares2022}. 
3/4 galaxies (NGC 193, NGC 1060 and NGG 5846) with radio jet-like morphology 
present filamentary ionised gas structures characteristic of ICM cooling.
The presence of misaligned or counter-rotating ionised gas discs with
respect to the stellar body is a strong indication of external accretion of gas. 
We find direct evidence of this, \cite{Olivares2022} found that most of the galaxies 
in our sample have ionised gas kinematically decoupled from the stellar component, which suggests 
an external origin of the gas. 
Observational evidence \citep[e.g.,][]{Sarzi2006,Gomes2016} and simulations 
\citep[][among others]{Starkenburg2019,Khim2021,Jung2022} indicate possible origins
of this misalignment in galaxies regardless of the morphology and environment, with early-type
having higher misaligned fractions. We argue that stellar mass loss is unlikely to be 
the dominant source of cold gas in our sample \citep[see][]{Olivares2022}.

\begin{table}
\centering
\begin{minipage}{67mm}
\caption{Comparison between Z (this work) and IGrM metallicities in units
of Z/Z$_{\odot}$.
Column (1) average value between the AGN N2 and interpolated
abundances from the AGN models. Column (2) abundances in the extended regions from DIG/LI(N)ERs O3N2
calibrator. Column (3) corresponds to IGrM metallicity from O'Sullivan et al. (2017).}      
\label{table:abundances_IGrM}                              
\begin{tabular}{l c c c c c c c} 
\hline
            &  nuclear 3"   & extended     & IGrM \\
            & AGN           & DIG/LI(N)ERs   &      \\
            & (1)           & (2)          & (3)  \\
\hline 
NGC 193      & 1.15 & \dots & 0.68\\
NGC 410      & 0.94 & \dots & 0.42\\
NGC 584      & 1.08 & 1.35  & $\dots$\\
NGC 677      & 0.84 & \dots & 0.38 \\
NGC 777      & 1.41 &$\dots$& 0.63\\
NGC 924      & 0.89 & 0.85  & $\dots$\\
NGC 940      & 1.57 & 1.07  & 0.06\\
NGC 978      & 0.82 & \dots & >0.29\\
NGC 1060     & 1.08 & \dots & 0.28\\
NGC 1453     & 1.30 & 1.29  & 0.42\\
NGC 1587     & 0.91 & 1.05  & 0.03\\
NGC 4008     & 0.84 &$\dots$& 0.32\\
NGC 4169     & 1.38 & 1.35  & 0.11\\
NGC 4261     & 1.58 &$\dots$& 0.23\\
ESO0507-G025 & 0.88 & 0.98  & $\dots$\\
NGC 5846     & 1.12 & 1.32  & 0.27\\
NGC 6658     & 1.27 &$\dots$& <0.18\\
NGC 7619     & 1.67 &$\dots$& 0.54\\
\hline                                   
\end{tabular}
\end{minipage}
\end{table}

If we assume that the metallicity of the warm gas, in our sample, is represented by a single 
calibrator (AGN N2 or DIG/LI(N)ERs O3N2), on average, the nuclear regions are more metal rich 
than their extended structures, i.e., $\langle$(O/H)$_{nuclear}\rangle \gtrsim \langle$(O/H)$_{extended}\rangle$. 
However, in Section \ref{subsec:morphology_2DBPT} we find that the ionisation sources are
having different impacts at different radii. Therefore, the abundance in the nuclear regions is well
represented by the 3" apertures and they can be obtained as the average between the AGN N2 and the
interpolated AGN abundances from the models (Figure \ref{fig:figures_BPT_int_krabbe}). 
While for the extended regions the average abundances are from the data points in Section \ref{subsec:abundances} 
(see Table \ref{table:4}) assuming the DIG/LI(N)ERs O3N2 calibrator.
This is a reasonable assumption given that most of the spaxels in these regions are in the composite 
area of the BPT diagrams. In Table \ref{table:abundances_IGrM} we summarize and compare 
the gas-phase metallicities (Z = [O/H] =  $\langle$log(O/H)$\rangle$ - log(O/H)$_{\odot}$) found 
in this work with those in the IGrM by \cite{OSullivan2017}.
In the nuclear regions the metallicities range from $\sim$0.9 to 1.7 Z$_{\odot}$. 
The metallicity in the extended structures often rise to values 
approaching the solar $\sim$1.0 Z$_{\odot}$ or higher ($\lesssim$0.3 dex), while the IGrM 
has metallicities down to  $\sim$0.1-0.7 Z$_{\odot}$.
In the case of NGC 940, NGC 4261 and NGC 7619 the nuclear metallicities are $\gtrsim$0.6 dex higher than 
the solar value. Interestingly, in four cases (NGC 584, NGC 1587, ESO0507-G025 and NGC 5846) we found 
a drop in the nuclear metallicity with respect to the extended regions of $\lesssim$0.2 dex 
in NGC 584 and NGC 5846 and $\sim$0.1 dex for NGC 1587 and ESO0507-G025. 
This suggest the accretion of metal-poor gas to the central AGN \citep[e.g.,][]{doNascimento2022}.
Since metallicity in the nuclear regions represent the average metallicity within
the 3" apertures and the uncertainties on the abundances are of the order of $\sim$0.1 dex,
we find in our sample of group-dominant galaxies that $Z_{nuclear}\gtrsim$Z$_{extended}$>Z$_{IGrM}$.

The mixing and dispersion of heavy elements in the ISM of galaxies, in general, should follow 
the "evolutionary" stages of disc growth at different spatial scales.
It might be suggested, in our case, by a correlation between gas-phase metallicity gradients 
$\nabla_{\text{O/H}}$ and H$\alpha$+[N\,{\sc ii}] morphology (see Figure \ref{fig:type_gradient}), 
since the effect of gas flows over the lifetime of the galaxies seems to produce 
the flattening of abundances out to large radii \citep[e.g.,][]{Kewley2010,LopezSanchez2015,Sanchez2014}, 
following the formation of the extended structures.
However, we find no correlation between the metallicity gradients and morphology 
(Section \ref{subsec:properties_gas}) in our sample of  BGGs. This is in agreement 
with the idea of relatively short time scales for the radial dispersion and mixing of metals 
to large spatial scales likely produced by the AGN/SF-driven outflows, gas accretion 
and mergers/interactions. Furthermore, some of these metals will be transport by these mechanisms 
from the galaxies into the IGrM/ICM. In particular, group-dominant galaxies often host radio AGN that are 
interacting with the surrounding gas by forming cavities and shock fronts 
\citep[see][for a description of these structures in our sample]{Olivares2022}. 
As seen in Section \ref{sub:velocity} we observe large gas velocity dispersion in the central 
regions of the galaxies, likely associated with the presence of AGN activity.
Therefore, group-dominant galaxies likely acquired their cold gas as a consequence of several
possible mechanisms, i.e., gas-clouds or satellite mergers/accretion and cooling flows which 
together with the AGN/SF activity are likely contributing to the growth of the ionized gas structures 
and flattening the metallicity gradients. 

\section{Conclusions}\label{sec:conclusions}

In this paper, we present archival MUSE observations for a sample of 18 group-dominant galaxies
from the CLoGS sample \citep{OSullivan2017}. We derive and removed the stellar continuum for all
galaxies by fitting the stellar SEDs using the spectral synthesis code FADO
\citep{GomesPapaderos2017}. We studied the properties (i.e., emission line ratios, chemical
abundances, etc) and structure of the warm gas, in each galaxy, in order to constrain the ionisation
processes, the origin of their gas and its chemical abundance distribution. 
We summarise our main results as follows:

\begin{itemize}

\item
We used the continuum-subtracted H$\alpha$+[N \,{\sc ii}] images (see Figure \ref{fig:figures_NB}) 
to classify the galaxies into three morphological groups or types: 
\textit{type i0} - strong or diffuse nuclear emission with (or without) unextended filamentary 
($\lesssim$1 kpc) structures connected to the nuclear region, \textit{type i} - strong or diffuse 
nuclear emission with extended (several kpc) filamentary structures beyond the nuclear region and
\textit{type ii} - i0 or i plus extranuclear H\,{\sc ii} regions (well-defined or in distorted 
ring-like structures).
We find that 5/18 (NGC 410, NGC 978, NGC 4008, NGC 4261 and NGC 6658) are type i0, 
9/18 of these objects are type i (NGC 193, NGC 584, NGC 677, NGC 777, NGC 1060, NGC 1453, 
NGC 1587, NGC 5846 and NGC 7619) and 4/18 galaxies are type ii (NGC 924, NGC 940, NGC 4169 and 
ESO0507-G025). 

\item
In order to distinguish between different ionisation mechanisms, in Section \ref{subsec:emission_ratios} 
we used the following emission line ratios [O \,{\sc iii}]/H$\beta$ 
and [N\,{\sc ii}]/H$\alpha$ and the equivalent width EW(H$\alpha$). 
The spatially resolved log [N \,{\sc ii}]/H$\alpha$ ratios decreases as the extent of the
emission line region increases, indicating that the sources of the ionisation are
acting at different spatial scales. The same decreasing pattern with
the distance is observed for the velocity dispersion $\sigma$([N \,{\sc ii}]), 
the 12 + log(O/H) abundances and in most cases the EW(H$\alpha$). 
Using emission-line diagnostic diagrams (or BPT diagrams) we find that all galaxies 
in our sample have a dominant LINER/AGN nuclear region. 
Extended LINER-like regions are observed in most galaxies with filamentary structures. 
In the same section, we studied the mechanisms (pAGBs, AGN and X--ray emission) responsible 
for the ionisation which produce the optical emission lines in our sample.
Although, AGN, pAGBs and X--ray emission models are able to reproduce the observational data,
we suggest that central regions are more influenced by a low-luminosity AGN, while extended
regions are ionised by other mechanisms with pAGBs photoionisation likely contributing significantly 
as suggested by their EW(H$\alpha$) values.

\item
We calculated the gas-phase metallicity (12 + log(O/H)) using linear interpolations between the 
AGN, pAGBs and X-ray emission models \citep{Krabbe2021} and their measured emission line ratios
(log([O\,{\sc iii}]/H$\beta$ and log([N \,{\sc ii}]]/H$\alpha$)). We also used the AGN N2 \citep{Carvalho2020} and DIG/LI(N)ERs O3N2-based \citep{Kumari2019} calibrators. 
Using a single calibrator (AGN N2 or DIG/LI(N)ERs O3N2), the 12 + log(O/H) in the nuclear and extended 
regions (see Figure \ref{fig:figures_OH_profiles_pp}) are 
$\langle$(O/H)$_{nuclear}\rangle \gtrsim \langle$(O/H)$_{extended}\rangle$.
We found that the metallicity gradients for the pixel-by-pixel data points are, in most cases,
negative in the innermost regions with a flat gradient for the extended areas beyond the centre,
which includes extended structures and some star-forming regions. 
In this sense, the morphological H$\alpha$+[N \,{\sc ii}] types defined in this study indicate 
that group-dominant galaxies with extended filamentary structures 
(type i) and S0 galaxies with extranuclear SF regions (type ii), on average, have shallow
metallicity gradients. Therefore, extended regions and ring-like structures of ionised gas can be
considered chemically homogeneous (nearly solar) within the uncertainties. 
If the ionisation sources have different impacts at different radii 
(as seen in Section \ref{subsec:emission_ratios}) we use the AGN N2 calibrator and AGN models to estimate 
the nuclear (3" aperture) abundances and the DIG/LI(N)ERs O3N2  calibrator for the extended regions. 
Therefore, we found in NGC 584, NGC 1587, ESO0507-G025 and NGC 5846 a slight drop in the nuclear metallicity 
with respect to the extended regions, suggesting the accretion of metal-poor gas to the central regions.
However, we find within the uncertainties that $Z_{nuclear}\gtrsim$Z$_{extended}$>Z$_{IGrM}$.

\item
We suggest that group-dominant galaxies likely acquired their cold gas in the past as 
a consequence of one or more external mechanisms where gas-clouds or satellite mergers/accretion 
(Section \ref{subsec:HI_gas}) and cooling flows are likely supplying the gas for the growth 
of the ionised gas structures and AGN/SF activity. Our results favor a scenario in which metals are 
distributed across the ISM of the galaxies on short timescales.

\end{itemize}

\section*{Acknowledgements}

We thank the reviewer for his/her careful reading
of the manuscript and helpful comments which substantially improved the paper.
PL (contract DL57/2016/CP1364/CT0010) and TS (contract DL57/2016/CP1364/CT0009) are supported 
by national funds through Funda\c{c}\~ao para a Ci\^encia e Tecnologia (FCT) and 
the Centro de Astrof\'isica da Universidade do Porto (CAUP).
SIL and KK are supported in part by the National Research Foundation of South Africa 
(NRF Grant Numbers: 120850). Opinions, findings and conclusions or recommendations 
expressed in this publication is that of the author(s), and that the NRF accepts 
no liability whatsoever in this regard. EOS ackowledges support for this work from the National Aeronautics and Space Administration through XMM-Newton award 80NSSC19K1056.
AB acknowledges support from NSERC through its Discovery Grant Program.
PL thanks Polychronis Papaderos for his very useful comments.
We thank Angela Krabbe for providing us with the CLOUDY models used in this work. 

\section*{Data availability}\label{sec:data_availability}
The data underlying this article will be shared on reasonable request
to the corresponding author.




\clearpage

\appendix

\section{Other emission line galaxies in our MUSE datacubes}\label{emission_galaxies_apen}

Other emission line galaxies in our MUSE FoVs. 
We found two objects in the FoV of NGC 677 and one in the FoV of NGC 777, NGC 924 and NGC 1453.
In Figure \ref{fig:emission_galaxies} we show the position of these objects in the FoV. 
Using the H$\alpha$ and [N \,{\sc ii}]$\lambda$6584 emission lines we calculated: the redshifts, 
H$\alpha$ SFRs and 12 + log(O/H) abundances using the H\,{\sc ii} calibrator. 
In Table \ref{table:emission_galaxies} we summarize their main properties.
\clearpage

\begin{figure*}
\caption{Emission line galaxies found in the FoV of the galaxies NGC 677, NGC 777
NGC 924 and NGC 1453.}
\label{fig:emission_galaxies}
\end{figure*}

\begin{table}
\centering
\begin{minipage}{70mm}
\caption{Properties of the emission line galaxies detected in the field of our sample galaxies. $^{(a)}$ z calculated using H$\beta$.}      
\label{table:emission_galaxies}                              
\begin{tabular}{l c c c c} 
\hline
  FoV      & z  & SFR(H$\alpha$)        & 12 + log(O/H)\\
           &    &(M$_{\odot}$yr$^{-1}$) & N2 \\
\hline 

NGC 677    & \\
 R1        & 0.283658 & 0.0154$\pm$0.0001 & 8.50$\pm$0.09  \\
 R2        & 0.282776 & 0.0030$\pm$0.0010 & 8.53$\pm$0.17  \\
NGC 777    &   \\
 R1        & 0.232878 & 0.0046$\pm$0.0001 & 8.48$\pm$0.15  \\
NGC 924$^{(a)}$   &   \\
 R1        & 0.491319 & \dots & \dots \\
NGC 1453   &  \\
 R1        & 0.118373 & 0.0003$\pm$0.0001 & 8.56$\pm$0.15 \\
\hline                                   
\end{tabular}
\end{minipage}
\end{table}
\clearpage

\section{H$\alpha$+[N \,{\sc ii}]$\lambda\lambda$6548,6584 emission line maps}\label{HaNII_maps_apend}

\begin{figure*}
\includegraphics[width=0.4\textwidth]{NGC193_Emi.eps}
\includegraphics[width=0.4\textwidth]{NGC0410_Emi.eps}\\
\includegraphics[width=0.4\textwidth]{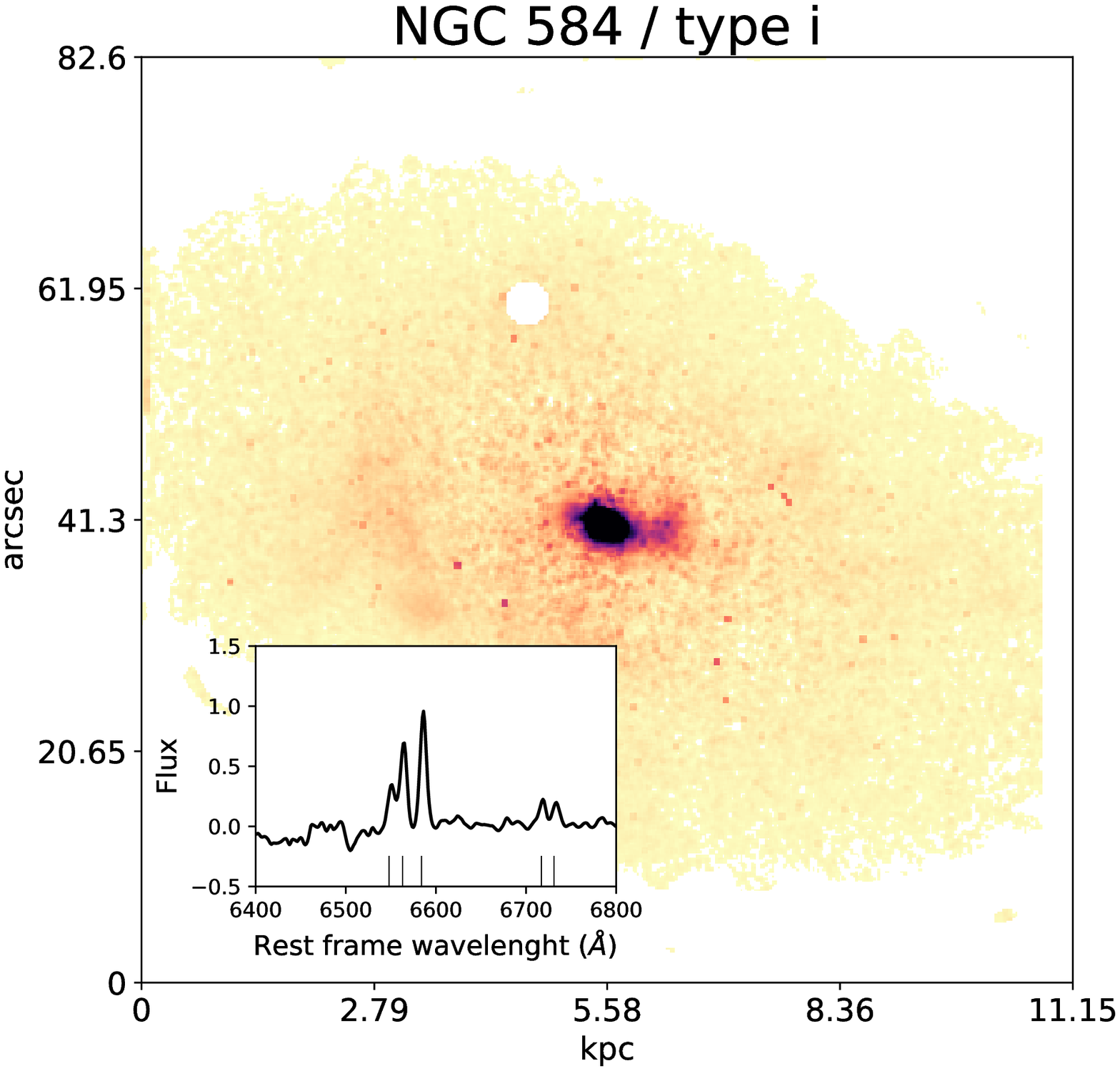}
\includegraphics[width=0.4\textwidth]{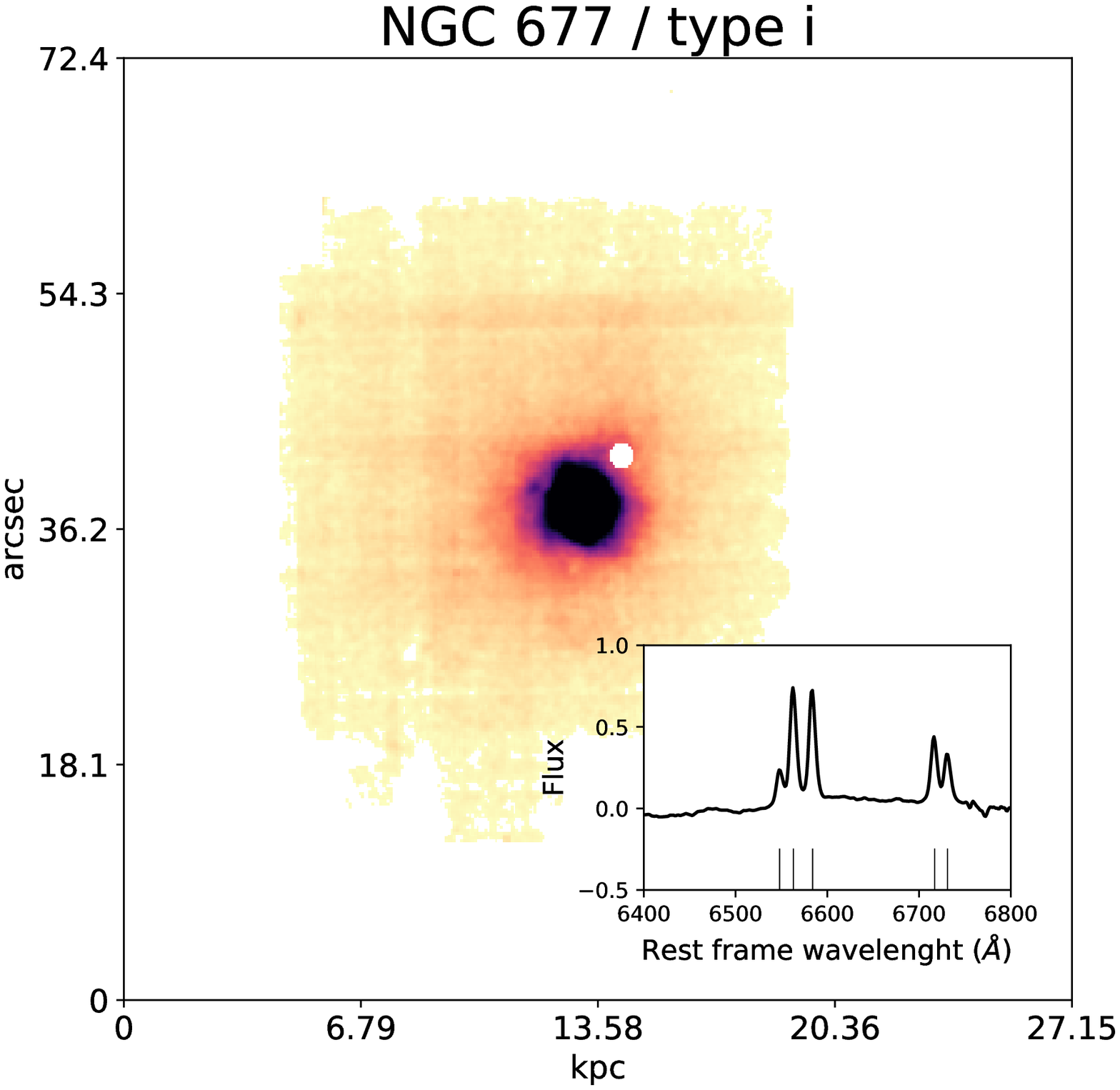}\\
\includegraphics[width=0.4\textwidth]{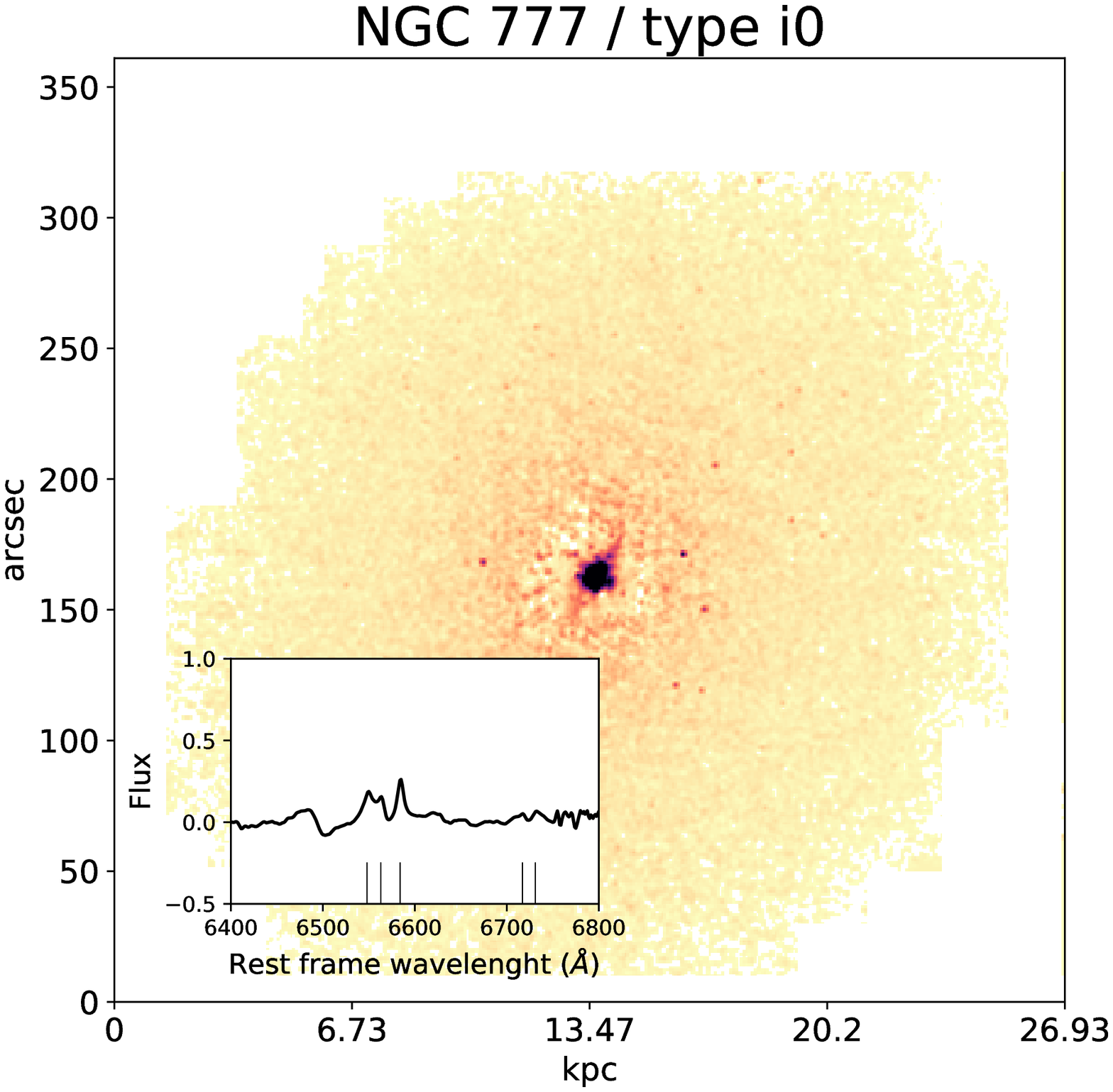}
\includegraphics[width=0.4\textwidth]{NGC0924_Emi.eps}\\
\caption{H$\alpha$+[N \,{\sc ii}]$\lambda\lambda$6548,6584 emission line maps from our sample.
We smoothed the emission line maps using a 3$\times$3 box filter. MUSE continuum-subtracted 
spectra from a nuclear 3" aperture covering the wavelength range from 6400 $\AA$ to 
6800 $\AA$ are shown in the inset panels. The vertical lines in the inset panels
indicate the wavelengths of the [N \,{\sc ii}]$\lambda$6548, H$\alpha$, 
[N \,{\sc ii}]$\lambda$6584, [S \,{\sc ii}]$\lambda$6717 and 
[S \,{\sc ii}]$\lambda$6731 emission lines. Fluxes in units of $\times$10$^{-15}$ 
    (erg s$^{-1}$ cm$^{-2}$ $\AA^{-1}$). North is to the top and East to the left.}
\label{fig:figures_NB_apend}
\end{figure*}

\begin{figure*}
\includegraphics[width=0.4\textwidth]{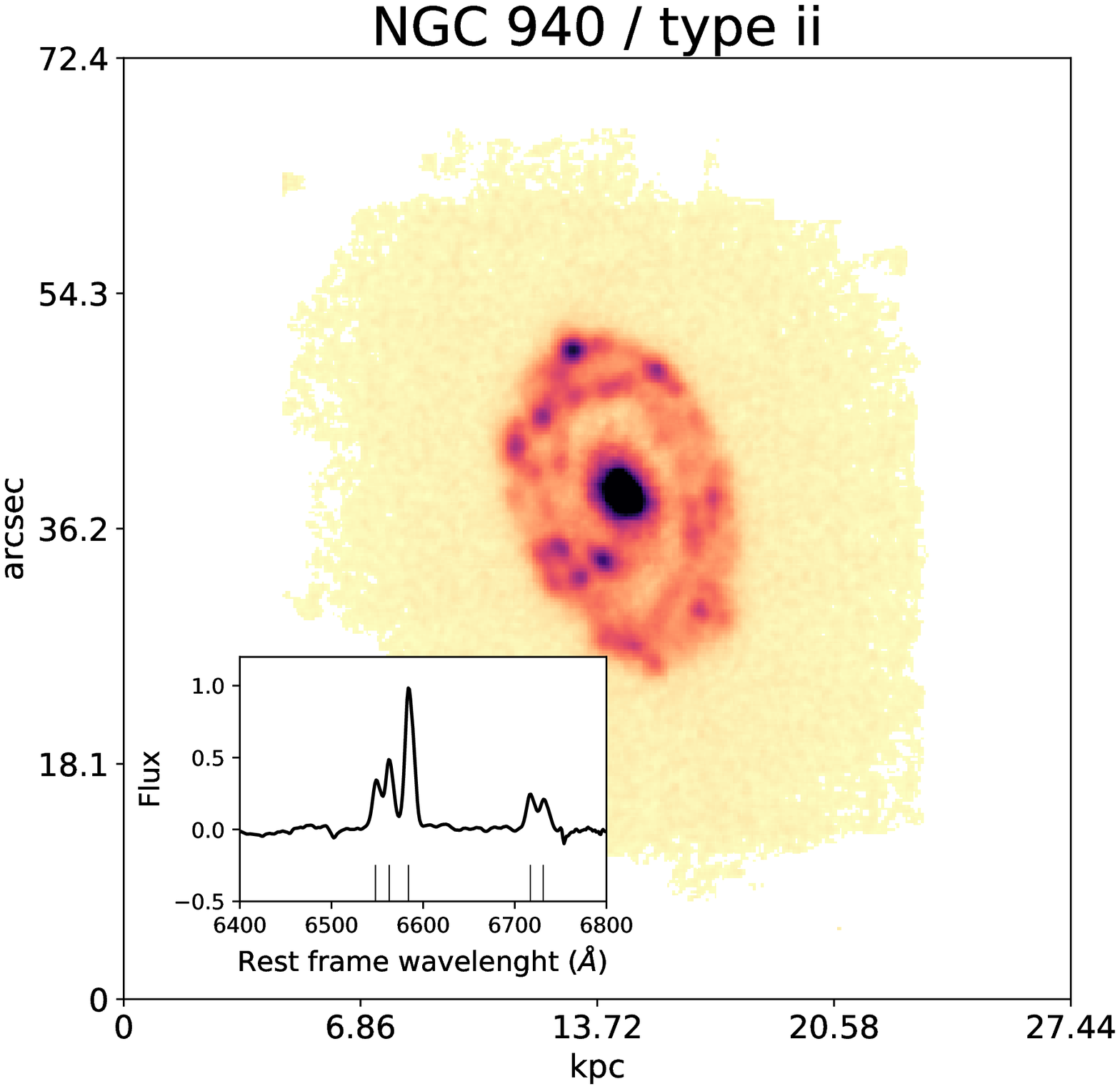}
\includegraphics[width=0.4\textwidth]{NGC0978_Emi.eps}\\
\includegraphics[width=0.4\textwidth]{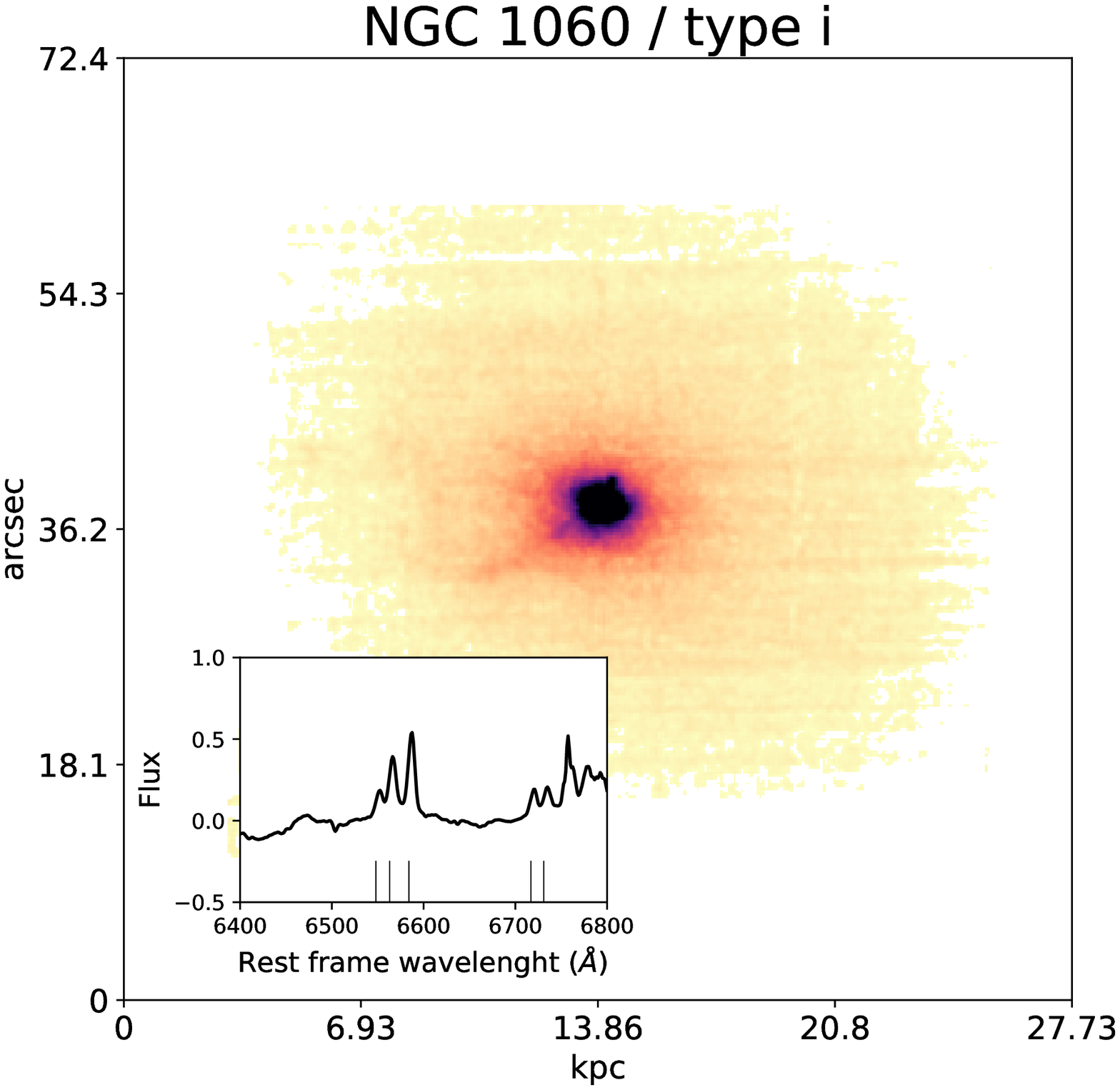}
\includegraphics[width=0.4\textwidth]{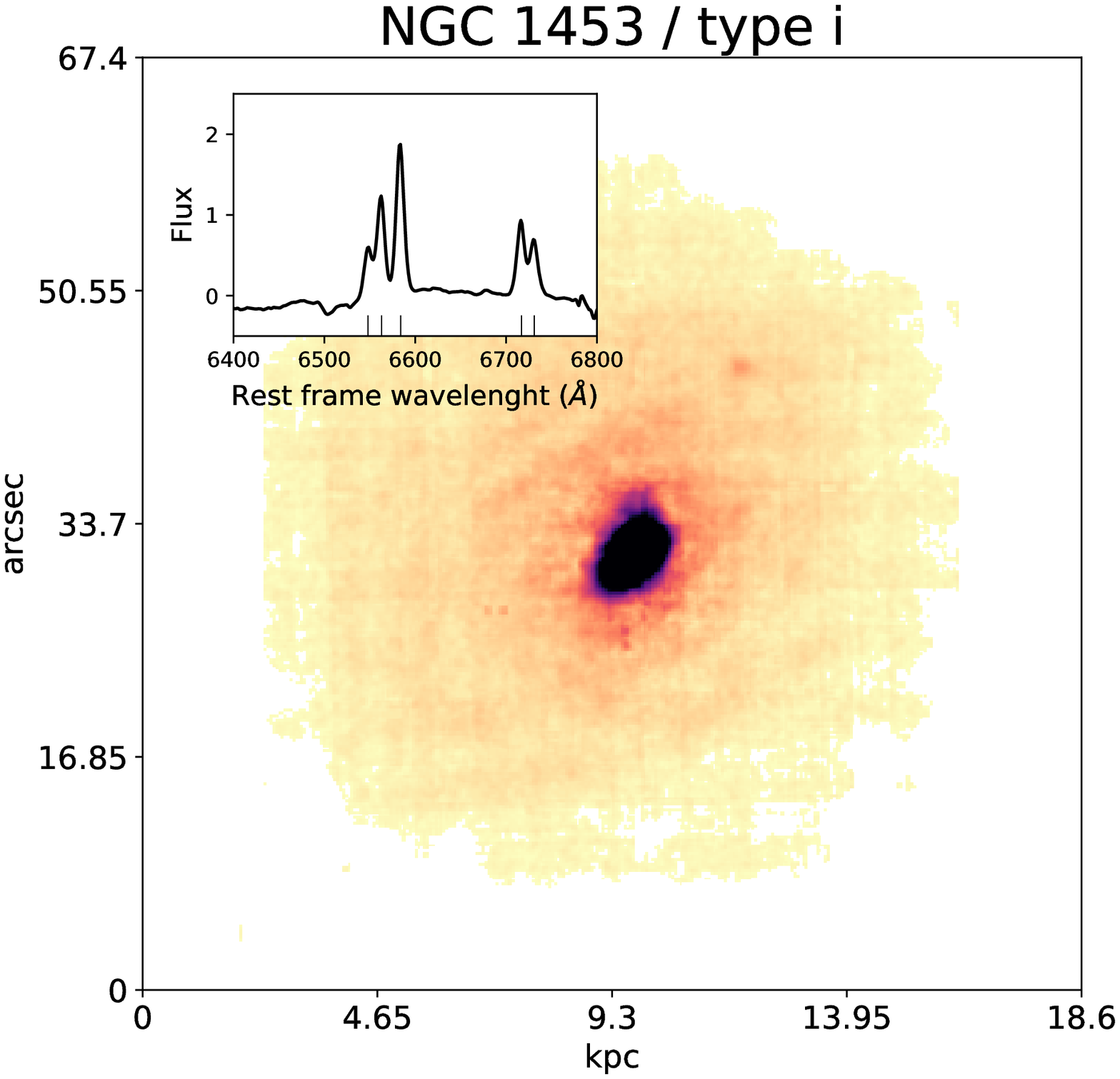}\\
\includegraphics[width=0.4\textwidth]{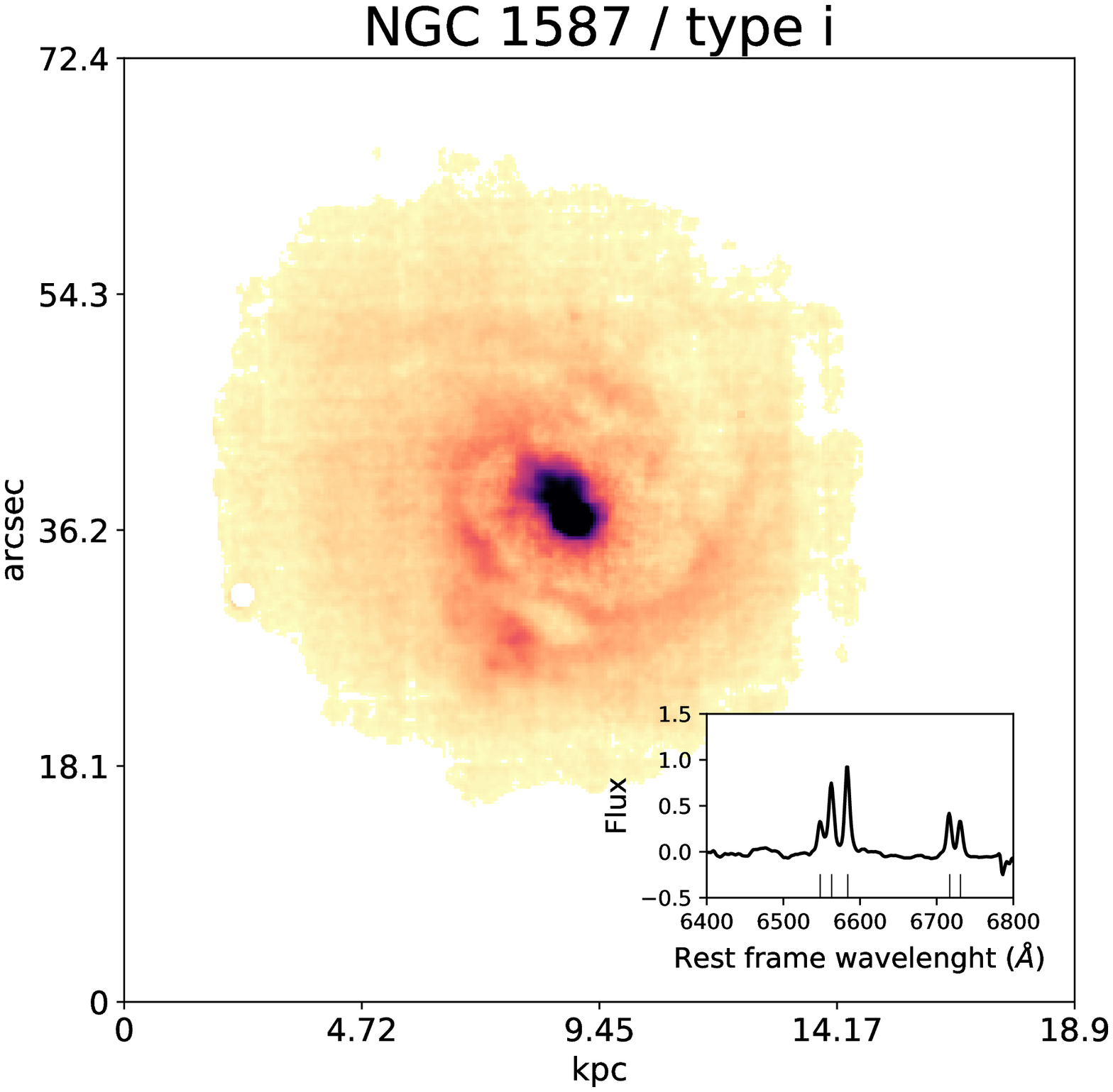}
\includegraphics[width=0.4\textwidth]{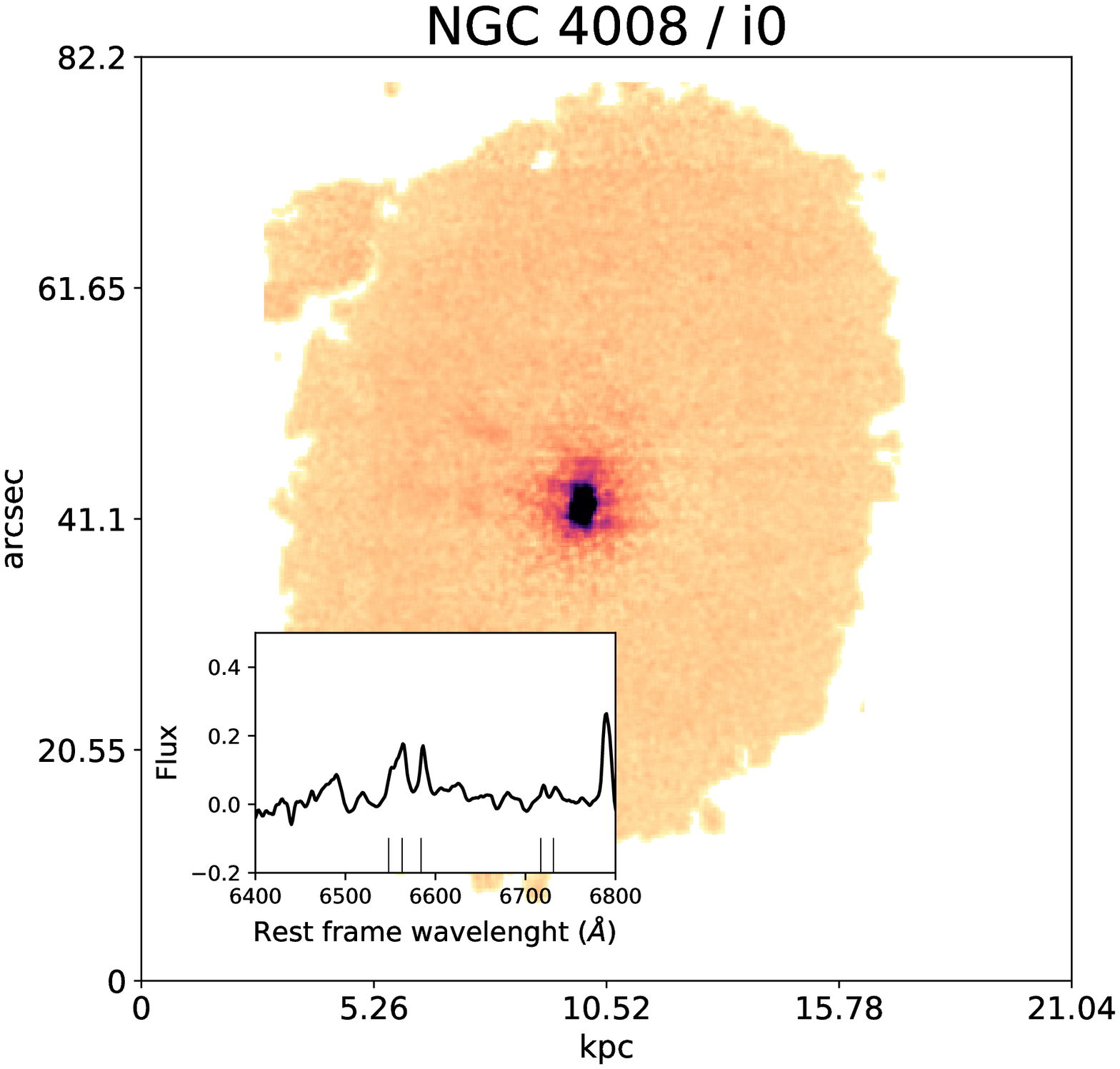}\\
\contcaption{}
\end{figure*}

\begin{figure*}
\includegraphics[width=0.4\textwidth]{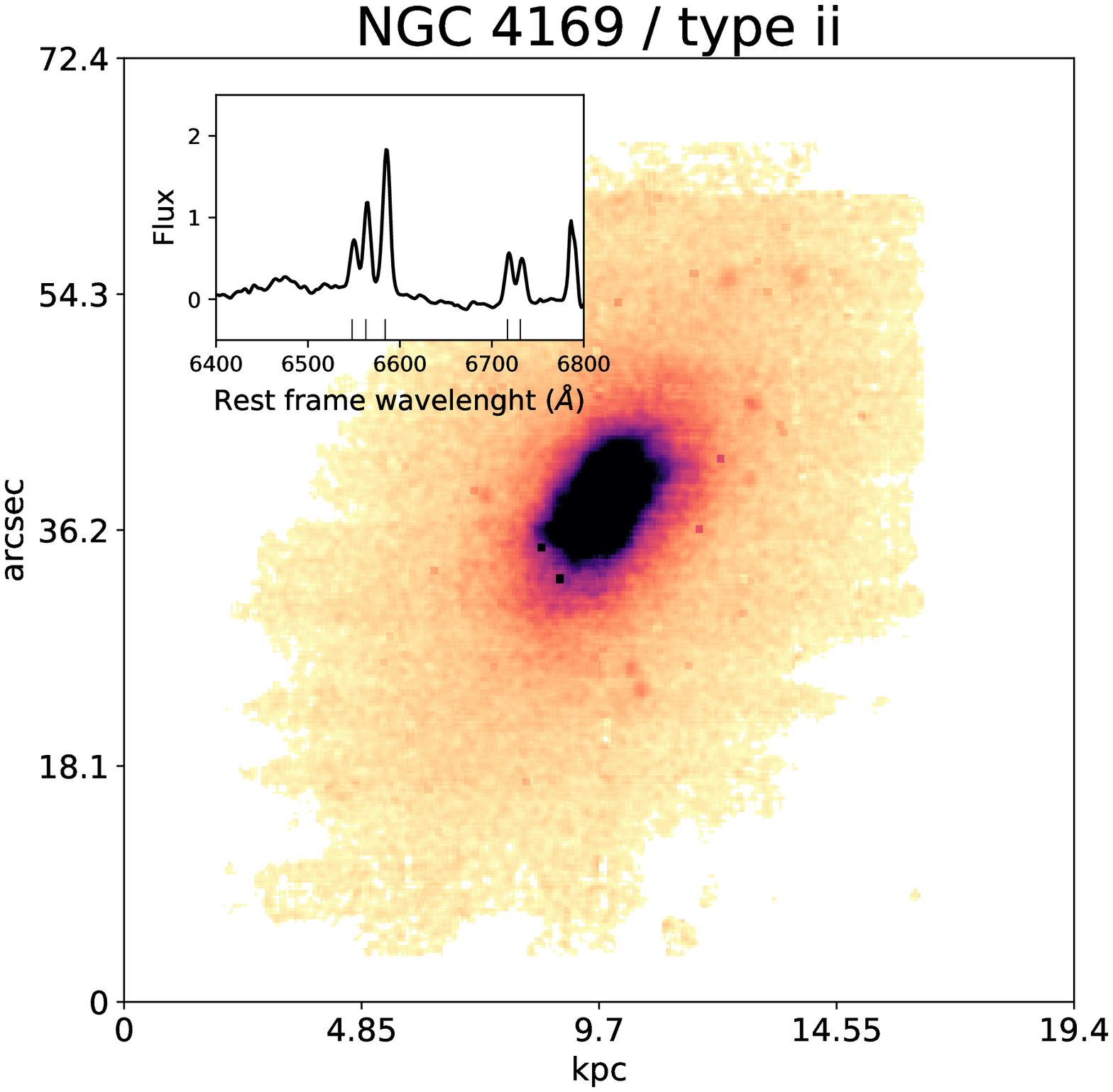}
\includegraphics[width=0.4\textwidth]{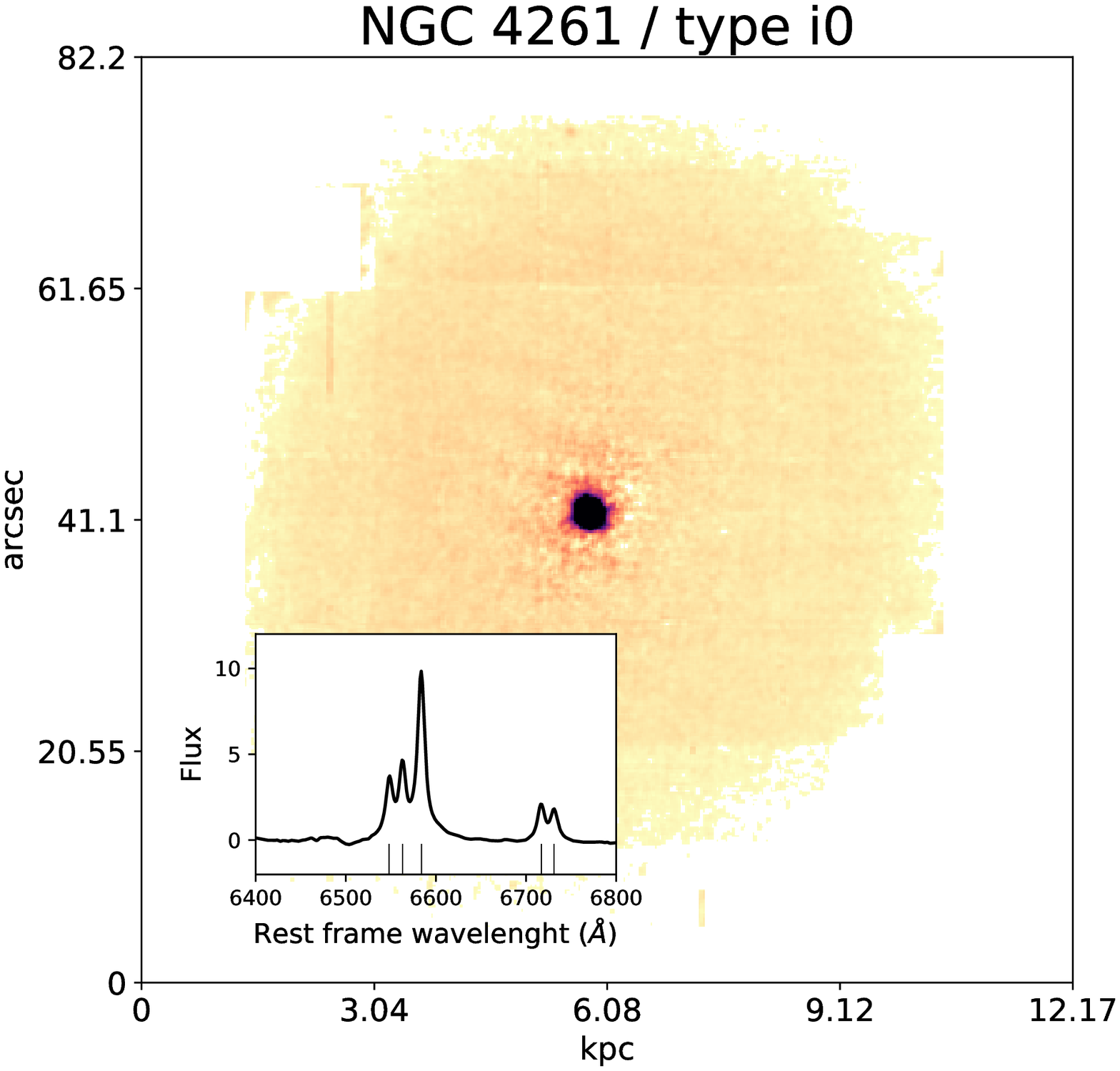}\\
\includegraphics[width=0.4\textwidth]{ESO0507_Emi.eps}
\includegraphics[width=0.4\textwidth]{NGC5846_Emi.eps}\\
\includegraphics[width=0.4\textwidth]{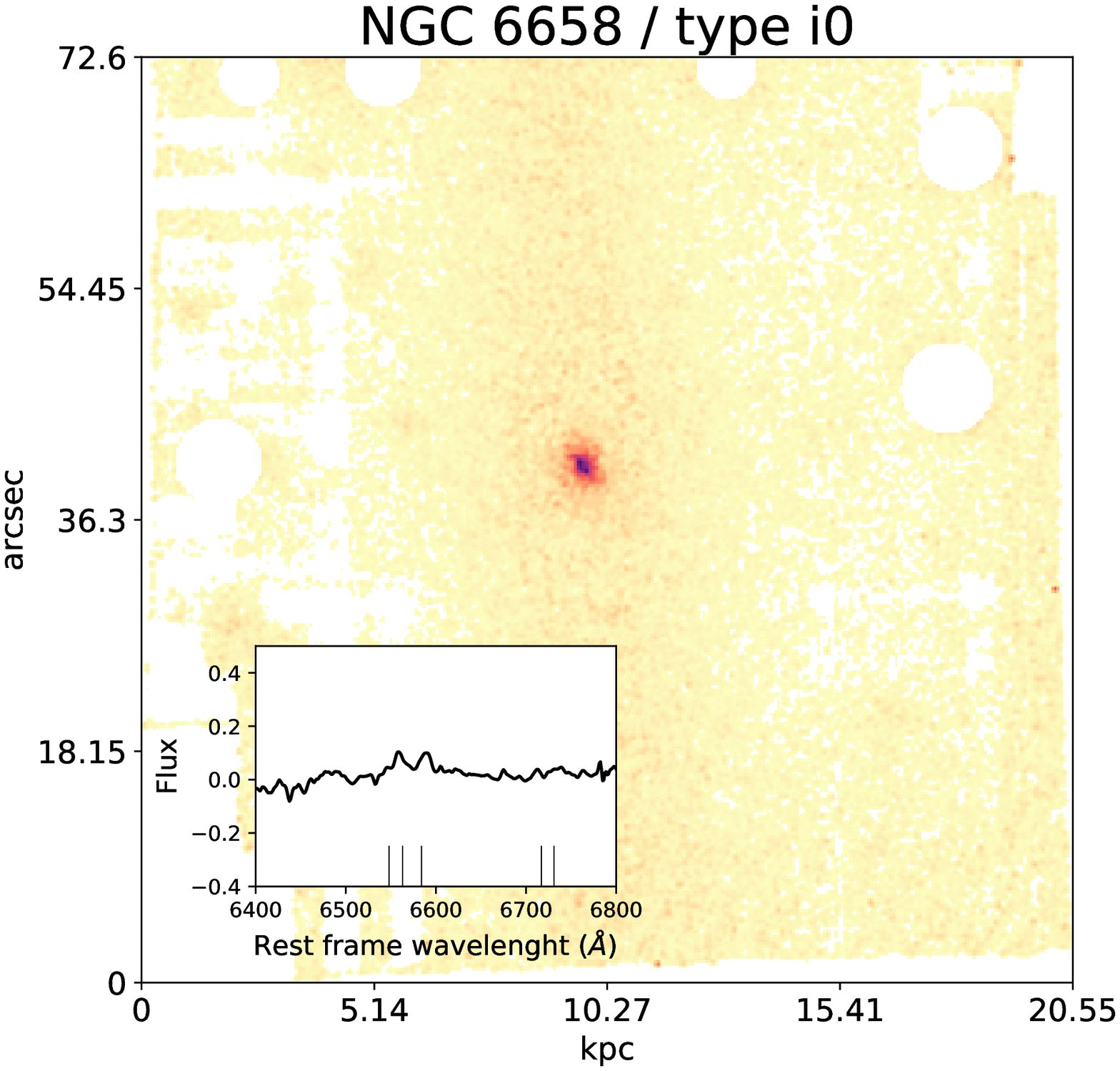}
\includegraphics[width=0.4\textwidth]{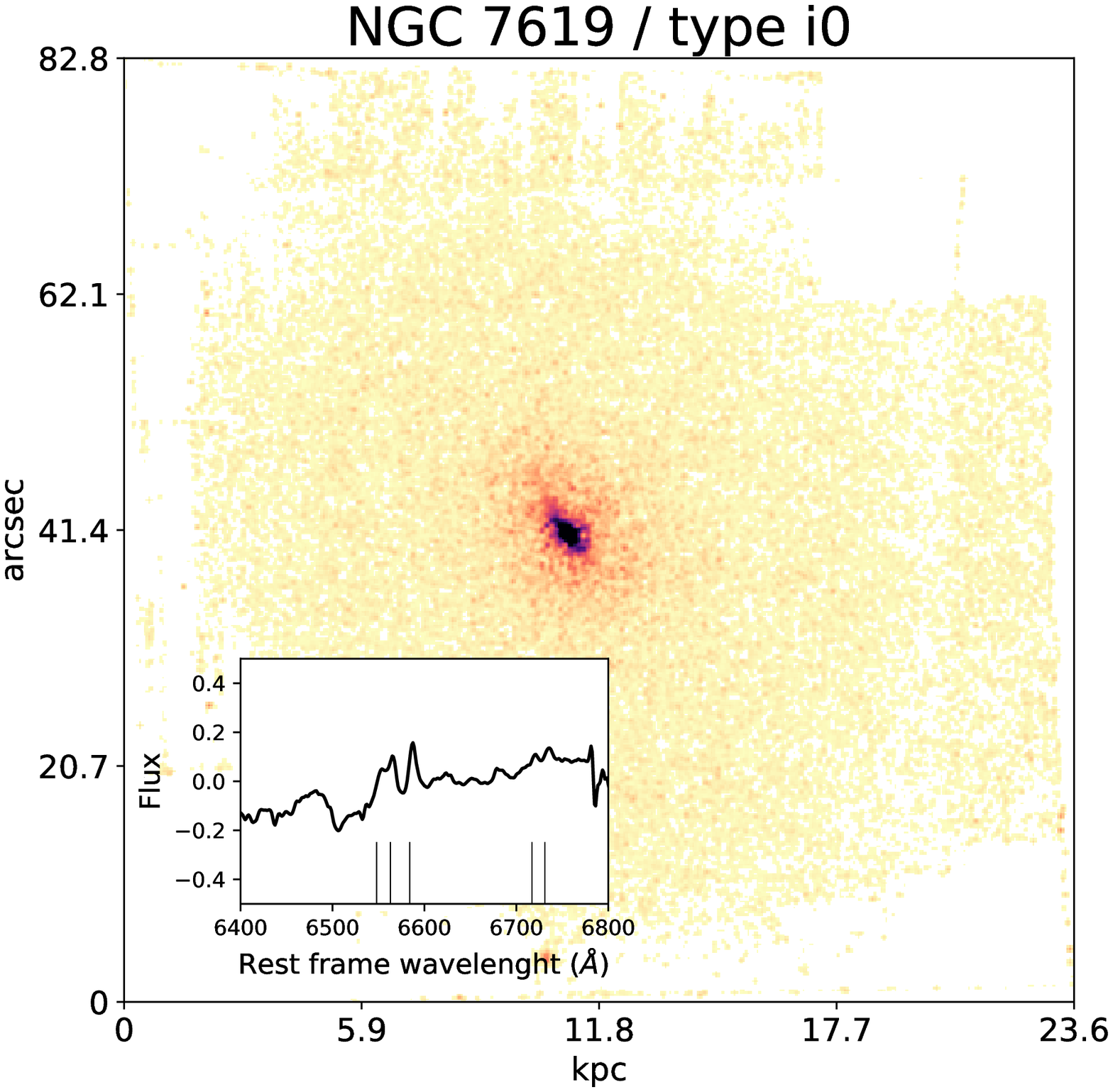}\\
\contcaption{}
\end{figure*}
\clearpage

\section{Emission line ratio maps and BPT diagrams of the galaxies}\label{Maps_sample}

\begin{figure*}
\caption{Emission line ratio maps log([N \,{\sc ii}]/H$\alpha$), 
log([O \,{\sc iii}]/H$\beta$ and log([O \,{\sc iii}]/H$\beta$ vs 
log([N \,{\sc ii}]/H$\alpha$) BPT diagram for emission lines with S/N $>$ 3. 
In the BPT diagram we include the Kewley et al. (2001; solid line) and Kauffmann et al. 
(2003; dashed line) lines to divide between regions dominated by H\,{\sc ii}
and AGN/LINERs. Filled black data points correspond to our measured values in the 3" aperture
enclosing the nuclear region.}
\label{fig:BPT_maps_1}
\end{figure*}

\begin{figure*}
\caption{Similar to Figure \ref{fig:BPT_maps_1}.}
\label{fig:BPT_maps_2}
\end{figure*}

\begin{figure*}
\caption{Similar to Figure \ref{fig:BPT_maps_1}.}
\label{fig:BPT_maps_3}
\end{figure*}

\begin{figure*}
\caption{Similar to Figure \ref{fig:BPT_maps_1}.}
\label{fig:BPT_maps_4}
\end{figure*}

\begin{figure*}
\caption{Similar to Figure \ref{fig:BPT_maps_1}.}
\label{fig:BPT_maps_5}
\end{figure*}

\begin{figure*}
\caption{Similar to Figure \ref{fig:BPT_maps_1}.}
\label{fig:BPT_maps_6}
\end{figure*}

\begin{figure*}
\caption{Similar to Figure \ref{fig:BPT_maps_1}.}
\label{fig:BPT_maps_7}
\end{figure*}

\begin{figure*}
\caption{Similar to Figure \ref{fig:BPT_maps_1}.}
\label{fig:BPT_maps_8}
\end{figure*}

\begin{figure*}
\caption{Similar to Figure \ref{fig:BPT_maps_1}.}
\label{fig:BPT_maps_9}
\end{figure*}

\begin{figure*}
\caption{Similar to Figure \ref{fig:BPT_maps_1}.}
\label{fig:BPT_maps_10}
\end{figure*}

\begin{figure*}
\label{fig:BPT_maps_11}
\end{figure*}

\begin{figure*}
\caption{Similar to Figure \ref{fig:BPT_maps_1}.}
\label{fig:BPT_maps_12}
\end{figure*}

\begin{figure*}
\caption{Similar to Figure \ref{fig:BPT_maps_1}.}
\label{fig:BPT_maps_13}
\end{figure*}

\begin{figure*}
\caption{Emission line ratio maps log([N \,{\sc ii}]/H$\alpha$) for galaxies with no H$\beta$
and/or [O\,{\sc iii}]$\lambda$5007 emission detection. North is to the top and East to the left.}
\label{fig:appendix_lNIIHa_maps}
\end{figure*}

\clearpage

\section{[N \,{\sc ii}]$\lambda$6584 velocity fields}\label{appendix_NII_velocity_maps}

\begin{figure*}
\caption{Radial velocity v([N\,{\sc ii}]) (left panel) and velocity dispersion 
    $\sigma$([N\,{\sc ii}]) (right panel) maps for our type ii galaxies NGC 924, NGC 940 and
    NGC 4169.  See also Figure \ref{fig:ESO0507_velocity} for ESO0507-G025. 
    The position of continuum maximum is indicated by an X symbol. North is to the top and East to the left.}
    \label{fig:appendix_NII_velocity_maps}
\end{figure*}
\clearpage

\section{[S \,{\sc ii}]$\lambda$6716 / [S \,{\sc ii}]$\lambda$6731 ratio maps}\label{appendix_SII_maps}

\begin{figure*}
\caption{[S \,{\sc ii}]$\lambda$6716 / [S \,{\sc ii}]$\lambda$6731 ratio maps. 
Using the temden IRAF STS package assuming t$_e$([O \,{\sc iii}])=10000 K, for reference, we find for [S \,{\sc ii}]$\lambda$6716 / [S \,{\sc ii}]$\lambda$6731 = 1.43, 1.32, 1.00, 0.51 and 0.46 an electron density n$_e$(S \,{\sc ii}) = $\sim$2, $\sim$100, $\sim$606, 10910 and 48893 cm$^{-3}$, respectively. North is up and east is to the left.}
    \label{fig:density}
\end{figure*}

\begin{figure*}
\contcaption{}
\end{figure*}
\clearpage

\section{Results of the linear fitting}\label{N2_fitting_apen}

\begin{landscape}
\begin{table}
\begin{minipage}{205mm}
\caption{Results of the linear fitting  of Figure \ref{fig:figures_OH_profiles_pp}
and statistics for all data points or spaxels using the H\,{\sc ii} N2 (Marino et al. 2013), 
AGN N2 (Carvalho et al. 2020) and AGN/LI(N)ERs O3N2 (Kumari et al. 2019) methods, respectively. 
The slope (metallicity gradient) from the linear fitting is indicated by $\nabla_{\text{O/H}}$.}      
\label{table:4}      
\centering                         
\begin{tabular}{l c c c c c c c c c c} 
\hline
             & \multicolumn{4}{c}{nuclear region} &   \multicolumn{4}{c}{extended region} &\\
Name         & intercept & slope $\nabla_{\text{(O/H)}}$ & mean & sd & intercept & slope $\nabla_{\text{(O/H)}}$ & mean & sd\\
             &           & (dex/arcsec)                              &      &  & & (dex/arcsec)   \\
\hline 

NGC 193      &  8.87/8.83/8.83  & -0.022/-0.026/-0.009 & 8.80/8.74/8.81  & 0.09/0.11/0.04  &  \dots/\dots/\dots &  \dots/\dots/\dots &  \dots/\dots/\dots &  \dots/\dots/\dots\\
NGC 410      &  8.87/8.82/\dots & -0.034/-0.027/\dots  & 8.80/8.76/\dots & 0.10/0.12/\dots &  \dots/\dots/\dots &  \dots/\dots/\dots &  \dots/\dots/\dots &  \dots/\dots/\dots\\
NGC 584      &  8.84/8.78/8.81 & -0.036/-0.042/-0.009 & 8.77/8.69/8.85 & 0.13/0.18/0.05 & 8.69/8.63/8.78 & 0.0030/0.0016/-0.0002 & 8.77/8.69/8.82 & 0.12/0.17/0.08 \\
NGC 677      &  8.76/8.68/8.80& -0.008/-0.008/-0.008 & 8.73/8.65/8.78 & 0.10/0.12/0.02 &  \dots/\dots/\dots &  \dots/\dots/\dots &  \dots/\dots/\dots &  \dots/\dots/\dots \\
NGC 777      &  8.85/8.81/\dots & -0.030/-0.027/\dots & 8.83/8.78/\dots  & 0.12/0.15/\dots &  \dots/\dots/\dots &  \dots/\dots/\dots &  \dots/\dots/\dots &  \dots/\dots/\dots\\
NGC 924      &  8.80/8.74/8.79 & -0.009/-0.009/-0.002 & 8.72/8.75/8.60 & 0.17/0.24/0.04 & 8.79/8.68/8.75 & -0.0060/-0.0025/0.0001 & 8.76/8.72/8.62 & 0.16/0.12/0.03  \\
NGC 940      &  8.97/8.96/8.85 & -0.055/-0.072/-0.017 & 8.60/8.51/8.70 & 0.04/0.06/0.02 & 8.63/8.53/8.82 & -0.0051/-0.0052/-0.0005 &  8.68/8.56/8.72 & 0.14/0.20/0.05 \\
NGC 978      & 8.75/8.66/8.79 & -0.005/-0.005/ -0.009 & 8.74/8.65/8.78 & 0.08/0.11/0.03 &  \dots/\dots/\dots &  \dots/\dots/\dots &  \dots/\dots/\dots &  \dots/\dots/\dots\\
NGC 1060     & 8.75/8.67/\dots &  -0.014/-0.006/\dots & 8.79/8.72/\dots & 0.05/0.06/\dots & 8.73/8.70/\dots & 0.0081/0.0010/\dots & 8.75/8.66/\dots & 0.04/0.06/\dots\\
NGC 1453     & 8.86/8.83/8.83 & -0.021/-0.028/-0.008 & 8.83/8.78/8.80 & 0.14/0.19/0.04 & 8.77/8.69/8.78 & -0.0011/0.0019/0.0013 & 8.80/8.74/8.80 & 0.13/0.18/0.04  \\
NGC 1587     &  8.79/8.72/8.78 & -0.016/-0.013/-0.004 & 8.75/8.67/8.72 & 0.08/0.11/0.04 & 8.74/8.73/8.80 & -0.0047/-0.0119/-0.0058 & 8.71/8.69/8.71 & 0.13/0.07/0.05 \\
NGC 4008     & 8.79/8.72/\dots & -0.017/-0.015/\dots & 8.76/8.69/\dots & 0.13/0.17/\dots &  \dots/\dots/\dots &  \dots/\dots/\dots &  \dots/\dots/\dots &  \dots/\dots/\dots\\
NGC 4169     &  8.88/8.85/8.83 & -0.042/-0.059/-0.006 & 8.71/8.61/8.78 & 0.05/0.07/0.03 &  8.81/8.74/8.81 & -0.0036/-0.0004/0.0004 & 8.74/8.69/8.82 & 0.2/0.25/0.06 \\
NGC 4261     & 8.90/8.88/8.82 & -0.010/-0.015/0.005 & 8.89/8.86/8.83 & 0.08/0.10/0.04 &  \dots/\dots/\dots &  \dots/\dots/\dots &  \dots/\dots/\dots &  \dots/\dots/\dots\\
ESO0507-G025 & 8.81/8.74/8.82 & -0.011/-0.014/-0.007 & 8.81/8.75/8.68 & 0.09/0.13/0.02 & 8.58/8.50/8.68 & -0.0010/-0.0019/-0.0001 &  8.74/8.66/8.68 & 0.06/0.09/0.01 \\
NGC 5846     &  8.91/8.89/8.85& -0.068/-0.094/-0.020 & 8.82/8.77/8.84 & 0.04/0.05/0.08 & 8.78/8.71/8.81 & 0.0002/0.0011/-0.0002 & 8.79/8.72/8.81 & 0.08/0.11/0.05 \\
NGC 6658     & 8.88/8.89/8.83 & -0.064/-0.105/-0.025 & 8.78/8.74/8.79 & 0.16/0.20/0.08 &  \dots/\dots/\dots &  \dots/\dots/\dots &  \dots/\dots/\dots &  \dots/\dots/\dots\\
NGC 7619     & 8.97/8.97/\dots & -0.065/-0.089/\dots & 8.88/8.85/ & 0.12/0.17/\dots &  \dots/\dots/\dots &  \dots/\dots/\dots &  \dots/\dots/\dots &  \dots/\dots/\dots \\
\hline                                   
\end{tabular}
\end{minipage}
\end{table}
\end{landscape}

\clearpage

\section{Cold gas content}\label{cold_gas_apen}

\begin{figure*}
\includegraphics[width=0.8\textwidth]{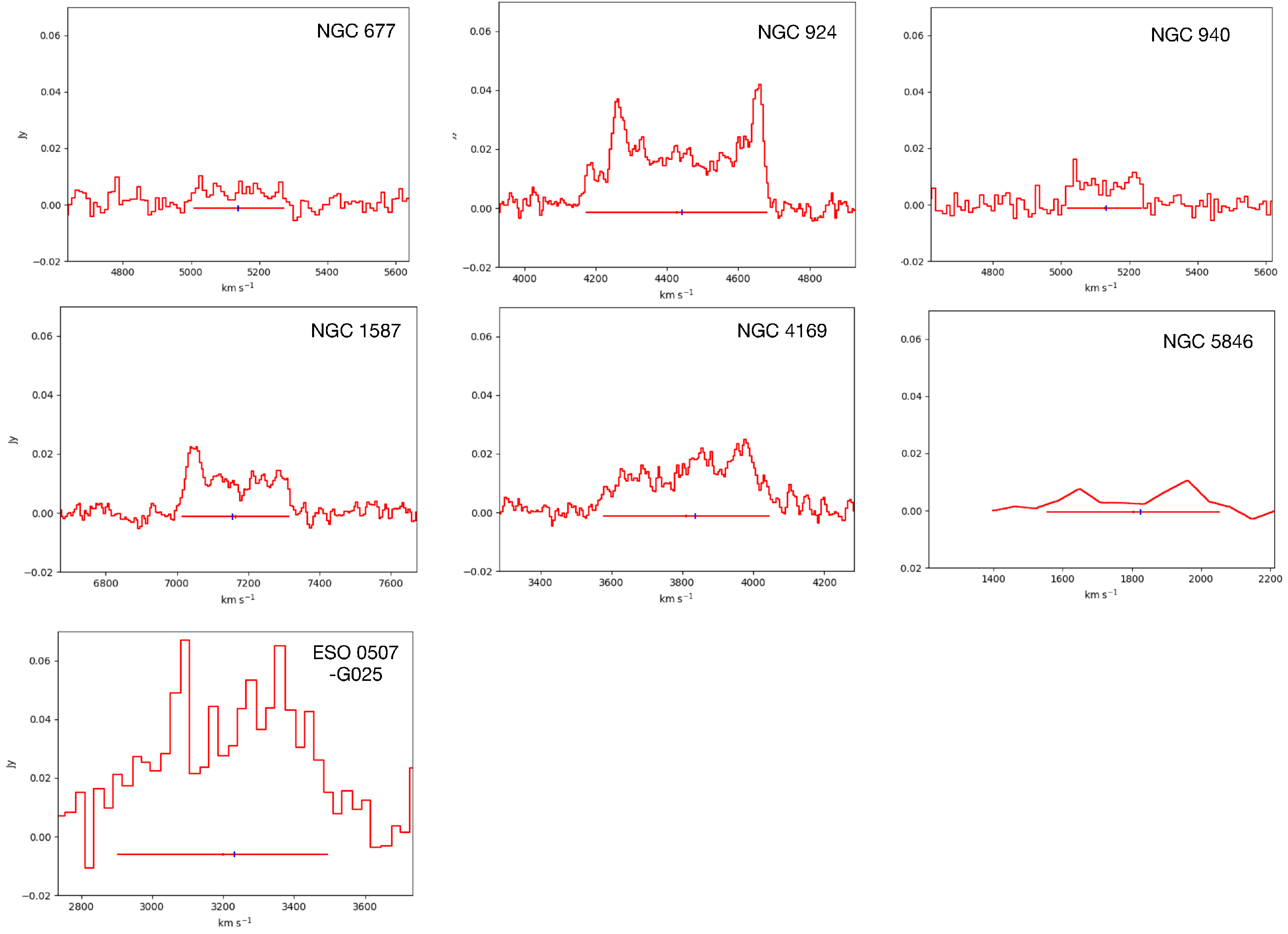}
    \caption{H\,{\sc i} profiles from NED with standardised x and y-axis. In the standardised
    plots, the NGC 677 and the ESO spectra velocities have been binned $\times$2.}
    \label{fig:figures_HI}
\end{figure*}

\begin{table*}
\begin{minipage}{180mm}
\caption{H\,{\sc i} and H$_2$ properties and M$_{*}$ available from the literature.}      
\label{table:properties_HI}      
\centering                        
\begin{tabular}{l c c c c c c c c c c c}
\hline
Name        &  H$\alpha$+[N\,{\sc ii}] & V(H\,{\sc i}) & W$_{20}$(H\,{\sc i}) & A flux(H\,{\sc i})\footnote{Asymmetry in the H\,{\sc i} profiles using the method from \cite{Espada2011}.} &   M(H\,{\sc i})\footnote{From H\,{\sc i} compilation in \cite{OSullivan2018}, although as noted by those authors H\,{\sc i} was not detected in NGC\,940 and NGC\,5846 during the ALFALFA survey as expected.}&M(H$_{2}$)\footnote{From \cite{OSullivan2015} and \cite{OSullivan2018}}& M$_{*}$\footnote{From \cite{Kolokythas2022}} & H\,{\sc i} spectrum \\
             & morphology & (km s$^{-1}$) & (km s$^{-1}$)&     & $\times$10${^9}$ M$_{\odot}$& $\times$10${^9}$ M$_{\odot}$& $\times$10${^{11}}$ M$_{\odot}$ & source\\
\hline
NGC 584     & i & \dots & \dots & \dots & 0.12& <0.01 & 2.13 & \dots \\
NGC 677     & i & 5138$\pm$6 & 272$\pm$12 & 1.10$\pm$0.07 & 1.70& <0.23 & 3.52 & \cite{Haynes2018}\\
NGC 924     & ii& 4428$\pm$5 & 509$\pm$10 & 1.07$\pm$0.08 & 9.12& 0.05 & 1.88 & \cite{Haynes2018}\\
NGC 940     & ii& 5127$\pm$4 & 218$\pm$8 & 1.02$\pm$0.08 & 9.14&6.10& 2.94 & \cite{Paturel2003}\\
NGC 1587     & i & 7163$\pm$1 &302$\pm$2& 1.29$\pm$0.04 & 2.51 & 0.23& 3.03 & \cite{Gallagher1981}\\
NGC 4169     & ii& 3811$\pm$7 & 470$\pm$13 & 1.56$\pm$0.07 & 10.71 &0.14& 1.27 & \cite{Haynes2018}\\
ESO0507-G025 & ii & 3248$\pm$8 & 450$\pm$16 & 1.26$\pm$0.51 &31.62 & 0.42 & 2.84 & \cite{Barnes2001}\\
NGC 5846     & i & 1804$\pm$14 &502$\pm$27 & 1.4$\pm$0.04  & 0.28& 0.01 & 3.39 & \cite{Bottinelli1979}\\
\hline                                  
\end{tabular}
\end{minipage}
\end{table*}


\bsp 
\label{lastpage}
\end{document}